\PassOptionsToPackage{dvipsnames}{xcolor} % xcolor used in aastex631; add this to fix package option-clash errors
\documentclass[twocolumn]{aastex631}

\usepackage{graphicx}	          % Including figure files
\usepackage{amsmath}	          % Advanced maths commands
\usepackage{upgreek}            % uptau
\usepackage{amssymb}	          % Extra maths symbols
\usepackage{bm}		              % Bold maths symbols, including upright Greek
\usepackage{CJKutf8}            % Chinese name
\usepackage[T1]{fontenc}        % for special char that needs \k command (e.g., Drazkowska)
\usepackage{comment}            % Comment big block
\usepackage{tabularx}           % Newer table package
\usepackage{threeparttable}     % Table notes smart width
\usepackage{booktabs}           % Better \hline
\usepackage{soul}               % better underline
\usepackage{ulem}               % better strikethrough
\usepackage{multirow}           % to enable multirow cells in tables
\usepackage{microtype}          % to disable ligatures -- don't like them
\DisableLigatures[f]{encoding = *, family = *} %% <- only disables f-ligatures
\usepackage[figuresleft]{rotating}  % to rotate figures
\allowdisplaybreaks             % allow page break in aligned equations 
\emergencystretch=\maxdimen     % to ease underfull/overfull lines.
\usepackage{tikz}

\usepackage{float}

\def\numx#1e#2{{#1}\mathrm{e}{#2}}

\defcitealias{zawadzki26}{ARKS III}
\defcitealias{marino26}{ARKS I}

\shorttitle{Debris disks, stirred and shaken}
\shortauthors{Chiang et al. (ARKS)}
\begin{document}

%\title{Vertical Density Profiles of Gravitationally Stirred Disks, with Application to Debris Disks}
%\title{Gravitationally Stirred Particle Disks and their Vertical Density Profiles, \\with Application to Debris Disks}
%\title{\red{The ALMA survey to Resolve exoKuiper belt Substructures (ARKS) XIII. \\Viscously Stirring Particle Disks into Lorentzians and Gaussians \\to Infer their Dynamical and Collisional Masses}}
\title{Viscously Stirring Particle Disks into Lorentzians and Gaussians \\ to Infer Dynamical and Collisional Masses  (ARKS XIII)}
%\title{Dynamical and Collisional Masses of Particle Disks:\\ Vertical Lorentzians and Gaussians from Viscous Stirring (ARKS XIII)}
%\title{Dynamical and Collisional Masses of Particle Disks \\via Vertical Lorentzians and Gaussians from Viscous Stirring (ARKS XIII)}

\author[0000-0002-6246-2310]{Eugene Chiang}
\affiliation{Department of Astronomy, University of California, Berkeley, 501 Campbell Hall, Berkeley CA 94720-3411, USA}
\affiliation{Department of Earth and Planetary Science, University of California, Berkeley, 307 McCone Hall, Berkeley, CA 94720-4767, USA}

\author[0000-0001-5653-5635]{Tim D.~Pearce}
\affiliation{Department of Physics, University of Warwick, Gibbet Hill Road, Coventry CV4 7AL, UK}

\author[0000-0001-6684-6269]{Marija R. Jankovic}
\affiliation{Institute of Physics Belgrade, University of Belgrade, Pregrevica
118, 11080 Belgrade, Serbia}

\author[0009-0003-3465-0271]{Alexander Jeffrey Backues}
\affiliation{Department of Physics, University of California, Berkeley, 366 Physics North MC 7300, Berkeley, CA 94720-7300, USA}
\affiliation{EECS Computer Science Division, University of California, Berkeley, 387 Soda Hall, Berkeley, CA 94720-1776, USA}

\author[0000-0002-2106-0403]{Yinuo Han}
\affiliation{Division of Geological and Planetary Sciences, California Institute of Technology, 1200 E. California Blvd., Pasadena, CA 91125, USA}

\author[0009-0009-4573-2612]{Alexander~V.~Krivov}
\affiliation{Astrophysikalisches Institut und Universit\"atssternwarte, Friedrich-Schiller-Universit\"at Jena, Schillerg\"a{\ss}chen 2-3, 07745 Jena, Germany}

\author{Margaret Pan}
\affiliation{Center for Astrophysics $\vert$ Harvard \& Smithsonian, 60 Garden Street, Cambridge, MA 02138, USA}

\author[0000-0001-9319-1296]{Brianna Zawadzki}
\affiliation{Department of Astronomy, Van Vleck Observatory, Wesleyan University, 96 Foss Hill Dr., Middletown, CT 06459, USA}

\author[0000-0002-4803-6200]{A.~Meredith Hughes}
\affiliation{Department of Astronomy, Van Vleck Observatory, Wesleyan University, 96 Foss Hill Dr., Middletown, CT 06459, USA}

\author{Krish Prakash Jhurani}
\affiliation{Department of Astronomy, University of California, Berkeley, 501 Campbell Hall, Berkeley CA 94720-3411, USA}
\affiliation{Department of Physics, University of California, Berkeley, 366 Physics North MC 7300, Berkeley, CA 94720-7300, USA}

\author[0000-0002-4248-5443]{Joshua B. Lovell}
\affiliation{Center for Astrophysics $\vert$ Harvard \& Smithsonian, 60 Garden Street, Cambridge, MA 02138, USA}

\author[0000-0002-5352-2924]{Sebasti\'an Marino}
\affiliation{Department of Physics and Astronomy, University of Exeter, Stocker
Road, Exeter EX4 4QL, UK}

\author[0000-0003-4623-1165]{Antranik A. Sefilian}
\affiliation{Department of Astronomy and Steward Observatory, University of Arizona, Tucson, AZ 85721, USA}

\author[0000-0003-1526-7587]{David J. Wilner}
\affiliation{Center for Astrophysics $\vert$ Harvard \& Smithsonian, 60 Garden Street, Cambridge, MA 02138, USA}

\author[0000-0001-9064-5598]{Mark C. Wyatt}
\affiliation{Institute of Astronomy, University of Cambridge, Madingley Road, Cambridge, CB3 0HA, UK}

\author[0000-0003-2953-755X]{Sebasti\'an P\'erez}
\affiliation{Departamento de F\'isica, Universidad de Santiago de Chile. Avenida Ecuador 3493, Estaci\'on Central, Santiago, Chile}
\affiliation{Millennium Nucleus on Young Exoplanets and their Moons (YEMS), Universidad de Santiago de Chile, Chile}
\affiliation{Center for Interdisciplinary Research in Astrophysics and Space Exploration (CIRAS), Universidad de Santiago de Chile, Chile}

\author[0000-0001-6015-646X]{P\'eter \'Abrah\'am}
\affiliation{Konkoly Observatory, HUN-REN Research Centre for Astronomy and Earth Sciences, MTA Centre of Excellence, Konkoly-Thege Mikl\'os \'ut 15-17, 1121 Budapest, Hungary}
\affiliation{Department of Astrophysics, University of Vienna, T\"urkenschanzstrasse 17, 1180 Vienna, Austria}

\author[0000-0001-7157-6275]{\'Agnes K\'osp\'al}
\affiliation{Konkoly Observatory, HUN-REN Research Centre for Astronomy and Earth Sciences, MTA Centre of Excellence, Konkoly-Thege Mikl\'os \'ut 15-17, 1121 Budapest, Hungary}

\author[0000-0002-1018-6203]{Patricia Luppe}
\affiliation{School of Physics, Trinity College Dublin, the University of Dublin,
College Green, Dublin 2, Ireland}

\correspondingauthor{E.~Chiang}
\email{echiang@astro.berkeley.edu}

%%%%%%%%%%%%%%%%%%%%%%%%%%%%%%%%%%%%%%%%%%%%%%%%%%%%%%%%%%%%%%%%%%%%%%%%%%%%%%%%
\begin{abstract}
Disks (Keplerian or otherwise, particulate or fluid) are often assumed to have densities that drop off vertically as Gaussians. Recent mm-wave imaging of circumstellar debris disks contradicts this assumption, revealing vertical profiles in dust that resemble Lorentzians. As part of the ARKS ALMA Large Program, we show how Lorentzians and Gaussians define an evolutionary sequence for disks of gravitationally scattering (viscously stirring) particles. When orbits are crossing and eccentricities $e \gg$ inclinations $i$, each scattering can change a particle's inclination by $\pm \,\Delta i \propto i$. A random walk with fixed steps in $\Delta i/i = \Delta \ln i$ produces a thick, log normal tail at large $i$ that leads to Lorentzian tails in density. This result holds independent of the origin of the large eccentricities, which may characterize either the stirrers or the objects being stirred; what matters is that relative motions parallel to the disk midplane are faster than perpendicular motions, and that vertical displacements during encounters are smaller than horizontal displacements. After enough scatterings, $i$ comes into equipartition with $e$,  $\Delta i$ stops exponentiating, and the vertical density relaxes to a Gaussian. We identify four regimes of dispersion-dominated viscous stirring, three of which are out-of-equipartition and where $i$ is stirred faster than $e$. 
%The sizes of perturbers needed to stir the $i$'s of Lorentzian, presumably out-of-equipartition disks from ARKS are smaller than equipartition predicts. 
The stirrers for ARKS debris disks may range from Pluto to a few times Mars in size, and be sufficiently few as to be collisionless. Or the stirrers may be even smaller, and be so numerous and collide so frequently that they source the collisional cascades that produce observable dust.
\end{abstract}


\section{Introduction}\label{sec:intro}
Astrophysical disks, particulate or fluid, form from objects that conserve their angular momentum and dissipate their energy, specifically their vertical kinetic energy (the energy in motions parallel to the angular momentum vector). A disk whose vertical kinetic energy damps to zero would have zero vertical thickness. In reality disks have non-zero thicknesses determined by processes that increase their vertical energy (e.g.~accretional/turbulent heating; irradiation; viscous stirring) and decrease it (e.g.~radiative loss; inelastic collisions between particles; dynamical friction cooling).

Disks, Keplerian and otherwise, are commonly assumed to have vertical density profiles that fall off as Gaussians with height above the midplane. Gaussian profiles characterize vertically isothermal gas disks (e.g.~\citealt{frank02}), and disks of solid particles whose mutual interactions (via collisions or gravitational scatterings) establish energy equipartition between random (epicyclic) motions parallel and perpendicular to the midplane (e.g.~\citealt{ida92}). Gaussians abound in nature, a consequence of the central limit theorem and statistical equilibrium. 

Surprisingly, \citet[][ARKS III]{zawadzki26} found that many of the brightest and youngest (10--1000 Myr old) debris disks in the mm-wave continuum exhibit vertical profiles with thick non-Gaussian tails. They made their observations as part of the ALMA survey to Resolve exoKuiper belt Substructures (ARKS; \citealt{marino26}, ARKS I) in circumstellar debris disks composed of optically thin dust. Debris disks are analogues of our Solar System's Kuiper belt, and ARKS is an ALMA (Atacama Large Millimeter Array) Large Program designed to image the brightest disks in the thermal continuum (0.8--1.4 mm wavelength) and $^{12}$CO and $^{13}$CO ($J$=3-2) at angular resolutions fine enough to resolve their heights and widths. When fitting the mm continuum visibilities of highly inclined disks, \citetalias{zawadzki26} found that  Lorentzian vertical density profiles, and in one case a double Gaussian profile, were preferred with high formal confidence over single Gaussians in 10 out of 13 disks. They employed parameterized three-dimensional disk models presumed capable of disentangling vertical from radial structures. \citet{ahmic09}, \cite{matra19}, and \citet{han26new} also reported vertical profiles broader than single Gaussians for the beta Pictoris debris disk.

The present paper, part of the ARKS series, seeks to understand the physical origin of vertical profiles that taper off more slowly than Gaussians. Non-Gaussian profiles were obtained in prior theoretical work by \citet{sefilian25} who examined secular vertical forcing/warping of a radially extended disk by a planet, accounting for the secular self-gravitational potential of the disk. Here, starting from first principles,  we consider the non-secular and radially local problem. We study how ``small bodies'' (the collisional progenitors of the presumed mm-sized particles responsible for the ALMA emission) are gravitationally stirred by one or more ``big bodies'' on crossing orbits, stochastically on close encounter (synodic) timescales. We will see how such ``viscous stirring'' produces broad Lorentzian vertical profiles within a narrow disk annulus, and how Lorentzians and Gaussians bracket an evolutionary continuum. The dynamics is simple, local, and does not require warping or varying disk aspect ratios with radius.

Throughout this paper, ``Lorentzian'' really means ``Lorentzian-like'', as vertical profiles must eventually deviate from strict Lorentzians far above the midplane. Our work does not rigorously prove a connection with Lorentzians; we only show that, under certain physical conditions, Lorentzians offer superior fits to numerically computed profiles, as compared to Gaussians. The same is true for the ALMA observations --- for certain disks, the data prefer Lorentzians over Gaussians, but do not rule out other functional forms. We will usually drop the modifier ``-like'' for convenience.

\subsection{Origin of Rayleigh inclination distribution and Gaussian vertical distribution}
To lay the groundwork for our analysis, recall how Gaussian vertical profiles arise in a disk of particles that are stochastically perturbed. Imparting random impulses to a particle's vertical velocity causes its angular momentum vector to random walk. The unit angular momentum vector random walks in 2D on the surface of the unit sphere; both the orbital inclination $i$ of the particle and its longitude of ascending node $\Omega$ change randomly with every ``kick'' in vertical velocity.

Consider an ensemble of particles that start with zero inclination, and have each particle take randomly directed steps in the 2D space $(p,q) = (i \cos \Omega, i \sin \Omega)$, with each step $(\Delta p, \Delta q)$ of fixed length $|\Delta i| = \sqrt{(\Delta p)^2 + (\Delta q)^2}$. The assumption of a strictly fixed length is made for simplicity and is not necessary for deriving a Gaussian profile; $|\Delta i|$ can be drawn from any probability distribution, as long as it is of finite variance and is independent of $i$. We will see later how $|\Delta i|$ is independent of $i$ when the particle inclinations $i$ and eccentricities $e$ are in equipartition, i.e.~when $i \simeq e/2$ (e.g.~\citealt{ida92,ida93}).

\begin{figure}
\centering
\includegraphics[width=\linewidth]{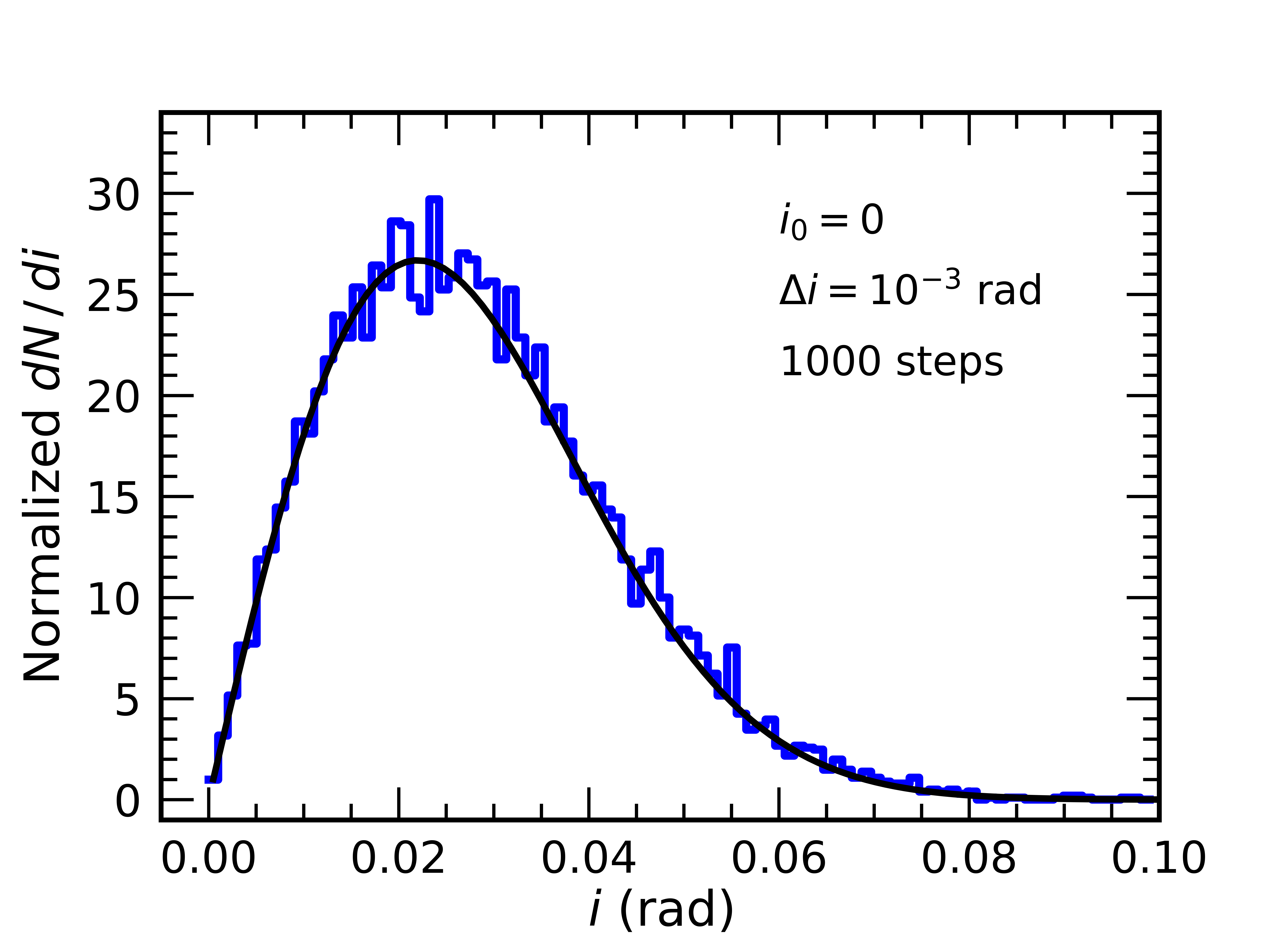}
  \caption{Fixed-length, randomly directed steps in $(p,q) \equiv (i \cos \Omega, i \sin \Omega)$ space, starting from the origin $i_0=0$, give rise to a Rayleigh distribution $\propto i \exp [-i^2 / (2\sigma_i^2)]$. Here 10000 particles each take 1000 random-walk steps of fixed length $\Delta i = \sqrt{(\Delta p)^2 + (\Delta q)^2} = 10^{-3}$ rad (normalized data in blue, best-fit Rayleigh distribution with $\sigma_i = 0.022$ in black).
  \label{fig:pq_dNdi}}
\end{figure}

Figure \ref{fig:pq_dNdi} shows the inclination distribution $dN/di$ of 10000 particles after they have taken $S=1000$ random-walk steps, each of length $|\Delta i| = 10^{-3}$ rad. The inclination distribution conforms to a Rayleigh distribution $\propto i \exp[-i^2/(2\sigma_i^2)]$ with $\sigma_i \sim \sqrt{S} |\Delta i|$. The Rayleigh distribution is what the central limit theorem predicts for a random walk on the surface of a sphere: the Rayleigh distribution is the product of a Gaussian and $i$ (technically $\sin i$), with the latter factor reflecting the available phase-space area near the polar cap of the unit sphere at small $i$.

\begin{figure}
\centering
\includegraphics[width=\linewidth]{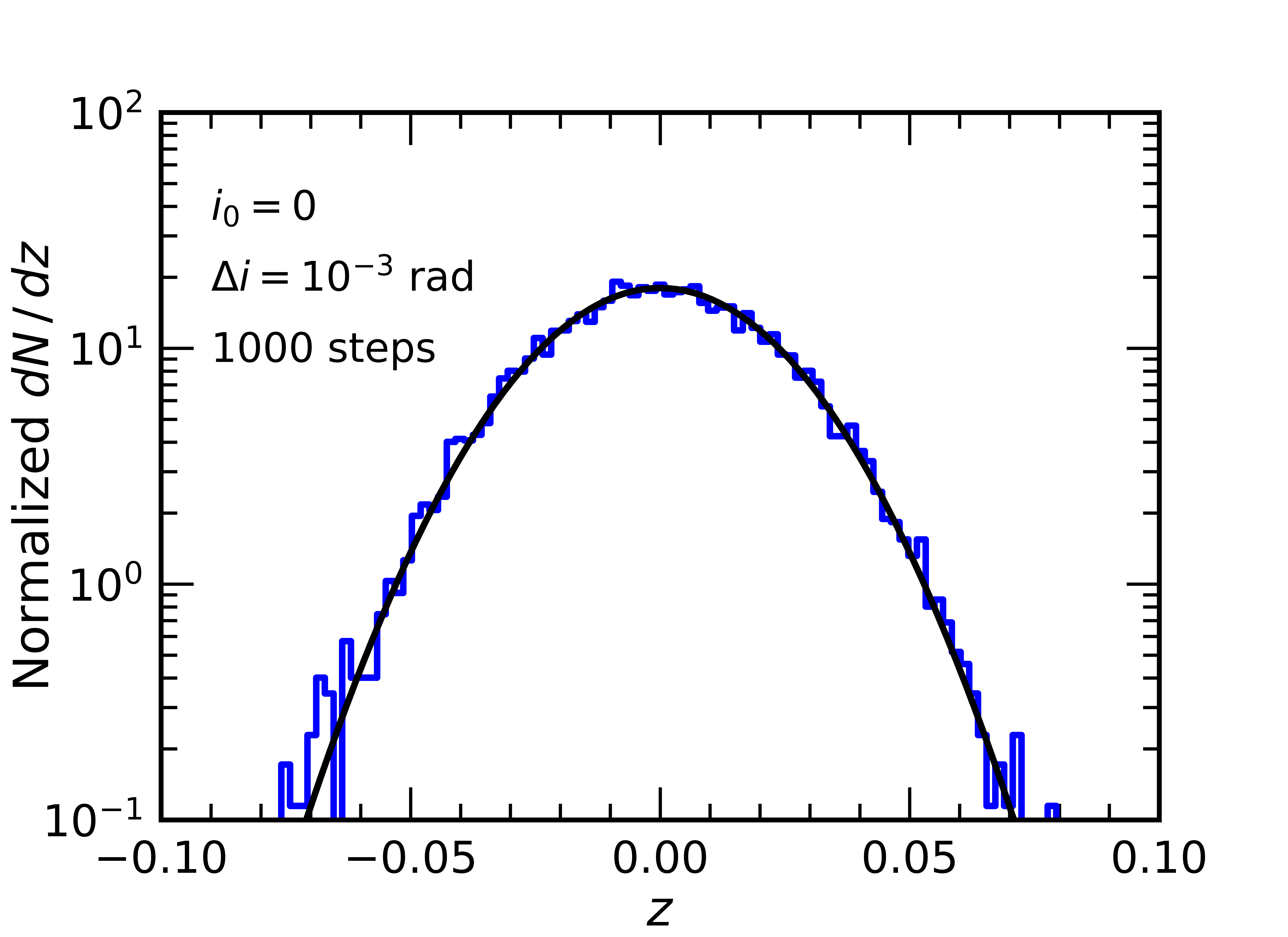} \caption{A Rayleigh distribution for $dN/di$ leads to a Gaussian distribution for the vertical density profile $dN/dz$. Here the 10000 $i$'s from Fig.~\ref{fig:pq_dNdi} are converted to vertical height $z = a \sin i \sin u$, where $a = 1$ and $u$ is drawn randomly as a uniform variate from 0 to $2\pi$, to generate $dN/dz$ (normalized data in blue, best-fit Gaussian with $\sigma_z = 0.022$ in black).
  \label{fig:pq_vert_profile}}
\end{figure}

Now convert each particle's $i$ to a height above the midplane $z = a \sin i \sin u$, assuming for simplicity zero eccentricity, semimajor axis $a = 1$ (arbitrary units), and $u$ (a particle's angle from the ascending node, a.k.a.~the argument of latitude) drawn randomly from a uniform variate $[0,2\pi)$. The resultant vertical density profile $dN/dz$, computed by Monte Carlo sampling, is shown in Figure \ref{fig:pq_vert_profile}. It is a Gaussian with $\sigma_z \sim a \sigma_i$. The same result is obtained analytically by \citet{matra19} by evaluating the sky-projection integral.

\subsection{Plan of this paper}
In the remainder of this paper we will see how the dynamics of viscous stirring can lead to a Lorentzian vertical profile as well as a Gaussian. The underlying idea is simple: the random-walk steps $|\Delta i|$ are not necessarily of fixed length, but can grow in proportion to $i$ in certain out-of-equipartition regimes where $i \ll e$. This exponentiation of the step size leads to broad non-Gaussian tails. Section \ref{sec:theory} details our semi-analytic (often order-of-magnitude) reasoning. Section \ref{sec:num} explores viscous stirring numerically with an $N$-body integrator. Section \ref{sec:app} applies our theory to Gaussian and Lorentzian disks from \citetalias{zawadzki26} to estimate how much  mass they contain in big body stirrers. The results in Section \ref{sec:app} complement and are informed by a parallel ARKS paper by Jankovic et al.~(2026, ARKS XII, submitted). Section \ref{sec:sum} summarizes and provides an outlook.

\vspace{0.2in}
\section{Theory of vertical stirring}\label{sec:theory}

We study how small bodies (test particles) are gravitationally stirred by big bodies, focussing on their vertical dynamics. We work in the dispersion-dominated (a.k.a.~super-Hill) regime where small and big bodies cross orbits, and consider both equipartition conditions when eccentricities are comparable to inclinations (rms eccentricity = $2 \times$ rms inclination; e.g.~\citealt{ida92}),
%``isotropic'' conditions when inclinations are comparable to eccentricities (the standard assumption), 
and out-of-equipartition (anisotropic) conditions when inclinations $\ll$ eccentricities. Much of our reasoning in this section may be extended to the non-crossing, a.k.a.~shear-dominated or sub-Hill case, as discussed in Appendix \ref{sec:appendix}. The shear-dominated case is less important than the dispersion-dominated case for real-life debris disks, whose present-day properties are inconsistent with shear-dominated self-stirring (Jankovic et al.~2026).

Also for simplicity in this section, we take big bodies to be on circular orbits in the reference plane. Many of our findings will still apply when big bodies are on eccentric orbits. What matters are the relative motions between big and small bodies in and out of the reference plane, i.e.~horizontal vs.~vertical relative velocities. Vertical relative velocities depend on mutual inclinations $i$. Horizontal relative motions depend on ``mutual'' eccentricities --- horizontal velocities can be large because small body eccentricities are large, or because big body eccentricities are large, or both. Thus the eccentricity $e$ below, though formally referring to the small bodies, may be reinterpreted to include the eccentricities of both big and small bodies --- roughly as their quadrature sum, assuming orbit orientations are uncorrelated. We will relax the assumption of zero big body eccentricity 
in the numerical simulations of Section \ref{sec:num}.

\subsection{How inclination changes in an encounter}\label{subsec:how0}
After a small body undergoes a close encounter with a big body, their mutual inclination, assuming it is not initially zero, increases or decreases. The small body's inclination $i$ (relative to a reference plane which we take to coincide with the big body's orbit plane) changes with time according to Gauss's perturbation equation (e.g.~\citealt{murray99}, equation 2.157):
\begin{align}\label{eq:gauss}
\frac{di}{dt} = \frac{r N \cos (\omega + f)}{\ell}
\end{align}
where $r$ is the radial distance of the small body from the central mass $m_\star$, $\omega$ is the argument of periapse, $f$ the true anomaly, $\ell = \sqrt{Gm_\star a (1-e^2)}$ the specific angular momentum, $a$ the semimajor axis, $e$ the eccentricity, $G$ the gravitational constant, and $N$ the perturbing acceleration felt by the test particle normal to its own orbit. As \citet{murray99} state, $r N \cos (\omega + f)$ is the component of the perturbing torque that rotates the angular momentum vector about the line of nodes, thereby changing $i$.

We re-cast equation (\ref{eq:gauss}) for our problem of a small body encountering a big body. For present purposes of illustration, we ignore $\mathcal{O}(e)$ terms as small compared to unity, and ask how much the inclination changes after the small body has been accelerated for some time interval:
\begin{align} \label{eq:delta_i}
\Delta i = \sqrt{\frac{a}{Gm_\star}} \int %_{t_a}^{t_b}
\frac{Gm_{\rm b}}{|\mathbf{r_{\rm b}}-\mathbf{r}|^2} \,f_N
%\frac{(\mathbf{r_{\rm b}}-\mathbf{r}) \cdot \hat{l}}{|\mathbf{r_{\rm b}}-\mathbf{r}|} 
\cos u \, dt
\end{align}
where $m_{\rm b}$ is the big body mass, $\mathbf{r} = x \mathbf{\hat{x}} + y \mathbf{\hat{y}}+z \mathbf{\hat{z}}$ and $\mathbf{r_{\rm b}}$ are the respective position vectors of the small and big bodies, and 
$u \equiv \omega + f$ is the angle between the small body and the ascending node (also called the argument of latitude). The factor $-1 \leq f_N \leq 1$ accounts for the component of the gravitational acceleration normal to the small body's orbit:
\begin{align}\label{eq:fN}
f_N \equiv \frac{(\mathbf{r_{\rm b}}-\mathbf{r}) \cdot \hat{l}}{|\mathbf{r_{\rm b}}-\mathbf{r}|} 
\end{align}
where $\hat{l} = \sin i \sin \Omega \,\mathbf{\hat{x}} - \sin i \cos \Omega \,\mathbf{\hat{y}} + \cos i \, \mathbf{\hat{z}} $ is the unit orbit normal, and $\Omega$ is the longitude of ascending node.
Thus $N = [Gm_{\rm b}/|\mathbf{r_{\rm b}}-\mathbf{r}|^2] \,f_N$ in eq.~(\ref{eq:delta_i}) is the vertical acceleration for small $i$. The factor of $\cos u$ in eq.~(\ref{eq:delta_i}) accounts for the fact that $N$ is most effective at changing $i$ when it acts near the ascending or descending node ($u \approx 0$ or $\pi$); at these orbital phases, $N$ can turn the velocity vector to change the magnitude of its vertical component, $v_z$. Put another way, near either node, $v_z$ is at an extremum, and $N$ can do work ($\mathbf{N} \cdot \mathbf{v_z} \neq 0$) to change $v_z$ (by contrast $v_z = 0$ at $u = \pm \pi/2$; here $N$ changes $\Omega$ but not $i$). Thus the integral in eq.~(\ref{eq:delta_i}) yields the change $\Delta v_z$, which normalized by the Keplerian velocity $v_{\rm K} = \sqrt{Gm_\star/a}$ gives $\Delta i = \Delta v_z / v_{\rm K}$.

\begin{figure}
\centering
\includegraphics[width=1.0\linewidth]{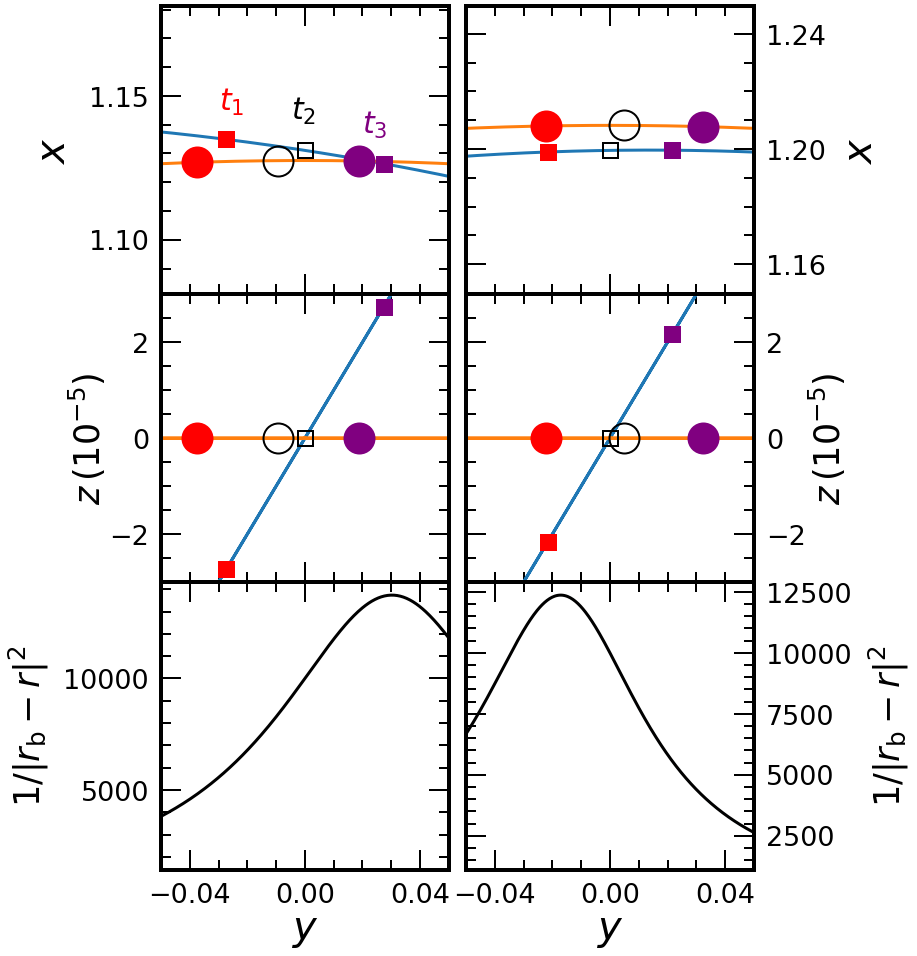}
  \caption{How close encounters between a small body (small square) and a big body (large circle) increase or decrease the small body's orbital inclination. The big body is assumed for simplicity to reside on a circular orbit (orange line) in the reference plane, while the small body traces an eccentric orbit (blue line) which crosses the big body's orbit (crossing is visible for the left column but not for the right), and whose ascending node is located on the $x$-axis at $y=z=0$. The unit of length is arbitrary. Top and middle rows show zoomed-in face-on ($x$-$y$) and edge-on ($z$-$y$) views of the encounter at three times (colored red, open, and purple, consecutively) with the open symbols marking the moment when the small body crosses its node. The bottom panel shows the inverse square distance (measure of gravitational attraction) between the bodies as a function of the small body's $y$-position. In the left column, the bodies are closest around $t_3$ (purple) after the small body ascends above the big body --- the net outcome is that the small body's inclination will decrease, because it is overall pulled downward by the big body (the change in $i$ is not shown). The opposite occurs in the right column, i.e.~the small-body inclination increases, because the interaction is strongest before node crossing.
  \label{fig:schematic}}
\end{figure}

The sign of $\Delta i$ can be positive or negative. Figure \ref{fig:schematic} illustrates both cases, for encounters near the crossing of the small body's ascending node. In the left column, the small body is pulled downward harder post-node than it is pulled upward pre-node, and therefore $\Delta i<0$. The right column shows the opposite case when $\Delta i > 0$.

\begin{figure}
\centering
\includegraphics[width=1.0\linewidth]{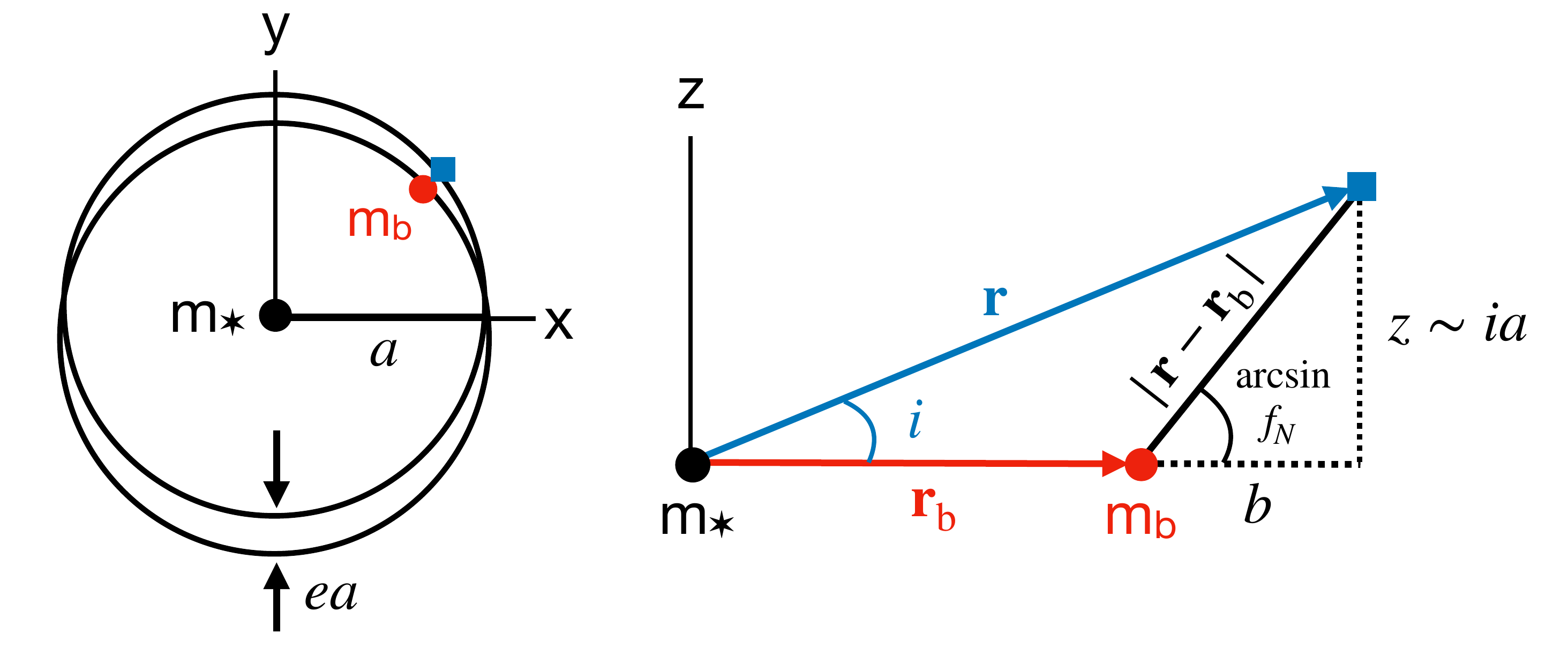}
  \caption{A big body of mass $m_{\rm b}$ (red circle) on a circular orbit in the $x$-$y$ reference plane encounters a small body (test particle, blue square) on a crossing orbit of eccentricity $e$ and inclination $i$. The bodies have similar semimajor axes $a$, and the in-plane impact parameter is $b \ll a$. While $\max b \sim ea$, there is otherwise no relation between $b$ and $ea$. At the moment of encounter (conjunction), the small body feels a vertical gravitational acceleration which is less than the full gravitational acceleration by a factor of $|f_N|$, by definition. To order of magnitude, $|f_N| \sim \min (|z|/b,1) \sim \min (ia/b,1)$.
  \label{fig:picture}}
\end{figure}

We may derive an order-of-magnitude version of equation (\ref{eq:delta_i}) for $\Delta i$ using the impulse approximation. Figure \ref{fig:picture} shows a small body with semimajor axis $a$ encountering a big body with an in-plane impact parameter $b \ll a$.
%formula for $|\Delta i|$. 
We approximate $Gm_{\rm b}/|\mathbf{r}_{\rm b}-\mathbf{r}|^2 \sim Gm_{\rm b}/b^2$, $\cos u \sim 1$, the encounter time $\Delta t \sim b / v_{\rm rel}$, the relative encounter velocity $v_{\rm rel} \sim \sqrt{e^2+i^2}\,v_{\rm K} \sim e v_{\rm K}$ (since $i$ grows to at most $e/2$ in equipartition; \citealt{ida92,ida93}), and $|f_N| \sim \min(|z|/b,1) \sim \min(ia/b,1)$. Thus eq.~(\ref{eq:delta_i}) simplifies to 
\begin{align} 
\Delta i &\sim \pm \frac{1}{v_{\rm K}} \frac{Gm_{\rm b}}{b^2} \frac{b}{e v_{\rm K}} \min \left( \frac{ia}{b},1\right) \nonumber \\ 
&\sim \pm \,\frac{m_{\rm b}}{m_\star} \frac{a}{b e} \min\left( \frac{ia}{b},1\right) \,.\label{eq:kickoom}
\end{align}
We see from eq.~(\ref{eq:kickoom}) that there are two possibilities: one where $ia < b$ and $\Delta i \propto \pm i$, and another when $ia > b$ and $\Delta i$ is independent of $i$. In Section \ref{subsec:time}, we will find that the first possibility $ia < b$ sub-divides further into two cases, which we will call the ``quasi-2D'' and ``2D-$i$/3D-$e$'' regimes. The second possibility $ia > b$ also sub-divides into its own two cases, the ``3D-anisotropic'' and ``equipartition'' regimes. \citet[][see also \citealt{petrovich14}]{rafikov10} found that inclinations amplify exponentially fast in the quasi-2D regime. We will find the same, though unlike \citet{rafikov10}, we do not assume that the inclination distribution follows a Gaussian in the quasi-2D regime (it does not).

\subsection{Random walking in log inclination}
In this paper we are most concerned with the case $ia < b$. Here eq.~(\ref{eq:kickoom}) implies that repeated scatterings cause the inclination to random walk, not with fixed steps of $\Delta i$, but with fixed steps of $\Delta i/i = \Delta \ln i$. 

We conduct a random walk experiment with such logarithmic steps, analogous to the one we performed in Section \ref{sec:intro} with linear steps that led to a Rayleigh inclination distribution. The random walk is in the 2D space $(p,q) = (i \cos \Omega, i \sin \Omega)$ because random perturbations in vertical velocity change both inclination $i$ and nodal longitude $\Omega$. We start with 10000 particles with the same initial inclination $i_0 = 10^{-2}$ rad and randomized nodes $\Omega_0 \in [0,2\pi)$. For each random-walk step, $(\Delta p,\Delta q) = (\Delta i \cos (\Omega - \Delta \Omega), \Delta i \sin (\Omega - \Delta \Omega))$, with $\Delta i/i = \Delta \ln i = 0.05$ and $\Delta \Omega$ drawn as a uniform variate between $0$ and $\pi$ (not $2\pi$). Our $\Delta \Omega$ prescription is such that the nodal longitude $\Omega$ decreases with every step, mimicking how an inclined particle nodally regresses when torqued by a perturber in the reference plane.\footnote{From Gauss's equation for $d\Omega/dt \propto N \sin u$, the particle's node changes most for conjunctions with the perturber that occur when the particle attains maximum height above or below the perturber reference plane ($u = \pm \pi/2$). At $u = \pi/2$ the particle is pulled down by an in-plane perturber ($N<0$), and at $u = -\pi/2$ the particle is pulled up ($N>0$). Either way,  $d\Omega/dt < 0$.} This prescription for the nodal evolution is just for illustration; other choices that preserve our assumed initial axisymmetry will give the same qualitative outcome, as we have tested. What is important is that the length of each step $\Delta i = \sqrt{(\Delta p)^2 + (\Delta q)^2}$ scales with the length $i = \sqrt{p^2+q^2}$, i.e.~the inclination changes by a fixed fractional (not absolute) amount per step.

Figure \ref{fig:pqi_dNdlogi} shows the inclination distribution of our 10000 particles after each has taken 1000 steps. Random walking in log space results in a log normal distribution, i.e.~a Gaussian in log space. This is just the central limit theorem at work in log space. As with a random walk in linear space, here in log space the width of the Gaussian (measured in dex) grows as the square root of the number of steps. For this simple experiment, the log normal distribution remains centered on the initial inclination $i_0$; this initial condition is not forgotten.

Unlike a Rayleigh distribution (Section \ref{sec:intro}), our log normal distribution does not reduce to $i$ as $i \rightarrow 0$. A more realistic computation would account for a non-zero inclination dispersion for the big bodies. If $i_{\rm b} \neq 0$, the actual inclination distribution of the small bodies when $i < i_{\rm b}$ (when the small bodies are immersed in the sea of big bodies) would differ from that computed here. In any case, it is the large $i > i_{\rm b}$ portion of the inclination distribution that matters for the wings of the vertical density profile. Our expectation of a log-normal shape to $dN/di$ at large $i$ will be confirmed in the $N$-body experiments of Section \ref{sec:num}.

%\vspace{0.2in}
\subsection{Vertical density profile \\for a log normal inclination distribution}\label{subsec:vdp_log}
A simple way to obtain the vertical density profile $dN/dz$ from the inclination distribution $dN/di$ is to assume that every particle has the same semimajor axis ($a=1$ in arbitrary units) and zero eccentricity. Then for each of our 10000 particles we compute $z = a \sin i \sin u$, where $u$ (the angular displacement of a particle from its ascending node) is drawn randomly from a uniform distribution between 0 and $2\pi$. The assumptions of fixed semimajor axis and zero eccentricity are made for simplicity; the numerical simulations in Section \ref{sec:num} will relax them, with no change in our conclusions.

\begin{figure}
\centering
\includegraphics[width=\linewidth]{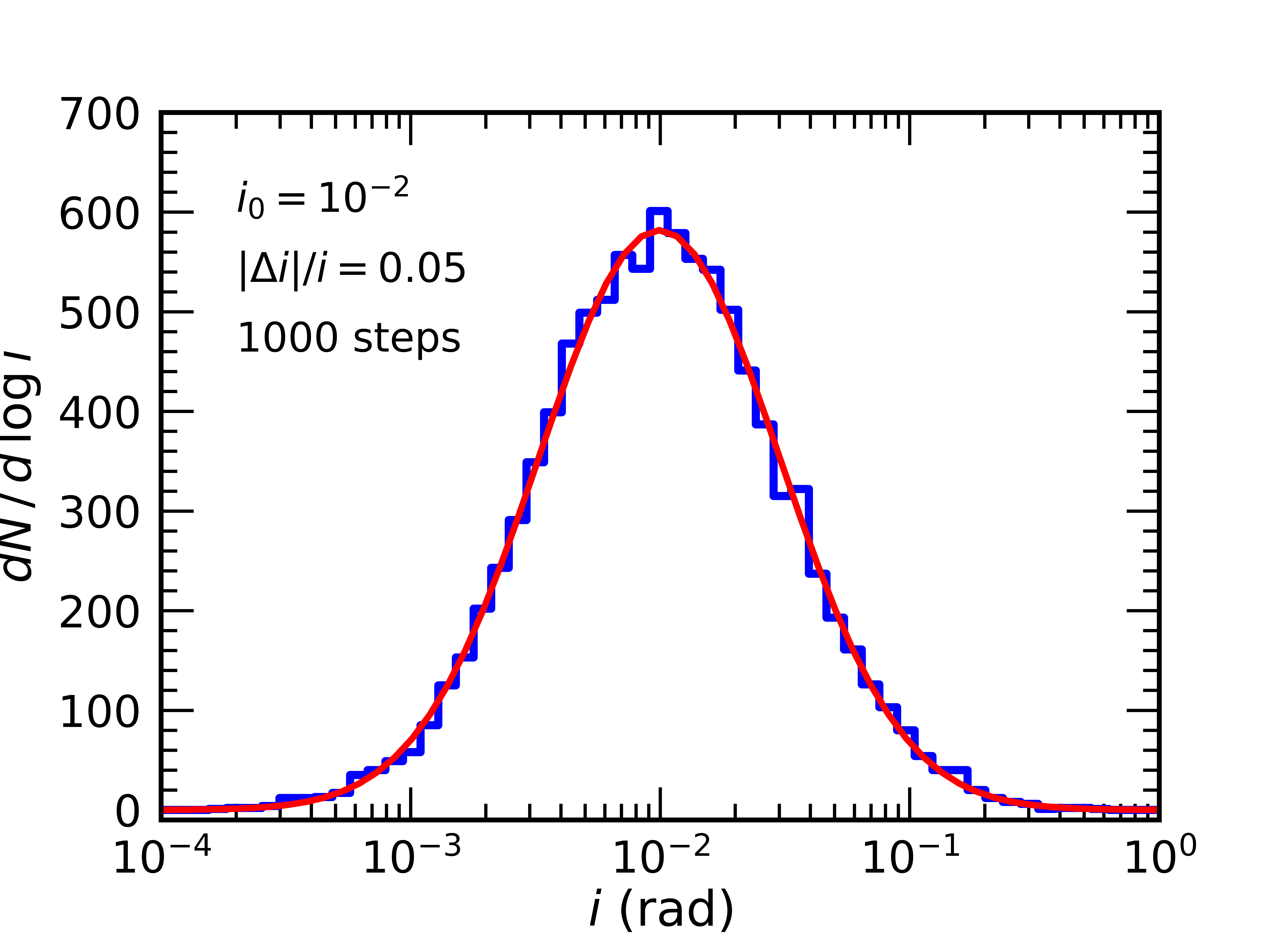}
  \caption{The inclination distribution of 10000 particles, initialized with $i_0 = 10^{-2}$ and randomized nodes $\Omega_0 \in [0,2\pi)$, after each has taken 1000 randomly directed steps in $(p,q)=(i\cos\Omega,i\sin\Omega)$ space. Each step moves a particle by $(\Delta p,\Delta q)=(\Delta i \cos (\Omega - \Delta \Omega), \Delta i \sin (\Omega - \Delta \Omega))$, with $\Delta i/i = \Delta \ln i = 0.05 = 0.021$ dex and $\Delta \Omega$ chosen randomly from a uniform distribution between 0 and $\pi$ (net regression; see main text). Data are shown as a blue histogram (logarithmically binned and not normalized), and the best-fit Gaussian in $\log i$ is shown in red. The log normal distribution is centered on $i_0$ (initial conditions are not forgotten) and we have verified that its width in $\log i$ scales as the square root of the number of steps ($\sqrt{1000} \times 0.021$ dex $\simeq 0.67$ dex).   Fixed $\Delta \log i$ kicks arise when small bodies and big bodies cross orbits and $i \ll e$ (Section \ref{sec:theory}), or when small bodies are radially separated from big bodies and $i \ll |a-a_{\rm b}|/a$ (see Appendix \ref{sec:appendix}).
  \label{fig:pqi_dNdlogi}}
\end{figure}

\begin{figure}
\centering
\includegraphics[width=\linewidth]{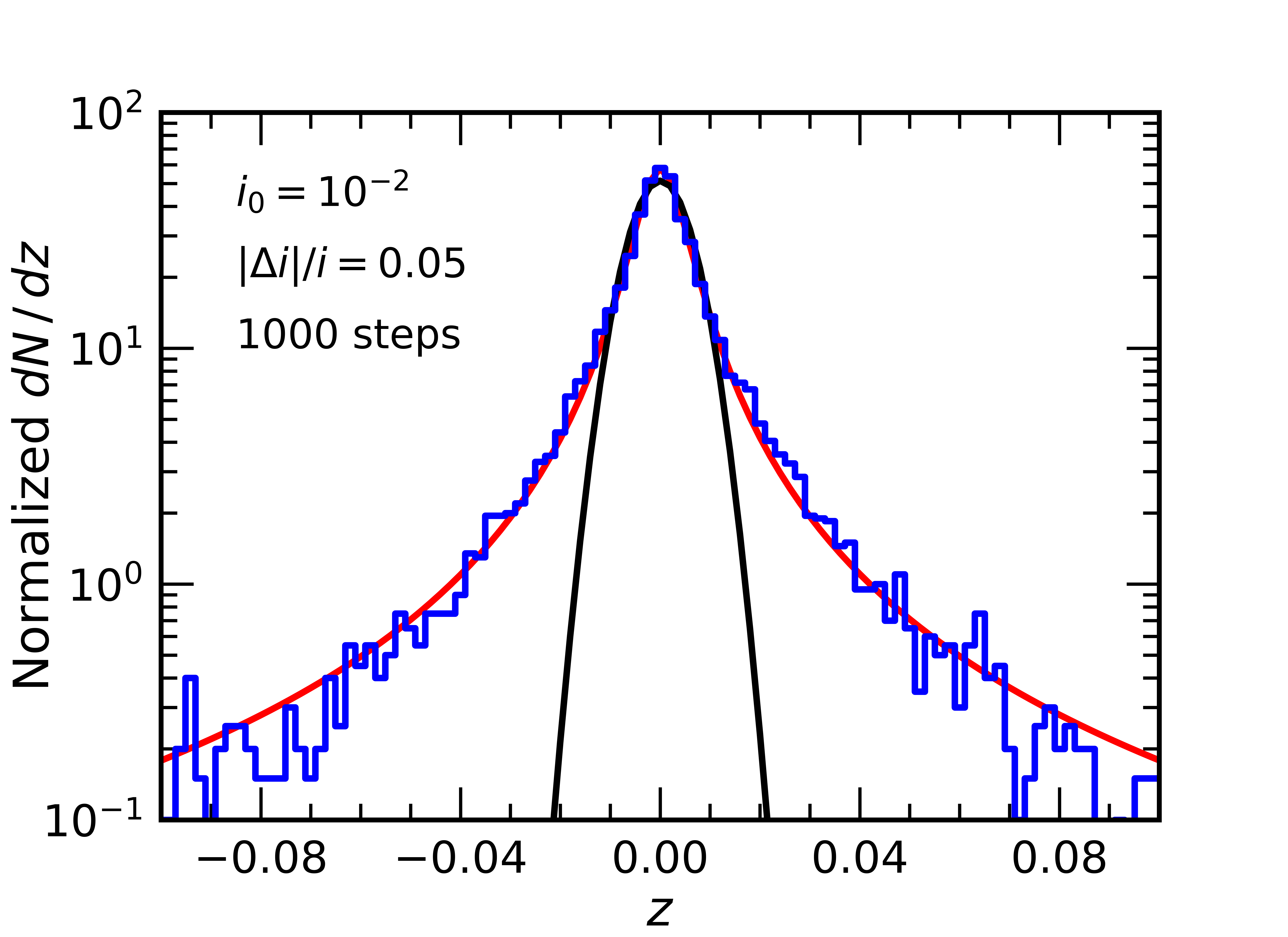}
  \caption{The log-normal $dN/d\log i$ distribution of Fig.~\ref{fig:pqi_dNdlogi} yields a vertical density profile $dN/dz$ (blue histogram, normalized) with broad wings. In computing $dN/dz$, we assume for simplicity that all particles have the same semimajor axis ($a = 1$ in arbitrary units) and zero eccentricity; every particle in this narrow ring has a vertical position $z = a \sin i \sin u$, with $u \in [0,2\pi)$.  The best-fit Lorentzian (red) does better than the best-fit Gaussian (black) in fitting the vertical density profile. Side note: our bin sizes in $z$ are small enough to resolve the central peak; the required bin size scales with the seed inclination $i_0$.
  \label{fig:pqi_vert_profile}}
\end{figure}

Figure \ref{fig:pqi_vert_profile} shows the vertical density profile $dN/dz$ derived from the log normal $dN/di$ of Fig.~\ref{fig:pqi_dNdlogi}, and fits $dN/dz$ to both a Lorentzian and a Gaussian. The Lorentzian with its broader wings fits better. The broader wings in $z$ follow from the log normal inclination distribution, whose large-$i$ tail is thicker than that of a linear Gaussian inclination distribution.

While we have so far only treated the case of big bodies on circular orbits, the same results should hold when $e_{\rm b} \neq 0$, as long as the mutual inclination $i \ll \max(e_{\rm b},e)$. We will confirm this expectation with numerical integrations in Section \ref{sec:num}. The anisotropic regime is generally defined as when the relative vertical (out-of-plane) motions are less than the relative horizontal (in-plane) motions. The former is controlled by the mutual inclination $i$, and the latter by a ``mutual'' eccentricity, which we can approximate as the quadrature sum of the small and big body eccentricities, or the maximum of the two, for random orbit orientations.

%\subsection{Timescales for self-stirring, and big body masses}
\subsection{Stirring rates for inclination and eccentricity}
\label{subsec:time}
We estimate the timescales over which   
small bodies %--- either small or big ---
have their $i$'s and $e$'s amplified by a disk of big bodies. For simplicity, big bodies are imagined to reside in the disk midplane on circular orbits, but this is not a major limitation since the small-body stirring rates thus derived should carry over with only order-unity modifications as big body random velocities approach those of small bodies. In the absence of damping, big bodies stir themselves as rapidly as they stir small bodies.

In Sections \ref{subsubsec:quasi2d}--\ref{subsubsec:eq} we work in natural units where $G = m_\star = a = 1$, and to order-of-magnitude accuracy dropping all order-unity coefficients. Readers interested only in our results may skip to Section \ref{subsubsec:numcoe} where we re-cap our formulae with physical units restored and numerical coefficients drawn from the literature.

We identify four regimes for stirring, three applying out-of-equipartition when $i\ll e$, and a fourth equipartition regime when $i \sim e$.  In all four regimes, relative velocities between small and big bodies are $v_{\rm rel} \sim e v_{\rm K} \sim e$. The big bodies have individual masses $m_{\rm b} = \mu_{\rm b} m_\star = \mu_{\rm b}$ and collective surface mass density $\Sigma_{\rm b}$.

Like \citet[][their Figure 6]{ida92}, \citet{rafikov10}, and \citet[][their section 4.4.2]{goldreich04}, we will find that when $i \ll e$ under dispersion-dominated conditions, stirring rates for $i$ are faster than for $e$. We extend the above works to derive analytic stirring rates for $i$ and $e$ separately, in each of four regimes. Our analytic stirring rates will be confirmed quantitatively in the numerical simulations of Section \ref{sec:num}. Some of our analytic relations will also be used in Section \ref{sec:app} to constrain the properties of big bodies in real-life disks.

\subsubsection{Quasi-2D: $i < \mu_{\rm b}/e^2 < \sqrt{\mu_{\rm b}/e} < e$}\label{subsubsec:quasi2d}
Consider a small body encountering a big body with impact parameter $b$. 
Repeating the derivation of eq.~(\ref{eq:kickoom}) but now in dimensionless units, the small body's inclination is changed by
\begin{align}\label{eq:deli}
\Delta i &\sim \frac{\mu_{\rm b}}{b^2} \cdot \frac{b}{v_{\rm rel}} \cdot \min\left( \frac{ia}{b},1 \right) \nonumber \\
& \sim \frac{\mu_{\rm b} i}{b^2e} 
\end{align}
where we have assumed $ia < b$. 
For $i$ to double in a single encounter, we require $\Delta i$ be as large as $i$, which in turn requires an impact parameter as small as
\begin{align}
b_i \sim \sqrt{\mu_{\rm b}/e} \,.
\end{align}
Since only in-plane forces change $e$, 
\begin{align}
\Delta e \sim \frac{\mu_{\rm b}}{b^2} \cdot \frac{b}{v_{\rm rel}} \sim \frac{\mu_{\rm b}}{be}
\end{align}
from which it follows that the impact parameter needed to double $e$ in a single encounter is
\begin{align}
b_e \sim \mu_{\rm b}/e^2 \,.
\end{align}
We will show at the end of this subsection that multiple less-than-doubling encounters are negligible.

Now $b_e < b_i$ since in the dispersion-dominated (super-Hill) regime, $e > \mu_{\rm b}^{1/3}$. We further assume $ia < b_e$: the disk is so vertically thin that encounters between small and big bodies occur as rapidly as they would in a 2D plane, at rate
\begin{align}\label{eq:2drate}
\mathcal{R}_{\rm 2D} \sim \frac{\Sigma_{\rm b}}{m_{\rm b}} \,b \,v_{\rm rel}
\end{align}
where $b$ is the in-plane impact parameter (encounter cross-sections in 2D have units of length). The defining condition of this quasi-2D regime is $ia < b$ (see also eq.~39 of \citealt{goldreich04}). The quasi-2D  regime is studied by \citet{rafikov10}, and in the strictly 2D simulations of \citet{ida90}.

Inserting $b=b_i$ into eq.~(\ref{eq:2drate}) gives the doubling time for inclination:
\begin{align}\label{eq:ti2d}
t_{i,{\rm quasi-2D}} \equiv \frac{i}{di/dt} \sim \left. \mathcal{R}_{\rm 2D}^{-1}\right|_{b = b_i} \sim \left( \frac{\mu_{\rm b}}{e} \right)^{1/2} \frac{1}{\Sigma_{\rm b}} \,.
\end{align}
Inserting $b=b_e$ into eq.~(\ref{eq:2drate}) gives the eccentricity doubling time
\begin{align}\label{eq:eee}
t_{e,{\rm quasi-2D}} \equiv \frac{e}{de/dt} \sim \left. \mathcal{R}_{\rm 2D}^{-1}\right|_{b = b_e} \sim \frac{e}{\Sigma_{\rm b}}
\end{align}
which is longer than the inclination doubling time by $\sqrt{e^3/\mu_{\rm b}} > 1$. Eq.~(\ref{eq:ti2d}) implies that for as long as $e$ is constant, $i$ grows exponentially with time. \citet{rafikov10} also found exponentially fast growth of $i$ in the quasi-2D regime. Based on the strictly 2D simulations of \citet{ida90}, \citet{ida92} cite an eccentricity ``relaxation'' time that scales linearly with $e$, in agreement with our eq.~(\ref{eq:eee}). %EC: I think Ida (1990) Figure 12 actually plots stirring time, not stirring rate

We have assumed in the above that single doubling encounters dominate the rate of doubling, not multiple less-than-doubling encounters with $b > b_i$ and $b > b_e$. This is justified because the latter would double the inclination over a diffusion time $\sim$$(i/\Delta i)^2 \mathcal{R}_{\rm 2D}^{-1} \propto b^3$, and the eccentricity over an analogous time $\sim$$(e/\Delta e)^2 \mathcal{R}_{\rm 2D}^{-1} \propto b$. In both cases, the doubling times are minimized for smallest $b$, i.e.~those single encounters with $b = b_i$ and $b = b_e$, as considered above.

\subsubsection{2D-$i$/3D-$e$: $\mu_{\rm b}/e^2 < i < \sqrt{\mu_{\rm b}/e} < e$}

Now we assume $b_e < ia < b_i$. Inclinations behave the same way as they do in the quasi-2D regime:
\begin{align}\label{eq:ti2da}
t_{i,{\rm 2D}-i/{\rm 3D}-e} \equiv \frac{i}{di/dt} \sim \left. \mathcal{R}_{\rm 2D}^{-1} \right|_{b=b_i} \sim \left( \frac{\mu_{\rm b}}{e} \right)^{1/2} \frac{1}{\Sigma_{\rm b}} 
\end{align}
but the impact parameter $b_e$ needed to double $e$ now fits comfortably inside the disk, so that $e$-doubling encounters occur at the 3D rate
\begin{align}\label{eq:rate3d}
\mathcal{R}_{\rm 3D} \sim \frac{\Sigma_{\rm b}/m_{\rm b}}{ia} \, b^2 \, v_{\rm rel} 
\end{align}
where the first factor on the right-hand side is the effective density (number per unit volume) of big bodies. Inserting $b=b_e$ into eq.~(\ref{eq:rate3d}) gives
\begin{align}\label{eq:te3da}
t_{e,{\rm 2D}-i/{\rm 3D}-e} \equiv \frac{e}{de/dt} \sim \left. \mathcal{R}_{\rm 3D}^{-1} \right|_{b=b_e} \sim \frac{e^3 i}{\Sigma_{\rm b} \mu_{\rm b}} 
\end{align}
which is longer than the inclination doubling time by $(ie^2/\mu_{\rm b} )\cdot (e^{3/2}/\mu_{\rm b}^{1/2})$ (both factors are $>1$).

Compared to the quasi-2D case, while single doubling encounters still dominate the rate of $i$-doubling, the rate of $e$-doubling is enhanced by multiple less-than-doubling encounters by a Coulomb logarithm.\footnote{It is called a Coulomb log because the same kind of logarithm appears when calculating bremsstrahlung emission from electron-proton scatterings in a plasma (e.g.~\citealt{rybicki86}). The commonalities are a force law that depends on the inverse square of the distance, and a 3D encounter rate.}  The eccentricity diffusion timescale $\sim$$(e/\Delta e)^2 \mathcal{R}_{\rm 3D}^{-1}$ is independent of $b$, which implies that encounters with various impact parameters (including $b_e$) contribute equally to doubling $e$. Thus eq.~(\ref{eq:te3da}) should be divided by a logarithmic factor, expected to be not much larger than unity.

\subsubsection{3D-anisotropic: $\mu_{\rm b}/e^2 < \sqrt{\mu_{\rm b}/e} < i < e$}\label{subsubsec:3dani}
Now $i > b_i > b_e$ so that $e$ and $i$-doubling encounters occur with random impact angles inside an effectively 3D disk, at rates given by eq.~(\ref{eq:rate3d}). Returning to eq.~(\ref{eq:deli}) for $\Delta i$, doubling encounters now occur with $f_N \sim \min(ia/b,1) \sim 1$:
%\footnote{Contrast this situation with the toy experiment in section \ref{subsec:how} where we restricted encounters to node crossings and argued that $f_N \sim i/e \ll 1$.}
\begin{align}\label{eq:nomorei}
\Delta i &\sim \frac{\mu_{\rm b}}{b^2} \cdot \frac{b}{v_{\rm rel}} \nonumber \\
& \sim \frac{\mu_{\rm b} }{be} 
\end{align}
(see also eq.~40 of \citealt{goldreich04} and note that they use $b \sim \mu_{\rm b}/e^2$). For inclinations to double in a single encounter (setting $\Delta i$ equal to $i$ in eq.~\ref{eq:nomorei}) now requires an impact parameter
\begin{align}
b_{i,{\rm 3D}} \sim \frac{\mu_{\rm b}}{ei}
\end{align}
which inserted into eq.~(\ref{eq:rate3d}) yields
\begin{align}\label{eq:ti3d}
t_{i,{\rm 3D-ani}} \equiv \frac{i}{di/dt} \sim \left. \mathcal{R}_{\rm 3D}^{-1} \right|_{b =b_{i,{\rm 3D}}} \sim \frac{i^3 e}{\Sigma_{\rm b} \mu_{\rm b}}\,.
\end{align}
The inclination doubling time is shorter than the eccentricity doubling time (which remains the same as in the 2D-$i$/3D-$e$ regime)
\begin{align}\label{eq:3danite}
t_{e,{\rm 3D-ani}} \equiv \frac{e}{de/dt} \sim \left. \mathcal{R}_{\rm 3D}^{-1} \right|_{b=b_e} \sim \frac{e^3 i}{\Sigma_{\rm b} \mu_{\rm b}}
\end{align}
by a factor of $(i/e)^2 \ll 1$. Eq.~(\ref{eq:ti3d}) implies that for as long as $e$ is constant, $i \propto t^{1/3}$.

Following the reasoning at the end of Section \ref{subsubsec:3dani}, eqs.~(\ref{eq:ti3d}) and (\ref{eq:3danite}) should be divided by Coulomb logarithms to account for more frequent but less-than-doubling encounters.

\subsubsection{Equipartition $i \sim e$}\label{subsubsec:eq}
The case where $e \sim i$ (technically $e \simeq 2i$) is the usual one assumed in the literature. The only difference between equipartition stirring and the 3D-anisotropic case considered above is in the rate equation (\ref{eq:rate3d}), where under equipartition we allow $i$ to cancel with $e$ to write
\begin{align}
\mathcal{R}_{\rm equi} \sim \frac{\Sigma_{\rm b}}{m_{\rm b}} \, b^2
\end{align}
from which it follows that
\begin{align}\label{eq:equi}
t_{i,{\rm equi}} \sim t_{e,{\rm equi}} \sim \frac{e^4}{\Sigma_{\rm b}\mu_{\rm b}} \sim \frac{i^4}{\Sigma_{\rm b}\mu_{\rm b}}
\end{align}
and $i\propto e \propto t^{1/4}$. Eq.~(\ref{eq:equi}) is missing a number of order-unity factors, including Coulomb logarithms that shorten the stirring times.

\subsubsection{Summary of stirring timescales}\label{subsubsec:numcoe}

We recapitulate the above results restoring physical units and attempting to calibrate numerical coefficients using the literature.

\paragraph{Equipartition} When $\mu_{\rm b}/e^2 \lesssim \sqrt{\mu_{\rm b}/e}$ (dispersion-dominated) and $i = e/2$ (equipartition), then $i$ and $e$ double over the same timescale:
\begin{align}\label{eq:eq}
t_{i,{\rm equi}} \sim t_{e,{\rm equi}} \sim \frac{e^4 m_\star}{40 \Sigma_{\rm b}a^2 \mu_{\rm b} n} \sim \frac{(2i)^4 m_\star}{40 \Sigma_{\rm b}a^2 \mu_{\rm b} n} 
\end{align}
where $n$ is the Keplerian orbital frequency and the factor of 40 is taken from the N-body simulations of \citet{ida93} (their equation 4.2 with $C_e \simeq 40$). In principle the factor of 40 includes a Coulomb logarithm among other quantities. Both $e$ and $i$ grow together as $t^{1/4}$.

\paragraph{3D-anisotropic} When $\mu_{\rm b}/e^2 \lesssim \sqrt{\mu_{\rm b}/e} \lesssim i < e/2$, the inclination doubles over timescale
\begin{subequations}\label{eq:subeq_3dani}
\begin{align}\label{eq:ti3dnumcoe}
t_{i,{\rm 3D-ani}} \sim \frac{(2i)^3 e m_\star}{40\Sigma_{\rm b} a^2 \mu_{\rm b}n}
\end{align}
which is shorter than the eccentricity doubling time
\begin{align}
t_{e,{\rm 3D-ani}} \sim \frac{e^3 (2i) m_\star}{40 \Sigma_{\rm b} a^2 \mu_{\rm b}n}
\end{align}
\end{subequations}
where the numerical coefficients are chosen to ensure continuity with equation (\ref{eq:eq}). Here $i \propto t^{1/3}$ if $e$ can be approximated as constant.

\paragraph{2D-$i$/3D-$e$} When $\mu_{\rm b}/e^2 \lesssim i \lesssim \sqrt{\mu_{\rm b}/e} < e/2$,
\begin{subequations}\label{eq:subeq_3de2di}
\begin{align}\label{eq:ti2danumcoe}
t_{i,{\rm 3D}-e/{\rm 2D-i}}\sim \left( \frac{\mu_{\rm b}}{e} \right)^{1/2} \frac{8m_\star}{40\Sigma_{\rm b}a^2n} 
\end{align}
which is shorter than
\begin{align}
t_{e,{\rm 3D}-e/{\rm 2D-i}} \sim \frac{e^3 (2i) m_\star}{40 \Sigma_{\rm b} a^2 \mu_{\rm b}n} \,.
\end{align}
\end{subequations}
The inclination grows exponentially with time if $e$ is approximately constant. While the numerical coefficient of 8/40 in eq.~(\ref{eq:ti2danumcoe}) is chosen to ensure continuity with eq.~(\ref{eq:ti3dnumcoe}) in the 3D-anisotropic regime, it is underestimated (i.e.~the inclination stirring time should be longer) because in the 2D-$i$ regime there is no Coulomb log to enhance the vertical stirring rate.

\paragraph{Quasi-2D} Finally when $i \lesssim \mu_{\rm b}/e^2 < \sqrt{\mu_{\rm b}/e} < e/2$,
\begin{subequations}\label{eq:subeq_quasi}
\begin{align}\label{eq:ti2dnumcoe}
t_{i,{\rm quasi-2D}} \sim \left( \frac{\mu_{\rm b}}{e} \right)^{1/2} \frac{8m_\star}{40\Sigma_{\rm b}a^2n} 
\end{align}
which is shorter than
\begin{align}
t_{e,{\rm quasi-2D}} \sim \frac{2em_\star}{40\Sigma_{\rm b}a^2n} \,.
\end{align}
\end{subequations}
The inclination grows exponentially fast if $e$ is approximately constant (in agreement with \citealt{rafikov10}). As with eq.~(\ref{eq:ti2danumcoe}), the numerical coefficients in eqs.~(\ref{eq:subeq_quasi}) are underestimated (the doubling times should actually be longer) because they should not include the Coulomb logs inherited by our insistence on continuity with the equipartition relation (\ref{eq:eq}).

\vspace{0.2in}
\section{Numerical experiments}\label{sec:num}
To test our analytic ideas, we conduct numerical experiments with the $N$-body code \texttt{REBOUND} \citep{rein12}, outfitted with the hybrid-symplectic \texttt{MERCURIUS} integrator that can treat close encounters \citep{rein19,tamayo20}. We simulate how one or two big bodies gravitationally stir a narrow annulus of small bodies (test particles), with the aim of finding out-of-equipartition situations where in-plane relative velocities greatly exceed out-of-plane velocities. Under such circumstances, according to the theory of Section \ref{sec:theory}, test particles should be stirred into vertical Lorentzians, or at least into profiles with broader wings than Gaussians. Such non-Gaussian profiles are also expected to morph into Gaussians, as particles eventually relax into equipartition. 

The simulation parameters (orbital radii, perturber masses, run times), though reminiscent of those in real-life debris disks, are chosen more for computational convenience, as our goal for this section is not to model observations, but to test theory.

%%%%%%%%%%%%%%%%%%%%%%%%%%%%%%%%%%%%%%%%%%%%%%%%%%%%%%%%%%%%
\begingroup % to localize the effect of the length modification below
\setlength{\medmuskip}{0mu} % to reduce the spacing around binary operators (\times)
\begin{deluxetable*}{cccccccccc}
  \tablecaption{Stirring simulations }\label{tab:paras}
  \tablecolumns{10}
  \tablehead{
    \colhead{(1)} &
    \colhead{(2)} &
    \colhead{(3)} &
    \colhead{(4)} &
    \colhead{(5)} &
    \colhead{(6)} &
    \colhead{(7)} &
    \colhead{(8)} &
    \colhead{(9)} & 
    \colhead{(10)} \\
    \colhead{\texttt{Run}} &
    \colhead{$\mu_{\rm b}$} &
    \colhead{$a_{\rm b}$} &
    \colhead{$a_{\rm init}$} &
 \colhead{$e_{\rm b}$} &
    \colhead{$e_{\rm init}$} &
    \colhead{$i_{\rm b}$} &
    \colhead{$i_{\rm init}$} &
    \colhead{L or G ($e_{\rm med}$/$i_{\rm med}$)} &
    \colhead{L or G ($e_{\rm med}$/$i_{\rm med}$)} \\
    \colhead{} &
    \colhead{($10^{-8}$)} &
    \colhead{(au)} &
    \colhead{(au)} &
    \colhead{} &
    \colhead{} &
    \colhead{(rad)} &
    \colhead{(rad)} & 
    \colhead{@ $t=1$ Myr} &
    \colhead{@ $t=20$ Myr}
  }
  \startdata
  \hline
  \texttt{A}%28--do not erase!
  & $1.5$  & $30$ & $30.154$ & $0.1$ & 
  $0$ & 0 & $10^{-5}$ & L  (0.009/0.001)\phantom{0}          & L (0.037/0.004)\phantom{0}   \\ 
\texttt{A-low}%29--do not erase!
& $1.5$  & $30$ & $30.154$ & $0.1$ & 
$0$ & 0 & $10^{-6}$ & L  (0.010/0.0001)         & L  (0.040/0.004)\phantom{0}   \\ 
\texttt{B}%16--do not erase!
& $1.5$  & $30$ & $30.154$ & $0.1$ & 
  $0.1$ & 0 & $10^{-5}$ & L  (0.100/0.001)\phantom{0}          & L  (0.101/0.005)\phantom{0}     \\
         \texttt{C}%12--do not erase!
         & $1.5$  & $30$ & $30.154$ & $0$ & 
         $0.1$ & 0 & $10^{-5}$ & L  (0.100/0.001)\phantom{0}          & L  (0.101/0.005)\phantom{0}   \\
       \texttt{D}%11--do not erase!
       & $1.5$  & $30$ & $30.154$ & $0$ & 
        $0.02$ & 0 & $10^{-5}$ & L  (0.024/0.002)\phantom{0}          & G  (0.021/0.010)\phantom{0}     \\ 
                \texttt{E}%27--do not erase!
                & $1.5$  & $30$ & $30.154$ & $0$ & 
                $0$ & 0 & $10^{-5}$ & ---            & L  (0.011/0.0002)     \\ \hline
    \texttt{AA}%30--do not erase!
    & $1.5,1.5$  & $30,30$ & $30.154$ & $0.1,0.1$ & 
    $0$ & $0,10^{-4}$ & $10^{-5}$ & L  (0.004/0.001)\phantom{0}          & L  (0.015/0.004)\phantom{0}     \\
    \texttt{AA-low}%31--do not erase!
    & $1.5,1.5$  & $30,30$ & $30.154$ & $0.1,0.1$ & 
    $0$ & $0,10^{-5}$ & $10^{-5}$ & L  (0.009/0.001)\phantom{0}          & L  (0.020/0.004)\phantom{0}     \\
    \texttt{BB}%17--do not erase!
    & $1.5,1.5$  & $30,30$ & $30.154$ & $0.1,0.1$ & 
    $0.1$ & $0,10^{-4}$ & $10^{-5}$ & L  (0.100/0.001)\phantom{0}          & L (0.101/0.004)\phantom{0}       \\
\texttt{CC}%32--do not erase!  
& $1.5,1.5$  & $29.959,30.349$ & $30.154$ & $0,0$ & 
        $0.1$ & $0,0$ & $10^{-5}$ & L  (0.100/0.001)\phantom{0} & L  (0.101/0.006)\phantom{0}     \\ 
        \texttt{DD}%18--do not erase! 
        & $1.5,1.5$  & $29.959,30.349$ & $30.154$ & $0,0$ & 
        $0.02$ & $0,0$ & $10^{-5}$ & L (0.024/0.003)\phantom{0}   & G (0.022/0.009)\phantom{0}     \\ 
                \texttt{EE}%26--do not erase!
                & $1.5,1.5$  & $29.959,30.349$ & $30.154$ & $0,0$ & 
                $0$ & $0,0$ & $10^{-5}$ & L  (0.011/0.003)\phantom{0}  & G  (0.016/0.008)\phantom{0}    \\ 
  \enddata
  \tablecomments{Columns: 
  (1) Run name. 
  (2) Ratio of big body mass to central mass (for reference, the mass of Pluto relative to the Sun is $6.6 \times 10^{-9}$). In the case of two big bodies, there are two entries.  
  (3) Initial big body semimajor axis or axes. 
  (4) Initial semimajor axis of test particles, typically $3$ big-body Hill radii away from the big body. Initial mean anomalies of test particles are drawn from a uniform variate $[0,2\pi)$. 
  (5) Initial big body eccentricity. In the case of 2 big bodies, both initial particle longitudes and initial periapse longitudes are 180$^\circ$ apart. 
  (6) Initial eccentricity of test particles. Initial periapse longitudes of test particles are drawn from a uniform variate $[0,2\pi)$. 
  (7) Initial big body inclination. 
  (8) Initial inclination of test particles. Initial nodal longitudes of test particles are drawn from a uniform variate $[0,2\pi)$. 
  (9) Whether the vertical distribution of test particles $dN/dz$ resembles more a Lorentzian (L) or a Gaussian (G), at simulation time $t = 1$ Myr. For \texttt{Run E}, test particle inclinations at $t = 1$ Myr remain mostly unchanged from their initial values and therefore the vertical density profile at this time is not usefully obtained. Also listed in parentheses are the median eccentricity $e_{\rm med}$ and median inclination $i_{\rm med}$ of test particles at $t = 1$ Myr, computed for $i > 10 \, i_{\rm init}$ to minimize contamination from initial conditions.
  (10) Same as (9) but for $t=20$ Myr.
  }
\end{deluxetable*}
\endgroup

\subsection{Numerical set-up}
The big bodies, numbering one or two in a given simulation, each have mass $m_{\rm b} = \mu_{\rm b} m_\odot = 1.5 \times 10^{-8}m_\odot$ (about $2\times$ Pluto's mass), and orbit the central $m_\star = m_\odot$ star with semimajor axis $a_{\rm b} \simeq 30$ au (similar to Neptune's orbit). The big bodies stir $N=10000$ nearby test particles, initialized with small non-zero inclinations relative to big bodies to seed the vertical stirring. Simulations are run for 20 Myr, comparable to the ages of the \citetalias{zawadzki26} debris disks studied in this paper (Section \ref{sec:app}).

Table \ref{tab:paras} compiles input parameters. Run names take a single letter for simulations with only one big body, and two letters for two big bodies. Between runs, the main parameters we focus on varying are the initial eccentricities, either of the big bodies ($e_{\rm b}$) or the small bodies ($e_{\rm init}$). For a given run, $e_{\rm b}$ and $e_{\rm init}$, and the inclinations $i_{\rm b}$ and $i_{\rm init}$ (relative to a reference plane coinciding with a big body orbit plane), are set to constants listed in Table \ref{tab:paras} --- in a given run, the test particle orbits are all initialized with the same $e_{\rm init}$ and $i_{\rm init}$, and differ only in their nodal and periapse longitudes, and mean anomalies. Large initial eccentricities and small inclinations set the stage for anisotropic, out-of-equipartition dynamics.

In line with the rest of this paper, we simulate dispersion-dominated, i.e.~orbit-crossing conditions. We focus on the orbital evolution of small bodies as they scatter off big bodies, whose own orbits are constructed to not change much, keeping the experiments as controlled as possible. In runs with only one big body (whose orbit cannot change), the test particle semimajor axes are all initialized with the same $a_{\rm init} = a_{\rm b} + 3 R_{\rm H}$, where $R_{\rm H} = (\mu_{\rm b}/3)^{1/3} a_{\rm b}$ is the big body Hill radius. These test particle semimajor axes lie just within the big body's chaotic zone (whose half-width is about $2\sqrt{3} R_{\rm H} \simeq 3.5 R_{\rm H}$; \citealt{gladman93,chambers96,pearce24};  cf.~\citealt{wisdom80}), which leads to strong scatterings and orbit crossings even when orbits are not initially crossing. In some runs the bodies start on eccentric and crossing orbits.  
When two (identical) big bodies are present with circular orbits, they are initialized with semimajor axes that are 6 mutual Hill radii apart ($6 \times 2^{1/3}R_{\rm H}$), with all test particles initialized with the same semimajor axis exactly midway. While the test particles scatter chaotically onto crossing orbits, the big bodies are far enough apart that they do not perturb one another much. For runs with two big bodies on eccentric crossing orbits, the initial eccentricities $e_{\rm b} = 0.1$ are large enough that eccentricity doubling times exceed the duration of the simulation.

\subsection{Results of numerical simulations for inclination and vertical density distributions}
We mostly review the runs with two big bodies, as their results are simpler to analyze. The results for runs with just one big body reinforce the results with two big bodies, but have some extra (mostly distracting) features, in retrospect stemming from our simplifying assumption of a single big body. We shunt these single big body technicalities to Appendix \ref{sec:extra_app}.

Table \ref{tab:paras} summarizes run outcomes. For runs with two big bodies, we find that $e_{\rm b} = 0.1$ or $e_{\rm init}=0.1$ (\texttt{Runs AA, AA-low, BB, CC}) yield $dN/dz$ profiles after 20 Myr that are better fitted to Lorentzians than Gaussians. By comparison, when $e_{\rm b}$ and $e_{\rm init}$ are $\leq 0.02$ (\texttt{Runs DD, EE}), the vertical profiles are more Gaussian at 20 Myr. When examining the vertical profiles, we omit particles with $i < 10\, i_{\rm init}$, as these have inclinations too close to their assumed initial value and are considered ``unstirred''. Our goal with these numerical experiments is to discover what vertical profiles result from viscous stirring (again, not to model observations), and to that end we need to sample only stirred (scattered) particles. For the simulations showcased in this section (\texttt{Runs BB} and \texttt{DD}), and most of the simulations in Table \ref{tab:paras}, the unstirred particles constitute less than a few percent of all particles at $t = 1$ Myr, and much less than a percent at $t = 20$ Myr (precise percentages are given in figure captions).

Notably, in all runs, including those where the outcomes are ultimately Gaussian, the vertical profiles sampled at an earlier time $t = 1$ Myr are Lorentzian. For all our initial conditions, test particle random motions start out of equipartition ($e \gg 2i$ $\rightarrow$ Lorentzian $dN/dz$). Particles then evolve toward equipartition ($e = 2i$ $\rightarrow$ Gaussian $dN/dz$), with runs starting closer to equipartition achieving it by their ends. 

Table \ref{tab:paras} lists median eccentricities $e_{\rm med}$ and median inclinations $i_{\rm med}$ of test particles at $t = 1$ and 20 Myr. We elect to tabulate the medians and not the rms (root-mean-squared) values, as the former are less sensitive to outlier events (ejections). Nevertheless we have also tabulated rms values (data not shown); they are larger than the medians, especially for inclination, but tell the same story as the median values: when $e \simeq 2 i$, we find vertical Gaussian profiles, and where $e \gg 2i$, we find vertical Lorentzians.

\begin{figure}[H]
\centering
\includegraphics[width=\linewidth]{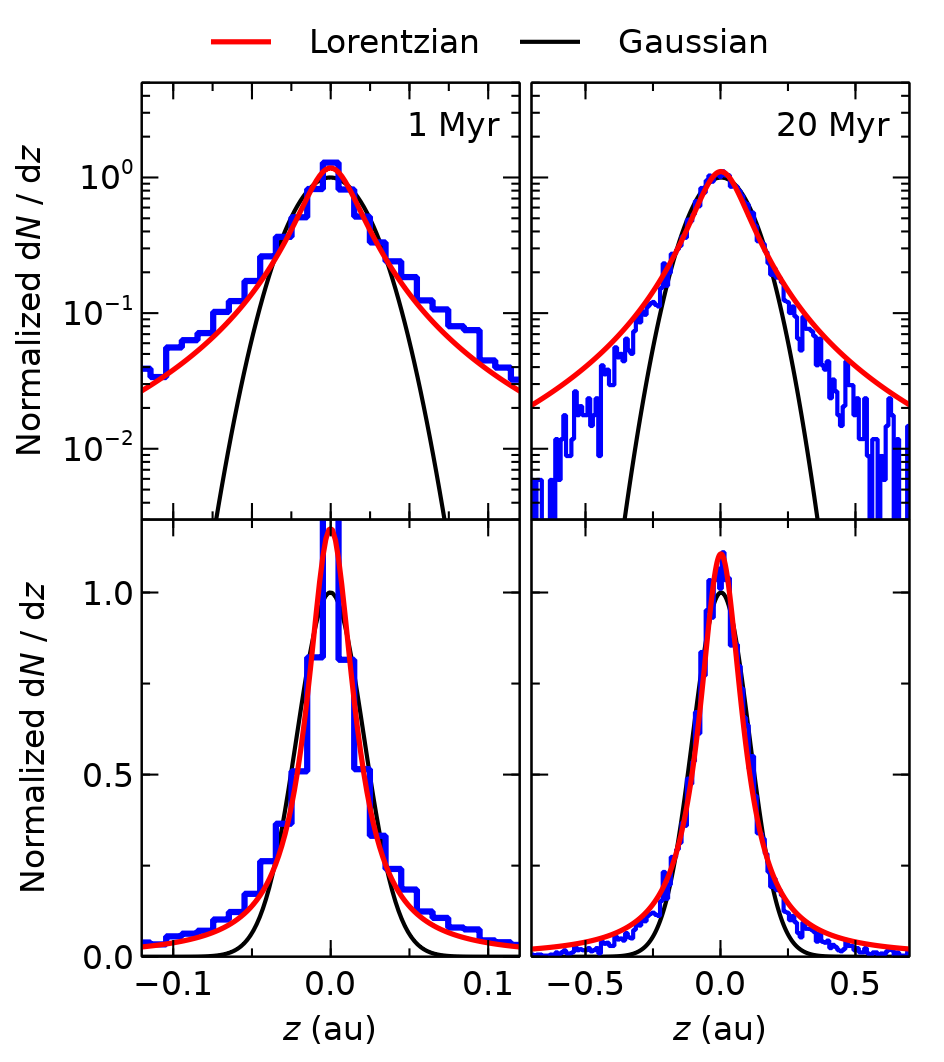}
\caption{Vertical density profiles of test particles (a.k.a.~small bodies) simulated with \texttt{REBOUND} in \texttt{Run BB}, sampled at $t = 1$ Myr (left column) and $20$ Myr (right). Log and linear scalings are shown on top and bottom rows, respectively; a linear scaling is more fairly compared to the ALMA observations for which measured disk surface brightness contrasts do not exceed a factor of $\sim$10 (see their Fig.~1). \texttt{Run BB} features small bodies (test particles) and two big bodies all with initial eccentricities = 0.1, much larger than their mutual inclinations of $10^{-5}$--$10^{-4}$ rad; accordingly, in-plane relative velocities between big and small bodies well exceed out-of-plane relative velocities most of the time. Under such circumstances big bodies force small bodies to take inclination steps $\Delta i \propto \pm \,i$ (Section \ref{sec:theory}), and small-body vertical distributions (blue histograms) should resemble Lorentzians (red curves) more than Gaussians (black curves). By $t = 20$ Myr the test particles have begun to transition from a Lorentzian to a Gaussian. Test particles having $i < 10\, i_{\rm init} = 10^{-4}$ (constituting 4.9\% of all test particles at $t = 1$ Myr, and 0.06\% at 20 Myr) have been omitted as these reflect too strongly our assumed initial inclinations. The horizontal axis changes scale by a factor of 5 between left and right columns as the disk thickens vertically with time.
  \label{fig:dNdz_twopanel_17}}
\end{figure}

\begin{figure}[H]
\centering
\includegraphics[width=\linewidth]{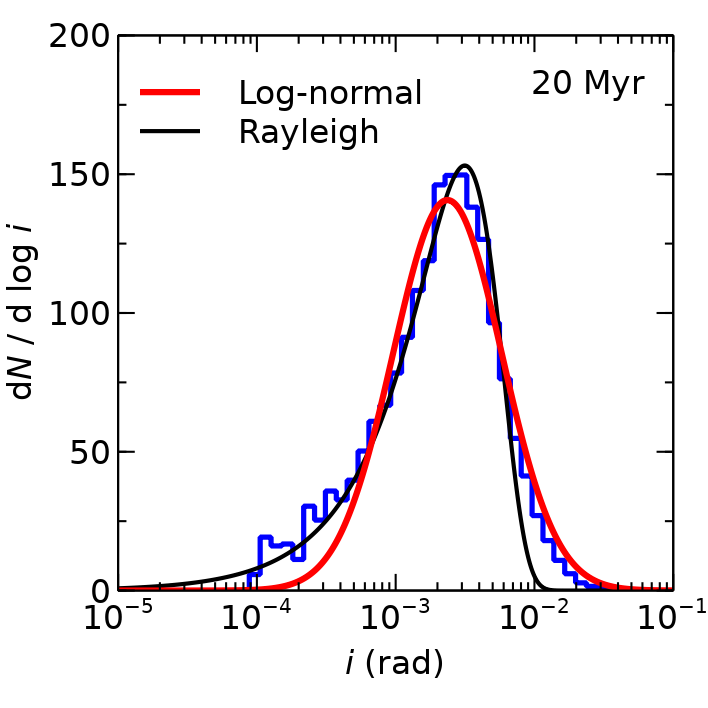}
\caption{Inclination distribution of test particles for \texttt{Run BB} (two big bodies with $e_{\rm b} = \{0.1, 0.1\}$, and test particles on initially eccentric $e_{\rm init}=0.1$ orbits with small seed mutual inclinations of $10^{-5}$--$10^{-4}$ rad relative to big bodies) at $t = 20$ Myr (blue histogram). Overlaid are a best-fit log normal distribution (red curve) and a Rayleigh distribution (black curve). The log normal fits better for large $i$, while Rayleigh fits better for small. 
%, as expected when in-plane relative velocities between the big body and test particles well exceed out-of-plane relative velocities, and small bodies random walk with fixed steps in $\Delta \log i$ (section \ref{sec:theory}).
%Initial particle inclinations are $i_{\rm init} = 10^{-5}$, and the low-$i$ tail reflects this initial condition which has not been completely forgotten.
  \label{fig:dNdi_17}}
\end{figure}

\begin{figure}[H]
\centering
\includegraphics[width=\linewidth]{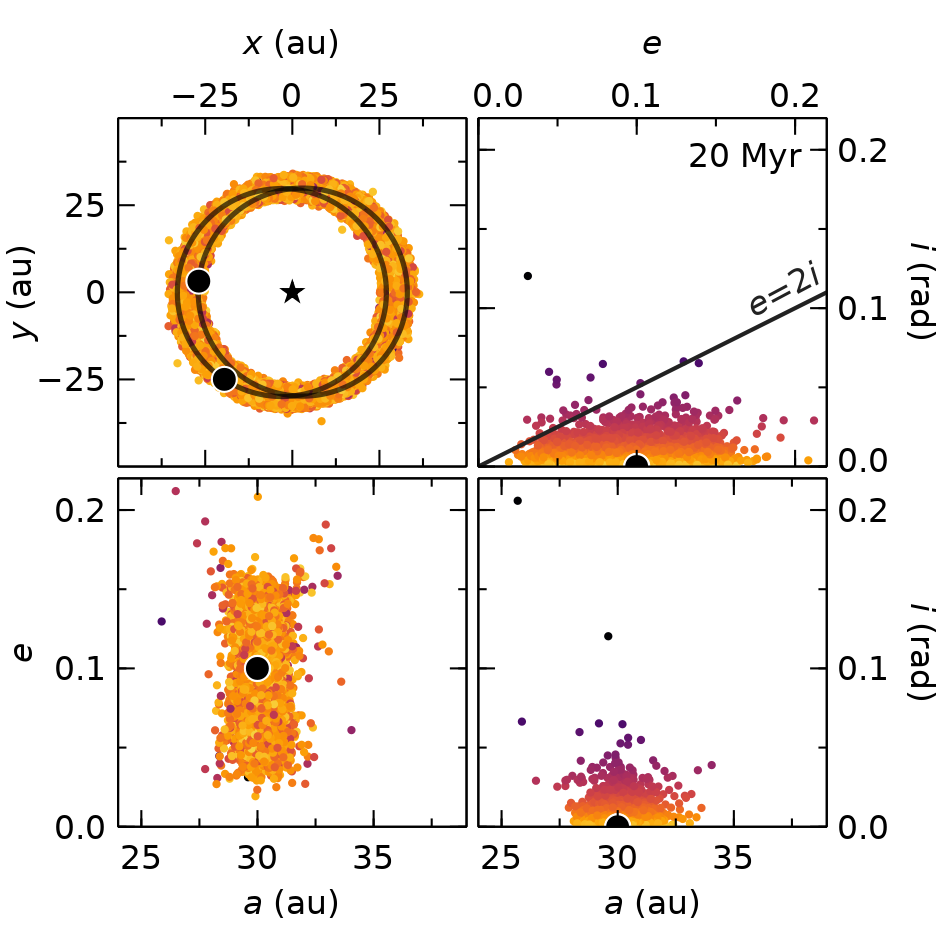}
\caption{Orbital elements for \texttt{Run BB} (two big bodies with $e_{\rm b} = 0.1, 0.1$, test particles on initially eccentric $e_{\rm init}=0.1$ orbits with small seed mutual inclinations of $10^{-5}$--$10^{-4}$ rad) at $t = 20$ Myr. In all panels, test particles are represented by colored points, and big bodies are represented by black discs. {\it Top left}: Snapshot of test particles, big bodies and their orbits, and the host star in the $x$-$y$ plane. {\it Top right}: Inclinations $i$ vs.~eccentricities $e$ for big bodies and test particles. The test particle points are colored according to their inclinations; the same color scheme is used for all panels. The $e=2i$ line marks equipartition between in-plane and out-of-plane motions; most bodies, including the big bodies, lie below this line (i.e.~$i < e/2$), as required for small-body kicks $|\Delta i|$ to scale as $i$ (Section \ref{sec:theory}). {\it Bottom left}: Eccentricities $e$ vs.~semimajor axes $a$. {\it Bottom right}: Inclinations $i$ vs.~semimajor axes $a$.
  \label{fig:params_17}}
\end{figure}

Figures \ref{fig:dNdz_twopanel_17}--\ref{fig:params_17} describe \texttt{Run BB} where equipartition is never reached. The vertical density profiles at $t = 1$ and 20 Myr are approximately Lorentzian, with the profile at the later time just starting to relax toward a Gaussian (Fig.~\ref{fig:dNdz_twopanel_17}). The inclination distribution at 20 Myr is intermediate between that of a log normal and a Rayleigh distribution (Fig.~\ref{fig:dNdi_17}), with the log normal better describing those large $i$ particles that dominate the thick tails of the vertical density profile. Eccentricities at 20 Myr remain considerably larger than inclinations (Fig.~\ref{fig:params_17}). 

\begin{figure}[H]
\centering
\includegraphics[width=\linewidth]{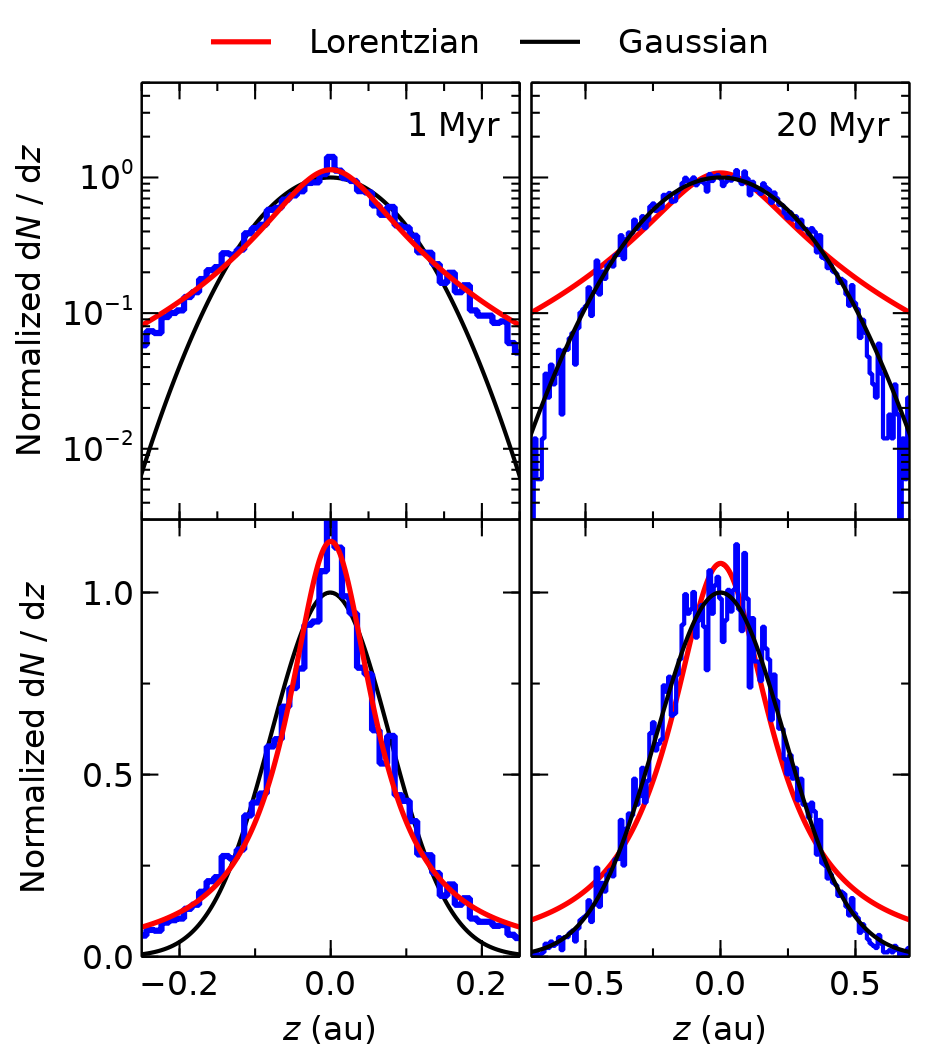}
\caption{Vertical density profiles of test particles simulated with \texttt{REBOUND} in \texttt{Run DD} which contains two big bodies on circular orbits ($e_{\rm b} = \{0, 0\}$) and initially low-eccentricity test particles ($e_{\rm init} = 0.02$). Compared to \texttt{Run BB} (Figs.~\ref{fig:dNdz_twopanel_17}--\ref{fig:params_17}) which has larger starting eccentricities, here for \texttt{Run DD} mutual inclinations become comparable to eccentricities at earlier times. At $t = 1$ Myr (left panel), the vertical profile of test particles (blue histogram) matches a Lorentzian (red curve). By $t = 20$ Myr (right panel), the vertical distribution has equilibrated to a Gaussian (black curve). Test particles having $i < 10\,i_{\rm init} = 10^{-4}$ (constituting 2.2\% of all test particles at $t = 1$ Myr, and 0.01\% at 20 Myr) have been omitted as these reflect too strongly our assumed initial inclinations. 
\label{fig:dNdz_twopanel_18}}
\end{figure}

\begin{figure}[H]
\centering
\includegraphics[width=\linewidth]{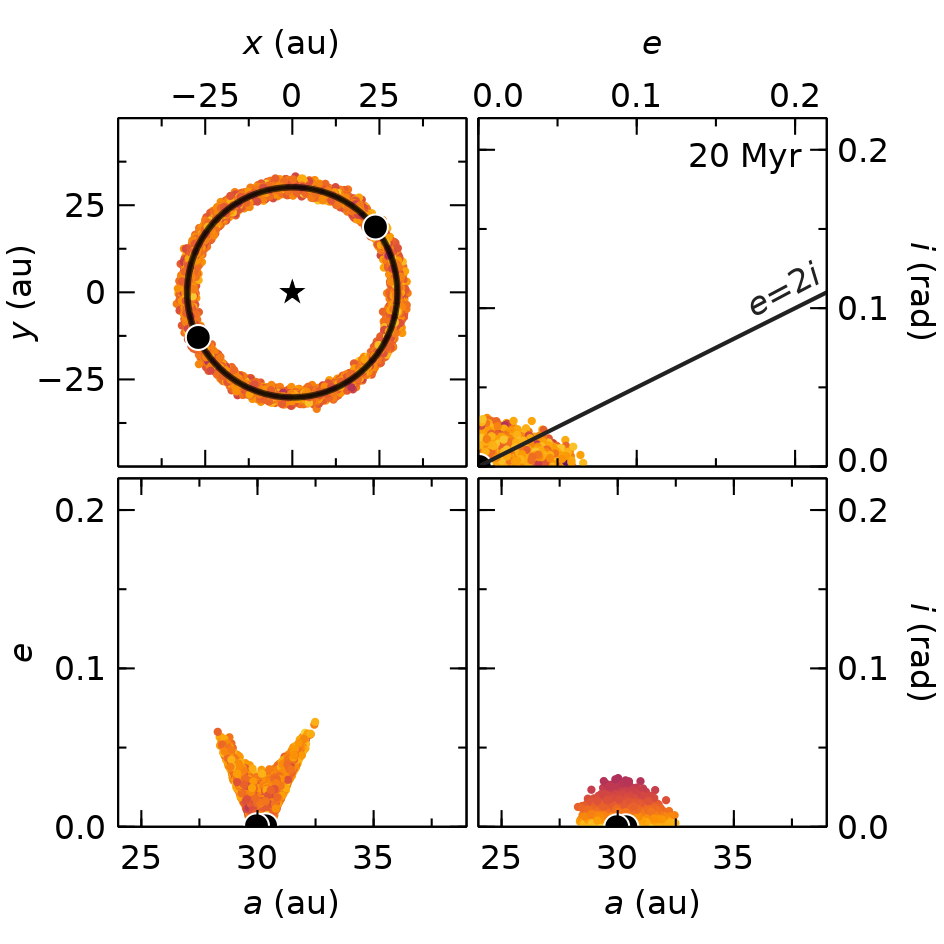}
\caption{Orbital elements for \texttt{Run DD} (two big bodies on initially circular orbits $e_{\rm b} = 0, 0$, test particles with initially low eccentricities $e_{\rm init} = 0.02$), sampled at $t = 20$ Myr. In all panels, test particles are represented by colored points, and the big bodies are represented by black discs. {\it Top left}: Snapshot of test particles, big bodies and their orbits, and the host star in the $x$-$y$ plane. {\it Top right}: Inclinations $i$ vs.~eccentricities $e$ of big bodies and test particles. The test particle points are colored according to their inclinations; the same color scheme is used for all panels. Most of the test particles are distributed about the $e=2i$ line indicating equipartition between in-plane and out-of-plane random motions. {\it Bottom left}: Eccentricities $e$ vs.~semimajor axes $a$. {\it Bottom right}: Inclinations $i$ vs.~semimajor axes $a$.
  \label{fig:params_18}}
\end{figure}

Figures \ref{fig:dNdz_twopanel_18} and \ref{fig:params_18} describe \texttt{Run DD} where a Lorentzian holds at $t = 1$ Myr, but not at $t=20$ Myr when test particles have relaxed into equipartition ($e \simeq 2 i$) and the vertical density profile is Gaussian.

%Compute $e_{\rm rms}$ vs. $i_{\rm rms}$.
Similar results are obtained for the single big body experiments: compare the all-Lorentzian runs initialized with $e_{\rm b}$, $e_{\rm init} = 0.1$ for which $e_{\rm b}, e> 2i$ (\texttt{Runs A, A-low, B, C}), against the early-Lorentzian, late-Gaussian \texttt{Run D} initialized with $e_{\rm b}=0$ and $e_{\rm init}=0.02$ for which $e > 2i$ at $t=1$ Myr, and $e \simeq 2i$ at $t = 20$ Myr. \texttt{Run A-low} is initialized with a smaller $i_{\rm init}$ than \texttt{Run A}, but the results at $t = 20$ Myr are practically indistinguishable. Additional technical details about \texttt{Runs A}, \texttt{B}, \texttt{C}, and \texttt{D} are relegated to Appendix \ref{sec:extra_app}.

\texttt{Runs E} and \texttt{EE} initialize both small and big bodies with zero eccentricity, and consequently $e$'s and $i$'s take longer than in other runs to grow. At $t = 1$ Myr in \texttt{Run E}, test particle inclinations remain largely at their initial values and a measurement of the vertical density profile at this time is not meaningful. At $t = 1$ Myr in \texttt{Run EE}, and $t = 20$ Myr in \texttt{Run E}, the $e$'s and $i$'s have grown, with $e \gg 2i$  and the vertical density profiles exhibiting Lorentzians (Table \ref{tab:paras}). From these results we infer that when eccentricities are initially zero and inclinations are non-zero but small, the eccentricities at first grow faster than and leapfrog over the inclinations; this is consistent with how viscous stirring under shear-dominated (sub-Hill) conditions excites eccentricities more than inclinations (\citealt{ida92}, their Figure 6; \citealt{goldreich04}). Eventually though, as the particles become dispersion-dominated (super-Hill), the inclinations catch up, as demonstrated by \texttt{Run EE} at $t = 20$ Myr (see also Sections \ref{subsec:time} and \ref{subsec:compare}).

%%% 

%\begin{figure}
%\centering
%\includegraphics[width=\linewidth]{sim18IHist20Myr.png}
%\caption{Inclination distribution of test particles at $t = 20$ Myr for \texttt{Run 18} (two big bodies with initially circular orbits $e_{\rm b} = 0, 0$, and test particles with initially low eccentricities $e_{\rm init} = 0.02$). A Rayleigh distribution (black curve) fits the data (blue histogram) better than a log normal (red curve), consistent with test particles random walking with fixed inclination steps $\Delta i =$ constant. As in other figures and fits, particles with $i < 10^{-4}$ have been omitted.
%  \label{fig:dNdi_18}}
%\end{figure}

%%%
\subsection{Numerically simulated  stirring rates compared with analytics}\label{subsec:compare}
\begin{figure}[H]
\centering
\includegraphics[width=\linewidth]{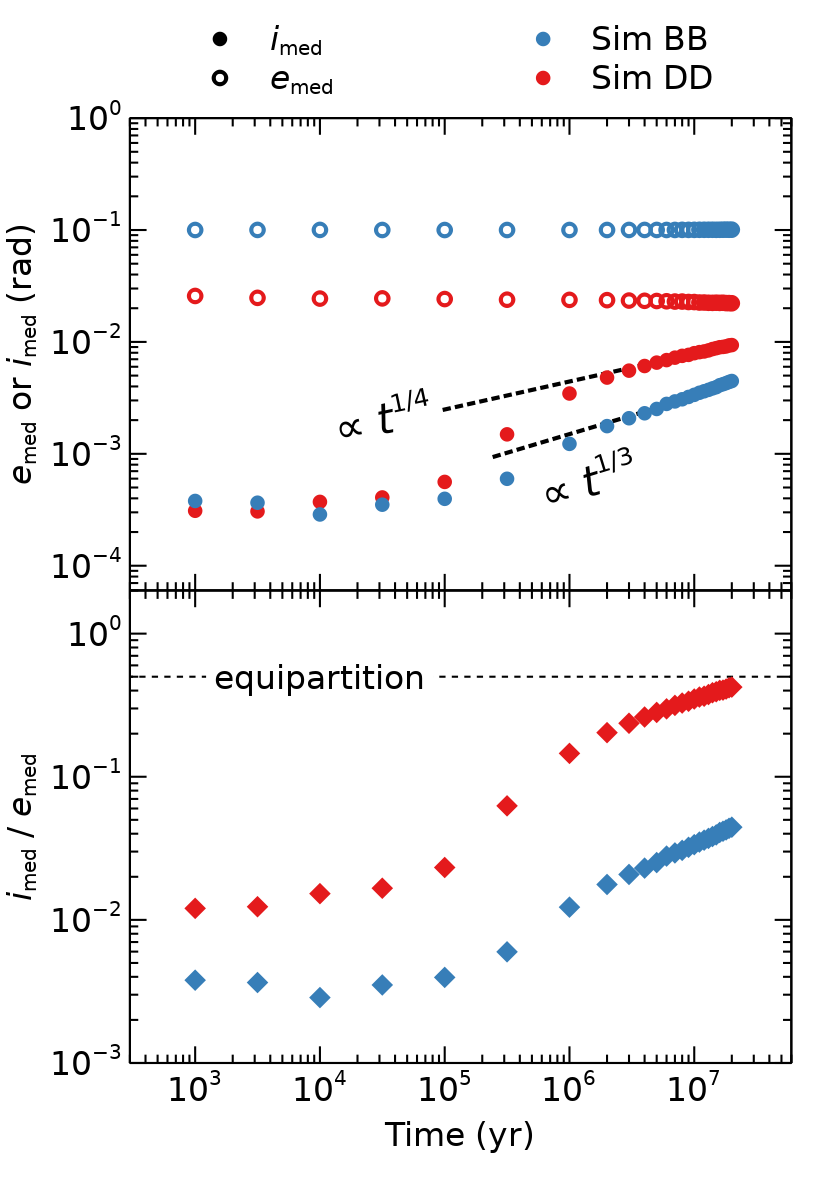}
\caption{Median $e$ and median $i$ of test particles vs.~time for the same runs shown in previous figures. 
%, plus \texttt{Runs A} and \texttt{A-low}.
%which starts at a lower $i_{\rm init}$ compared to \texttt{Run A}. 
Inclinations rise faster than eccentricities. In \texttt{Run DD}, $i$ catches up to within a factor of 2 of $e$ and eventually grows as $t^{1/4}$ as $i/e$ nears the equipartition value of 1/2 (dotted line); accordingly, the particles in \texttt{DD} conform to a vertical Gaussian at late times (Fig.~\ref{fig:dNdz_twopanel_18}). The median $e$ decreases as the median $i$ increases, reflecting how relative motions parallel to the disk midplane are re-directed into perpendicular motions as equipartition is approached. Inclinations in \texttt{Run BB} also grow --- slightly faster toward the end of the simulation as $t^{1/3}$, as predicted in the 3D-anisotropic regime (Section \ref{subsubsec:3dani}) --- but fall short of eccentricities by more than a factor of 2; the corresponding vertical profiles are Lorentzian-like (Fig.~\ref{fig:dNdz_twopanel_17}).  
%The eccentricities in \texttt{Runs A} and \texttt{A-low} rise at late times nearly linearly, we suspect because of secular driving (non-viscous stirring) by the single eccentric planet in these runs. This growth in $e$ is not seen in the other two-planet runs, presumably because the planet eccentricities are zero (\texttt{Run DD}) or because secular driving is muted when the two planets are apsidally anti-aligned (\texttt{Run BB}).
  \label{fig:time}}
\end{figure}

Figure \ref{fig:time} shows how the median eccentricities and inclinations of our test particles evolve with time, for the same runs featured in previous figures. As noted above, while both medians and rms values yield qualitatively similar outcomes, we found the medians to be less sensitive to ejections and to better represent the majority of particles making up the bulk of the vertical profile. As anticipated from Section \ref{subsec:time}, inclinations (either median or rms) grow faster than eccentricities. In \texttt{Run DD}, which exhibits a vertical Gaussian at late times, $i$ catches up to $e$ to within a factor of 2, eventually growing as $t^{1/4}$ as expected in equipartition. As the energies in motions parallel and perpendicular to the midplane are exchanged, inclinations grow at the expense of eccentricities --- the decrease in $e$ is most visible in \texttt{Run DD} at late times when $i$ is largest. In \texttt{Run BB} which features vertical Lorentzians at all times, inclinations also grow but remain smaller than eccentricities by more than a factor of 2.

We can compare the analytic estimates of stirring timescales made in Section \ref{subsec:time} to those in Figure \ref{fig:time}. For \texttt{Run BB}, we have $\mu_{\rm b} = 1.5\times 10^{-8}$ and $e \simeq 0.1$, and estimate $\Sigma_{\rm b} \sim 2 \times m_{\rm b} / (2\pi a \times 2ea) \sim 2 \times 10^{-4}$ g/cm$^2$; it follows that $\mu_{\rm b}/e^2 \simeq 1.5 \times 10^{-6} < \sqrt{\mu_{\rm b}/e} \simeq 4 \times 10^{-4}$. At the earliest times in \texttt{Run BB}, $i \simeq 4 \times 10^{-4} \sim \sqrt{\mu_{\rm b}/e}$; thus the system starts on the border between the 2D-$i$/3D-$e$ and 3D-anisotropic regimes. From eqs.~(\ref{eq:subeq_3dani})--(\ref{eq:subeq_3de2di}),
\begin{subequations}
\begin{align}
t_{i,{\rm 2D}-i/{\rm 3D}-e} &\sim t_{i,{\rm 3D-ani}} \sim 10^5 \, {\rm yr} \\
t_{e,{\rm 2D}-i/{\rm 3D}-e} &\sim t_{e,{\rm 3D-ani}} \sim 10^9 \, {\rm yr} \,.
\end{align}
\end{subequations}
The estimated inclination doubling time is shorter than that shown in Figure \ref{fig:time}, but only by factor of several ($i$ doubles from $t = 10^5$ yr to $t = 7 \times 10^5$ yr). At least part of the discrepancy can be attributed to our implicit inclusion of a Coulomb log in the 2D-$i$ regime when there should be none (see discussion following eq.~\ref{eq:ti2danumcoe}). The estimated eccentricity doubling time is consistent with Figure \ref{fig:time} ($e$ does not change).

At the end of \texttt{Run BB}, at $t = 2\times 10^7$ yr, $\sqrt{\mu_{\rm b}/e} < i \simeq 4 \times 10^{-3} < e$; now the system is squarely in the 3D-anisotropic regime, and
\begin{subequations}
\begin{align}
t_{i,{\rm 3D-ani}} &\sim 10^8 \, {\rm yr} \\
%\end{align}
%\begin{align}
t_{e,{\rm 3D-ani}} &\sim 10^{10} \, {\rm yr} 
\end{align}
\end{subequations}
both consistent with Figure \ref{fig:time}. We also see in Figure \ref{fig:time} how $i \propto t^{1/3}$ as predicted for the 3D-anisotropic regime.

Similarly good agreement between analytics and numerics is obtained for \texttt{Run DD}, which features a smaller $e \simeq 0.02$ and therefore faster stirring timescales overall. Here $t_{e,{\rm equi}} \sim 6 \times 10^7$ yr from eq.~(\ref{eq:eq}), within a factor of 3 of the numerical time to reach equipartition, $t \simeq 2\times 10^7$ yr.

In \texttt{Run BB}, the system ends at $t = 20$ Myr in the 3D-anisotropic regime where individual inclination kicks $\Delta i$ do not scale with $i$ (eq.~\ref{eq:nomorei}). The vertical density profile has dropped at this time from a pure Lorentzian, but still retains broad non-Gaussian wings (Fig.~\ref{fig:dNdz_twopanel_17}). The persistence of non-Gaussian wings in the 3D-anisotropic regime helps to justify our assumption in Section \ref{sec:app} that real-life debris disks are in this regime.

%%%%%%%%%%%%%%%%%%%%%%%%%%%%%%%%%%%%%%%%%%%%%%%%%%%%%%%%%%%%
\begingroup % to localize the effect of the length modification below
\setlength{\medmuskip}{0mu} % to reduce the spacing around binary operators (\times)
\begin{deluxetable*}{ccccccccccc}[ht]
  \tablecaption{Radially narrow debris disks from \citetalias{zawadzki26}}\label{tab:disks}
  \tablecolumns{9}
  \tablehead{
    \colhead{(1)} &
    \colhead{(2)} &
    \colhead{(3)} &
    \colhead{(4)} &
    \colhead{(5)} &
    \colhead{(6)} &
    \colhead{(7)} &
    \colhead{(8)} &
    \colhead{(9)} &
    \colhead{(10)} &
    \colhead{(11)} \\
    \colhead{{Disk ID}} &
    \colhead{Vertical form} & 
    \colhead{$M_\star$} & 
    \colhead{$t_{\rm age}$} & 
    \colhead{$R$} &
    \colhead{$\Delta R_{\rm FWHM}$} &
    \colhead{$\Delta R_{\rm FWHM}/R$} &
    \colhead{$h_{\rm HWHM}$} &
    \colhead{$i$} &
    \colhead{$e$} & 
    \colhead{$\max m_{\rm b} = \min M_{\rm b,tot}$} \\
    \colhead{} &
    \colhead{} & 
    \colhead{($M_\star$)} &
    \colhead{(Myr)} &
    \colhead{(au)} &
    \colhead{(au)} &
    \colhead{} &
    \colhead{} &
    \colhead{} &
    \colhead{} & 
    \colhead{($M_\oplus$)} 
  }
  \startdata
  \hline
  HD 15115 & vL  & $1.426$  & $45$ & $80.7$ & $14$ & 0.17 & 
  $0.021$ & 0.025 & 0.1 & 0.4  \\ 
    HD 32297  & vL  & $1.57$  & $30$ & $115.9$ & $37$ & 0.32 & 
  $0.0098$ & 0.012 & 0.047--0.16 & 0.2--0.4  \\ 
      HD 61005 & vL  & $0.95$  & $45$ & $72.4$ & $36$ & 0.50 & 
  $0.0129$ & 0.015 & 0.062--0.25 & 0.2--0.4  \\ 
        HD 197481 (AU Mic) & vL   & $0.614$  & $23$ & $33.6$ & $13$ & 0.39 & 
  $0.0026$ & 0.003 & 0.012--0.193 & 0.004--0.01  \\ 
  HD 109573 & vG & 2.14 & 10 & 75.9 & 7 & 0.09 & 0.0065 & 0.0078 & 0.016 & 0.06  \\
  HD 131488 & vG & 1.804 & 16 & 89.7 & 13 & 0.14 & 0.0048 & 0.0058 & 0.012 & 0.03 \\
  \enddata
  \tablecomments{Columns: 
  (1) Disk name.  (2) Best-fitting vertical profile from \citetalias{zawadzki26}, vL = Lorentzian, vG = Gaussian. 
  (3) Host stellar mass (\citetalias{marino26}).  
  (4) System age (\citetalias{marino26}).  
  (5) Disk radius, as determined from fits to mm-wave continuum images assuming the dust surface density follows a single Gaussian in orbital radius (\citetalias{marino26}). For HD 15115, two Gaussians (separated in radius by $\sim$30\%; see e.g.~\citetalias{zawadzki26} Table A.1) are required to fit the visibilities, and the tabulated $R$ is the average of their centers (\citetalias{marino26}).
  (6) Disk annular width, as measured by the full width at half-maximum (FWHM) of an assumed Gaussian dust surface density in radius (\citetalias{marino26}). For HD 15115, two Gaussians are needed to fit the data, and the reported $\Delta R$ is the sum of their widths. (7) Disk fractional width in radius. 
  (8) Best-fit disk aspect ratio $h_{\rm HWHM} \equiv H_{\rm HWHM}/R$ from \citetalias{zawadzki26} (their Table 2), where $H_{\rm HWHM}$ is the vertical height where the dust density falls to half its midplane value. (9) Inclinations of disk particles estimated as $i = h_{\rm HWHM}/\sqrt{\ln 2}$ (formally valid for rms $i$ and vertical Gaussians)  (10) Eccentricities of disk particles (free components only), either equal to the equipartition value of $2i$ if the disk is vertically Gaussian (vG), or ranging from $e_{\rm min} = 4i$ (eq.~\ref{eq:emin}) to $e_{\rm max} = 0.5 \, \Delta R_{\rm FWHM}/R$ (eq.~\ref{eq:emax}) for vertically Lorentzian (vL) disks. For HD 15115, $e_{\rm min} =0.1$ and $e_{\rm max} = 0.09$ and so we set $e = 0.1$. 
  (11) Big body mass $m_{\rm b}$ in the case where there is only one big body stirring the entire disk. This is also the same as the minimum total mass in big bodies $M_{\rm b,tot}$, since $M_{\rm b,tot} \propto 1/m_{\rm b}$ (eqs.~\ref{eq:mbtot} and \ref{eq:mbtot_3dani}). Ranges in $\max m_{\rm b} = \min M_{\rm b,tot}$ are shown for those Lorentzian disks with a range of eccentricities between $e_{\rm min}$ and $e_{\rm max}$.
  }
\end{deluxetable*}
\endgroup

\section{Application to real-life disks}\label{sec:app}
We apply the stirring theory of Section \ref{sec:theory} to estimate the individual and collective masses of big bodies within \citetalias{zawadzki26} debris disks, as observed in the mm continuum. 

\subsection{Sample}
We restrict consideration to radially narrow disks having annular widths $\Delta R$ less than 0.5 $\times$ their mean orbital radius $R$, since their dynamics should be local and simpler to interpret (see also Section \ref{subsec:alt}). Table \ref{tab:disks} lists those disks with measured aspect ratios $h_{\rm HWHM}$ from \citetalias{zawadzki26}, and $R$ and $\Delta R = \Delta R_{\rm FWHM}$ from \citetalias{marino26}. The latter derive from parametric fits to dust surface densities assumed to be radially Gaussian. While more complicated radial structures were considered by ARKS II \citep{han26} and \citetalias{zawadzki26}, all fitting methods agree that the sources listed in Table \ref{tab:disks} can be described as radially narrow rings. The two neighboring rings detected in HD 15115 have been merged into one for simplicity (\citetalias{marino26}); our conclusions about this system do not change qualitatively were we to keep the rings separate.

\subsection{Procedure and rationale}\label{subsec:procrat}
The stirring rate formulae in Section \ref{subsec:time} provide joint constraints on the big body surface densities $\Sigma_{\rm b}$ and  individual masses $m_{\rm b} = \mu_{\rm b}m_\star$. We assume the two Gaussian (``vG'') disks in Table \ref{tab:disks} (HD 109573 and HD 131488) have been self-stirred into equipartition, and accordingly use eq.~(\ref{eq:eq}), with the left-hand side set equal to the system age $t_{\rm age}$, to solve for $\Sigma_{\rm b}$ as a function of $m_{\rm b}$. The surface density is used to evaluate the total mass in big bodies in a disk annulus of area $2\pi R \Delta R$: 
\begin{align}
 M_{\rm b,tot}|_{\rm equi} \simeq & \,\,2\pi \Sigma_{\rm b} R \Delta R \sim \frac{2\pi e^4}{40} \,\frac{\Delta R}{R} \frac{1}{nt_{\rm age}} \frac{m_\star}{m_{\rm b}} m_\star \nonumber \\
\sim  & \,\,500 \,M_\oplus \left( \frac{e}{0.1} \right)^4 \left( \frac{\Delta R / R}{0.1} \right) \left( \frac{30 \, {\rm Myr}}{t_{\rm age}} \right) \nonumber  \\
&  \,\,\,\,\,\,\, \times \left( \frac{R}{100 \,{\rm au}} \right)^{3/2} \left( \frac{0.002\,m_\oplus}{m_{\rm b}} \right) \left( \frac{m_\star}{m_\odot} \right)^{3/2}  \label{eq:mbtot}
\end{align}
where for the sample numerical evaluation we used Pluto-like big bodies with $m_{\rm b} = 0.002 \, m_\oplus$. Eccentricities are evaluated assuming equipartition:
\begin{align} \label{eq:equipe}
e = 2 i = 2\sqrt{2} \,h_{\sigma} = \frac{2}{\sqrt{\ln 2}} \,h_{\rm HWHM} \,\,\,\,\,\, ({\rm vertical \, Gaussian})
\end{align}
where $e$ and $i$ are here formally their rms values, and we have distinguished between the disk scale height measured as a standard deviation ($h_\sigma$), and as a half width at half maximum ($h_{\rm HWHM}$) \citep{matra19,ida92} --- see Table \ref{tab:disks}. The largest possible big-body mass, $\max m_{\rm b}$, is the same as the smallest possible total mass, $\min M_{\rm b,tot}$, and corresponds to having a single big body in the disk annulus: 
\begin{align}
\max m_{\rm b}|_{\rm equi} & = \min M_{\rm b,tot}|_{\rm equi} \nonumber \\
& \sim \left( \frac{2\pi e^4}{40}\frac{1}{nt_{\rm age}} \frac{\Delta R}{R}  \right)^{1/2} m_\star \nonumber \\
& \sim \,\, 1 \,M_\oplus \left( \frac{e}{0.1} \right)^2 \left( \frac{\Delta R / R}{0.1} \right)^{1/2} \left( \frac{30 \, {\rm Myr}}{t_{\rm age}} \right)^{1/2} \nonumber \\
& \,\,\,\,\,\,\,\, \times 
\left( \frac{R}{100 \,{\rm au}} \right)^{3/4} \,.\label{eq:square}
\end{align}

The procedure described so far is largely identical to the one carried out by Jankovic et al.~(2026; see also \citealt{matra19} and \citealt{pearce25}). Their equation (20) (taken from \citealt{ida93}) matches our equation (\ref{eq:mbtot}) to within an order-unity factor. All of these studies assume equipartition stirring.

Where our study diverges from others is in our interpretation of the vertically Lorentzian (``vL'') disks in Table \ref{tab:disks} (HD 15115, HD 32297, HD 61005, AU Mic). These disks we assume are vertically stirred in the out-of-equipartition, 3D-anisotropic regime (Sections \ref{subsubsec:3dani}, \ref{subsubsec:numcoe}). There are three out-of-equipartition regimes, with the 3D-anisotropic regime opening the largest parameter space ($\mu_{\rm b} \lesssim i^2e \sim 10^{-5}$; an individual big body can be anything less massive than a super-Earth). Although 3D-anisotropic stirring acts to convert the Lorentzians produced in earlier out-of-equipartition stirring into Gaussians, the numerical simulations of Section \ref{sec:num} show that the conversion does not necessarily complete and can leave intact broad non-Gaussian profiles (e.g.~Fig.~\ref{fig:dNdz_twopanel_17}).

Accordingly we use the 3D-anisotropic eq.~(\ref{eq:ti3dnumcoe}), with the left-hand side set equal to $t_{\rm age}$, to solve for $\Sigma_{\rm b}$ as a function of $m_{\rm b}$, and from there the total disk mass:
\begin{align}
 M_{\rm b,tot}|_{\rm 3D-ani} \simeq & \,\,2\pi \Sigma_{\rm b} R \Delta R \sim \frac{2\pi (2i)^3e}{40} \,\frac{\Delta R}{R} \frac{1}{nt_{\rm age}} \frac{m_\star}{m_{\rm b}} m_\star \nonumber \\
\sim  & \,\,4 \,M_\oplus \left( \frac{e}{0.1} \right) \left( \frac{i}{0.01} \right)^3 \left( \frac{\Delta R / R}{0.1} \right) \nonumber \\
&  \times \left( \frac{30 \, {\rm Myr}}{t_{\rm age}} \right) \left( \frac{R}{100 \,{\rm au}} \right)^{3/2} \left( \frac{0.002\,m_\oplus}{m_{\rm b}} \right) \nonumber  \\
&   \times \left( \frac{m_\star}{m_\odot} \right)^{3/2}  \,.\label{eq:mbtot_3dani}
\end{align}
%Contrast the lower out-of-equipartition disk mass in eq.~(\ref{eq:mbtot_3dani}) with the equipartition disk mass in eq.~(\ref{eq:mbtot}). 
Under out-of-equipartition $e > 2i$ conditions, we still take $i = \sqrt{2}h_\sigma$,\footnote{Technically the $\sqrt{2}$ applies only to Gaussian profiles, but we adopt it anyway for convenience and so that our Lorentzian and Gaussian treatments share a common reference.} but now consider a range of eccentricities between 
\begin{align} \label{eq:emin}
e_{\rm min} \equiv 2 \times 2i & = 4\sqrt{2} \, h_\sigma \nonumber \\ &= \frac{4}{\sqrt{\ln 2}} \,h_{\rm HWHM} \,\,\,\,\,\, ({\rm vertical \, Lorentzian}) 
\end{align}
and
\begin{align} \label{eq:emax}
e_{\rm max} = 0.5 \times \Delta R_{\rm FWHM} / R  \,\,\,\,\,\, ({\rm vertical \, Lorentzian})
\end{align}
(Table \ref{tab:disks}). For HD 15115, $e_{\rm min} = 0.10$ and $e_{\rm max} = 0.09$, and so for this disk we set $e = 0.1$. From eq.~(\ref{eq:mbtot_3dani}) the mass of the largest possible (single) big body is
\begin{align}
\max m_{\rm b}|_{\rm 3D-ani} & = \min M_{\rm b,tot}|_{\rm 3D-ani} \nonumber \\
& \sim \left( \frac{2\pi (2i)^3e}{40}\frac{1}{nt_{\rm age}} \frac{\Delta R}{R}  \right)^{1/2} m_\star \nonumber \\
& \sim \,\, 0.1 \,M_\oplus \left( \frac{e}{0.1} \right)^{1/2} \left( \frac{i}{0.01} \right)^{3/2}  \nonumber \\
& \,\,\,\,\,\,\,\, \times \left( \frac{\Delta R / R}{0.3} \right)^{1/2}
\left( \frac{30 \, {\rm Myr}}{t_{\rm age}} \right)^{1/2} \nonumber \\
& \,\,\,\,\,\,\,\, \times \left( \frac{R}{100 \,{\rm au}} \right)^{3/4} 
\,.\label{eq:square_3dani}
\end{align}

Under our out-of-equipartition  interpretation, the observed inclinations of vertically Lorentzian disks are viscously stirred by big bodies, but the eccentricities are not; the latter are somehow generated by another mechanism (see Section \ref{subsec:ideas} for some ideas).
%The constraint in eq.~(\ref{eq:joint}) depends strongly on the present-day eccentricity $e$, interpreted as the quadrature sum of the small and big body eccentricities (see discussion between eqs.~\ref{eq:mbtot} and \ref{eq:square}). 

Our procedure assumes that collisional damping of eccentricities and inclinations is negligible. The no-damping assumption is valid when collisional strengths $Q^\ast$ are small compared to the square of relative velocities $v_{\rm rel}^2$ in the collisional cascade, so that bodies of a given mass fragment from collisions with projectiles of much smaller mass.  Such high-mass-ratio collisions do not damp relative velocities efficiently (\citealt{jankovic24}; Jankovic et al.~2026). Whether the no-damping condition holds is model-dependent, and damping might actually be significant for the smallest eccentricities (smallest $v_{\rm rel}$) we consider. Nevertheless insofar as collisional damping is expected to drive disks to equipartition and thereby relax into vertically Gaussian profiles, the Lorentzian disks in our sample evidently escaped damping.

%Allowing for a range of eccentricities from $e_{\rm min}$ to $e_{\rm max}$ is where we differ from Jankovic et al.~(2026), who instead use the equipartition relation (eq.~\ref{eq:equi}) for all disks regardless of whether they better fit vertical Lorentzians or Gaussians. (Their sample size is also larger, as they do not select for radially narrow disks.) 
 
%Our use of $e_{\rm max}$ to derive a maximum $M_{\rm b,tot}$ as a function of $m_{\rm b}$  is the same procedure used by \citet{pearce25} to constrain the big body properties in the Fomalhaut debris disk; our resultant formulas (their eq.~5 vs.~our eq.~\ref{eq:mbtot} with eq.~\ref{eq:emax} inserted) differ only by an order-unity factor.

A final note on $e$ as it has been used throughout this paper: these are free eccentricities, not forced eccentricities. The latter refer to mean values characterizing globally eccentric disks such as Fomalhaut (e.g.~\citealt{kalas05,macgregor17,lovell25}). For the disks in our sample, Lovell et al.~(\citeyear{lovell26}, ARKS VI; and references therein) measured forced eccentricities of $\sim$0.1 for HD 109573, $> 0.03$ for HD 32297, and $> 0.02$ for HD 15115. These forced eccentricities are not directly relevant here because they refer to mean orbital streamlines that do not cross. We are concerned instead with gravitational scatterings in the dispersion-dominated, orbit-crossing regime, where relative velocities are controlled by free (a.k.a.~proper) eccentricities (e.g.~\citealt{murray99}).

\begin{figure*}[t]
\centering
\includegraphics[width=\textwidth]{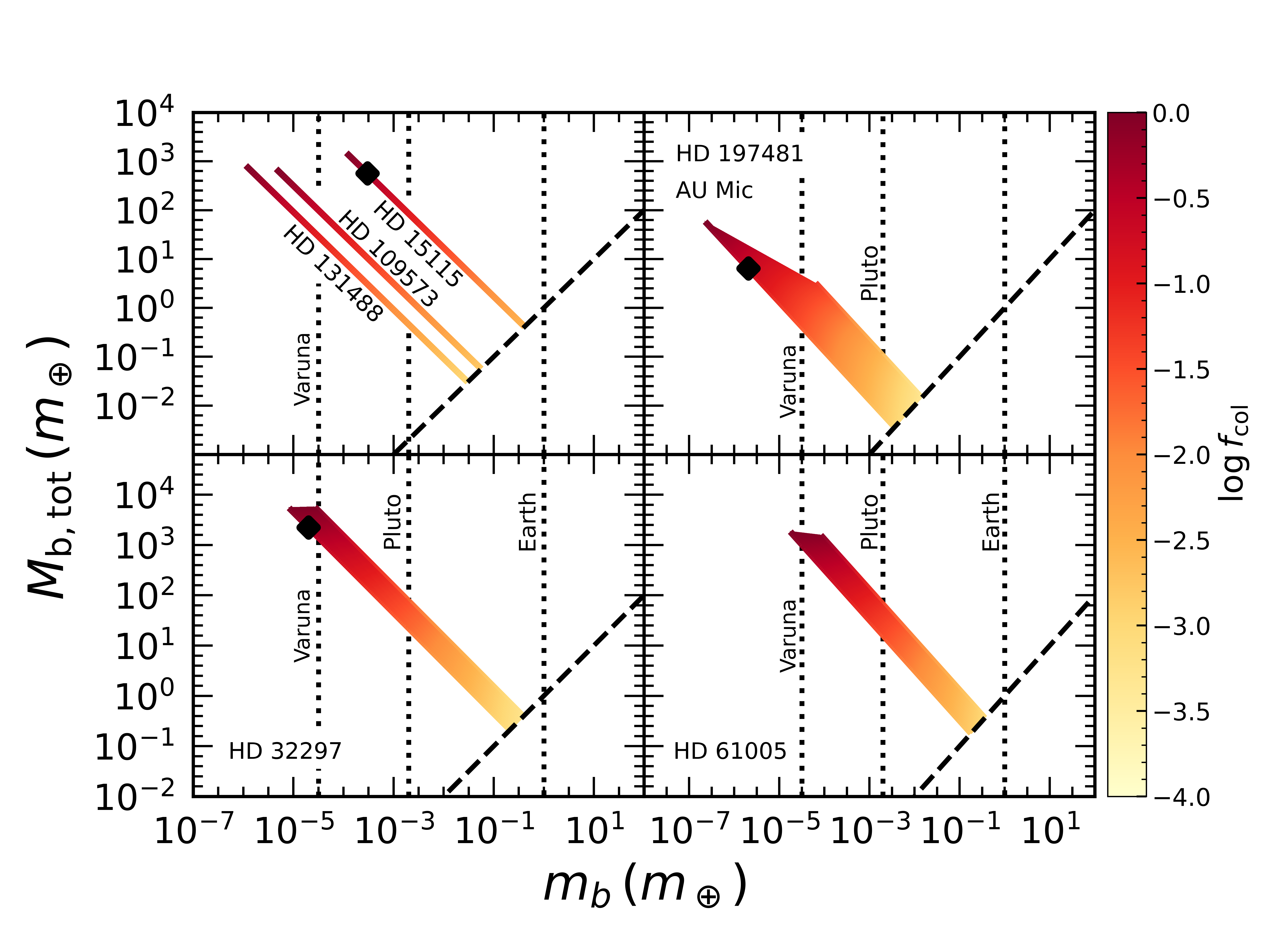}
\caption{Possible individual big body masses $m_{\rm b}$, and total masses in all big bodies $M_{\rm b,tot} = 2\pi \Sigma_{\rm b} R \Delta R_{\rm FWHM}$, for the six radially narrow rings in Table \ref{tab:disks}. The loci are computed from either the equipartition eq.~(\ref{eq:mbtot}) applied to the vertically Gaussian disks HD 109573 and HD 131488, or the out-of-equipartition relation (\ref{eq:mbtot_3dani}) applied to the vertically Lorentzian disks HD 15115, HD 32297, HD 610005 HD 197481 (AU Mic); in all cases a given disk's present-day inclinations $i$ are assumed to result from viscous self-stirring over the disk age $t_{\rm age}$, with no collisional damping. The free eccentricity $e$ is either estimated from its equipartition value ($2i$, where $i=h_{\rm HWHM}/\sqrt{\ln 2}$ and $h_{\rm HWHM}$ is an observationally fitted disk aspect ratio), or bounded by the out-of-equipartition values $e_{\rm min}$ (eq.~\ref{eq:emin}) and $e_{\rm max}$ (eq.~\ref{eq:emax}). In the case of HD 15115 for which $e_{\rm min} = 0.1$ and $e_{\rm max} = 0.09$, we set $e = 0.1$. For each of the three disks with definite $e$ (top left panel), the loci of possibilities is a line of slope -1, since $M_{\rm b,tot} \propto m_{\rm b}^{-1}$ by eq.~(\ref{eq:mbtot}). Each line starts at top left at the value of $m_{\rm b}$ for which the fraction $f_{\rm col}$ of big bodies colliding over $t_{\rm age}$ equals one (see color bar), and ends at bottom right at the $m_{\rm b} = M_{\rm b,tot}$ dashed line corresponding to a single big body. Each of the other disks (top right and bottom panels) features a continuum of slope -1 lines, each line corresponding to a different $e > 2i$. The disks HD 15115, HD 32297, and AU Mic are claimed to exhibit the remains of recent catastrophic disruptions of dwarf planets \citep{jones23,chiang17}. For these disks we indicate with black diamonds the combinations of $m_{\rm b}$, $M_{\rm b,tot}$ that generate enough debris in erosive collisions to supply the mass being ground down in standard collisional cascades, as constrained by disk brightnesses at infrared to mm-wave wavelengths (Jankovic et al.~2026). See Table \ref{tab:match} for the full set of parameters, and Figure \ref{fig:locus_Ncol} for a companion plot. 
\label{fig:locus}}
\end{figure*}

\subsection{Results for individual and total big body masses in  \citetalias{zawadzki26} disks}\label{subsec:results_app}

Figure \ref{fig:locus} shows our results for total disk mass $M_{\rm b,tot}$ vs.~individual big body mass $m_{\rm b}$. For each disk with a definite $e$ (HD 15115, plus the two vG disks assumed to be in equipartition), the possible combinations of $M_{\rm b,tot}$ vs.~$m_{\rm b}$ trace a line of slope -1 ($M_{\rm b,tot} \propto 1/m_{\rm b}$). For each Lorentzian disk having a range of possible non-equipartition $e$'s, the possibilities for $M_{\rm b,tot}$ vs.~$m_{\rm b}$ fill a 2D region that is bounded from above by the slope -1 line for $e = e_{\rm max}$, and from below by the slope -1 line for $e = e_{\rm min}$.

Our Figure \ref{fig:locus} should be compared with the colored lines in the left panel (no collisional damping) of Figure 5 of Jankovic et al.~(2026), who assume all disks are in equipartition whether they exhibit vertical Gaussians or Lorentzians. The differences between our figure and theirs are modest. The similarity arises partly because we assume a 3D anisotropic stirring timescale for inclination (eq.~\ref{eq:ti3dnumcoe}) that differs from the equipartition stirring time (eq.~\ref{eq:eq}) by only a factor of $e/(2i)$, which mostly stays $< 10$ (our study and theirs use the same $i$, while ours allows for larger $e$). Jankovic et al. (2026) argue that collisional damping of $i$ and $e$ may be significant for HD 32297 and HD 197481 = AU Mic, in which case the stirring masses for these disks become larger (compare the left and right panels of their Figure 5). It is not clear how collisional damping, which we would expect to result in equipartition and Gaussian vertical profiles, can be compatible with the ARKS III finding of vertical Lorentzians. In any case, our respective studies overlap in predicting stirring bodies ranging in size from Varuna to Mars, with total masses between $\sim$0.1 and $\sim$1000 Earth masses (see also their Figure 7).
%Our vertically Gaussian disks (HD 109573, HD 131488) are not in their sample. 
%For the vertically Gaussian disks that both our studies assume to be in equipartition (HD 109573, HD 131488), there is agreement. 
%For the vertically Lorentzian disks (HD 15115, HD 32297, HD 61005, HD 197481 = AU Mic), our out-of-equipartition interpretation yields lower individual big body masses than in Jankovic et al.~(2026). For example, for HD 32297, big bodies range in size from Varuna to Mars assuming out-of-equipartition, whereas assuming equipartition they range from Earth to Neptune. The difference arises because vertical stirring under out-of-equipartition 3D-anisotropic conditions is faster (for the same disk surface density) than equipartition stirring, by a factor of $\sim$$(2i/e)^3$ (eq.~\ref{eq:eq} vs. \ref{eq:ti3dnumcoe}). For out-of-equipartition inclination stirring, the big bodies need not be so big.

At large $m_{\rm b}$, 
every slope -1 line in Fig.~\ref{fig:locus} is capped on the right at the $M_{\rm b,tot} = m_{\rm b}$ dashed line, corresponding to the case where a given disk contains just a single big body. This single big body mass ranges from $4\times 10^{-3} \, m_\oplus$ (twice Pluto's mass; the bottom-most point for AU Mic) to $\sim$$0.4 \,m_\oplus$ (between Mars and Earth; the right-most points for HD 15115, HD 32297, and HD 61005).

Each slope -1 line in Fig.~\ref{fig:locus} is capped on the left at small $m_{\rm b}$ such that the time it takes a given big body to collide with another big body 
%collision time between big bodies 
equals the system age. Including gravitational focussing, this mean free time is estimated as
\begin{align}\label{eq:tcol}
t_{\rm col} = \frac{1}{\eta_{\rm b} \times \pi (2s_{\rm b})^2 \left[1 + 4Gm_{\rm b}/(s_{\rm b}v^2_{\rm rel}) \right] \times v_{\rm rel}} 
\end{align}
where the volumetric number density of big bodies is 
\begin{align}
\eta_{\rm b} = \frac{M_{\rm b,tot}}{m_{\rm b}} \frac{1}{2\pi R \Delta R_{\rm FWHM} \times iR} \,,
\end{align}
the big body radius is
$s_{\rm b} = [3m_{\rm b}/(4\pi \rho_{\rm b})]^{1/3}$ for internal density $\rho_{\rm b} = 3$ g/cm$^3$, and the relative velocity between big bodies is 
\begin{align}\label{eq:vrel}
v_{\rm rel} = \sqrt{(5/4)e^2 + i^2} \, \sqrt{\frac{Gm_\star}{R}}
\end{align}
\citep{lissauer93}.\footnote{Equation (\ref{eq:vrel}) for $v_{\rm rel}$ is derived for big bodies in equipartition. For bodies that are out of equipartition, corrections to eq.~(\ref{eq:vrel}) should only alter the coefficients for $e^2$ and $i^2$ by factors of order unity.} Gravitational focussing is important for our parameter space ($4Gm_{\rm b}/s_{\rm b} \gtrsim v_{\rm rel}^2$). Dropping the $+1$ in the brackets in eq.~(\ref{eq:tcol}), we find that along each slope -1 line, the collision time $t_{\rm col} \propto m_{\rm b}^{2/3}$. A lower bound on $m_{\rm b}$ is obtained by requiring $t_{\rm col} > t_{\rm age}$, since otherwise the big body stirrers would completely obliterate themselves (mostly in mergers; see Section \ref{subsec:source}). Thus the loci of possibilities plotted in Fig.~\ref{fig:locus} are capped at low $m_{\rm b}$ (high $M_{\rm b,tot}$) such that the collision probability 
%fraction of big bodies that have collided within the system age 
\begin{align}
f_{\rm col} = \frac{t_{\rm age}}{t_{\rm col}} 
\end{align}
equals 1. How $f_{\rm col} \propto 1/m_{\rm b}^{2/3}$ decreases from 1 as $m_{\rm b}$ increases is encoded in color in Fig.~\ref{fig:locus}. 
%We have found that the lower bound on $m_{\rm b}$ derived from $f_{\rm col} = 1$ is larger (more constraining) than the one usually adopted in the literature, that the surface escape velocity from the big body equals $v_{\rm rel}$ (e.g.~\citealt{daley19}).
%EC: possible footnote here saying how Daley used the wrong equation to estimate the big body mass

\subsection{Sourcing the collisional cascade by eroding big body stirrers}\label{subsec:source}

The self-stirring constraints in  Fig.~\ref{fig:locus} allow for collective big body masses $M_{\rm b,tot}$ approaching $\sim$$10^3 \, m_\oplus$, up to the limit of what is plausible on cosmogonic/planet formation grounds \citep{krivov21}. These same large values of $M_{\rm b,tot}$ also have the largest collisional probabilities $f_{\rm col}$. Thus it seems possible that in this corner of parameter space, big bodies colliding over the system age may contribute significantly to the collisional cascades that must be present in debris disks. Bodies at the top of ``standard'' cascades have estimated sizes of $\sim$10 m to 10 km, and continuously grind down to produce the small bodies observable from $\mu$m to mm wavelengths (e.g.~\citealt{pan05,wyatt11,krivov21}; Jankovic et al.~2026). The total cascade mass $M_{\rm cascade}$ (i.e.~the amount of mass ground down over the system age = the mass at the top of the cascade) can range from $\sim$$0.01 \, m_\oplus$ (AU Mic) to $\sim$$10^2 \, m_\oplus$ (Jankovic et al.~2026, the data points in their Figure 7).

We consider here the possibility that collisionally eroding some of the big body stirrers can feed the observable small body cascade. By ``eroding'' we mean that only a small fraction of the mass in a given pair of colliding big bodies will be released as debris. In the allowed parameter space in Fig.~\ref{fig:locus}, big body collisions are gravitationally focussed: the velocity at impact is comparable if not equal to the mutual surface escape velocity. Assuming the big bodies are big enough to be held together by their own internal gravity (as opposed to their ``strength'' in intermolecular bonds), such focussed collisions result predominantly in mergers, i.e. accretion. Only a small fraction $f_{\rm debris}$ of the colliding mass escapes as ejecta; as the impact angle ranges from 0$^\circ$ (head-on collisions) to 90$^\circ$ (grazing), $f_{\rm debris}$ varies from $\sim$10\% to 0 (e.g.~\citealt{emsenhuber20}, their Figure 5, bottom of right panel; \citealt{emsenhuber24}, their Figure 9b). We assume $f_{\rm debris} = 3\%$ in our calculations below.

\begin{figure*}[t]
\centering
\includegraphics[width=\textwidth]{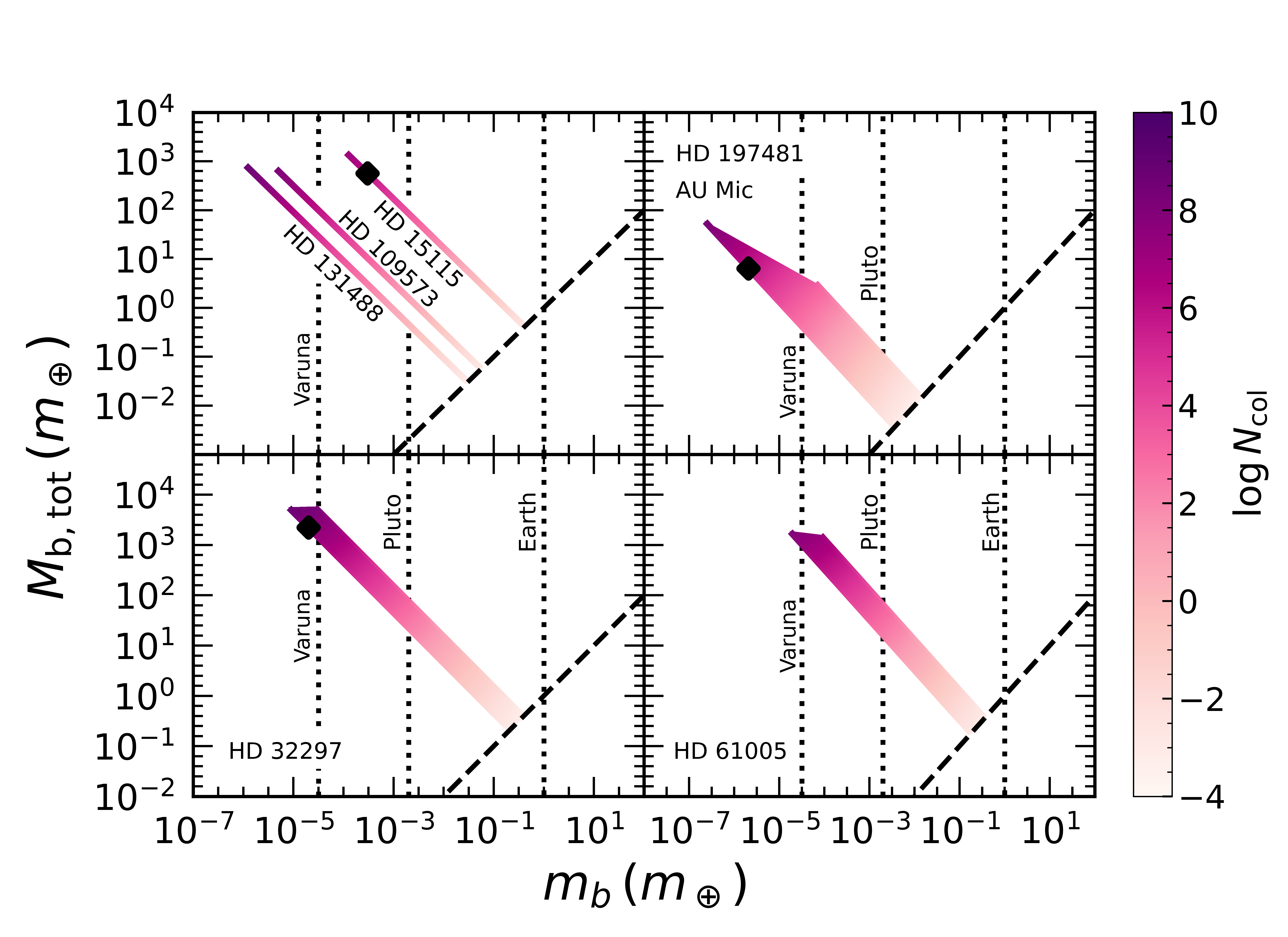}
\caption{Same as Fig.~\ref{fig:locus}, but now color-coded according to $N_{\rm col}$, the number of big body collisions occurring over $t_{\rm age}$. Black diamonds mark where there are sufficient collisions ($N_{\rm col}$ of order $10^6$ for HD 15115 and AU Mic, and $10^8$ for HD 32297; see Table \ref{tab:match} for more precise values) that the mass $M_{\rm big\,body\,debris}$ eroded from big bodies over the system age approaches the mass $M_{\rm cascade}$ ground down into small dust grains observed at $\mu$m to mm wavelengths (Jankovic et al.~2026). Similar estimates hold for the other disks in the figure, but we highlight HD 15115, HD 32297, and AU Mic because they are candidate disks for hosting the remains of recent giant impacts \citep{jones23,chiang17}.
\label{fig:locus_Ncol}}
\end{figure*}

The rate at which the big body stirrers collisionally inject debris into the disk is $f_{\rm debris} \times M_{\rm b,tot}/t_{\rm col}$ --- this formula holds even when $t_{\rm col} > t_{\rm age}$, which is the regime in which we are using it --- and can in principle match the rate at which mass is ground into observable small bodies in the standard cascade. Indeed three of the disks in our sample --- HD 197481 (AU Mic), HD 32297, and HD 15115 --- have been argued based on their scattered-light morphologies to contain the recently shattered remains of dwarf planets \citep{chiang17,jones23}. In this vein we present Figure \ref{fig:locus_Ncol}, which is the same as Figure \ref{fig:locus} except that it color-codes the allowed parameter space by the total number of big body collisions occurring within the system age:
\begin{align}
N_{\rm col} = f_{\rm col} \frac{M_{\rm b,tot}}{m_{\rm b}} \,.
\end{align}
For each of the three aforementioned candidate giant impact disks, we mark with a diamond in Fig.~\ref{fig:locus_Ncol} the specific  parameter space where the mass in debris released over the system age
\begin{align}
M_{\rm big \, body\, debris} = f_{\rm debris} N_{\rm col} m_{\rm b}
\end{align}
approaches the cascade mass $M_{\rm cascade}$ calculated by Jankovic et al.~(2026, their Figure 7). We assume throughout $f_{\rm debris} = 3\%$. For AU Mic, $M_{\rm cascade} \sim 0.025$--$0.05 \, m_\oplus$ in ``boulders'' $\sim$3--20 m in radius (Jankovic et
al.~2026; see also \citealt{strubbe06} for a similar cascade mass estimate). We find that if $M_{\rm b,tot} =
6 \, m_\oplus$ and the individual sizes of the big bodies is $s_{\rm b} = 100 {\rm km}$, then $N_{\rm col} \sim 5 \times 10^5$ and $M_{\rm
  big\,body\,debris} \sim 0.035 \, m_\oplus$, comparable to 
$M_{\rm cascade}$. %The above value for $m_{\rm b}$ is comparable to that of the Kuiper belt object Varuna (volume-equivalent diameter of $700$ km and internal density of 1 g/cc;  \citealt{fv19}).\footnote{The recent destruction of a Varuna-like body of mass $\sim$$10^{-4} \,m_\oplus$ was proposed by \citet{chiang17} to explain AU Mic's escaping dust clouds as seen in scattered light \citep{boccaletti15,boccaletti18}. But the colliding Varunas that we are presently considering result in mergers and percent-level erosion of mass, not the kind of catastrophic disruption envisioned by \citet{chiang17}. It might be possible to modify their scenario to create a Varuna's worth of dust from the collisional merger of bodies $1/f_{\rm debris} \sim 30\times$ more massive than Varuna. Such larger bodies would be more numerous and collide more frequently in a viscous stirring model that included collisional damping of small body eccentricities and inclinations (see Jankovic et al.~2026 for such a model; our models assume no damping).\label{foot:au}} In other words, in this region of allowed parameter space, collisionally merging $N_{\rm col} \sim 10^4$ viscously stirring Varunas over $t_{\rm age} \sim 23$ Myr produces $\sim$$0.02 \, m_\oplus$ of boulders that can feed the collisional cascade in AU Mic. See Table \ref{tab:match} for a summary of these parameters.
%\citet{chiang17} proposed that a Varuna-like body of mass $\sim$$10^{-4} \,m_\oplus$ was catastrophically disrupted $\sim$$3 \times 10^4$ yr ago (thereby providing the source material for dust avalanches triggered by the stellar wind).

For HD 15115 and HD 32297, analogous agreement between $M_{\rm big\,body\,debris}$ and $M_{\rm cascade}$ (now ranging from $\sim$5 to $60 \,m_\oplus$) can be obtained for big bodies 200--500 km in radius (Table \ref{tab:match}). These calculations illustrate the potential for the big body stirrers to source a healthy fraction and perhaps even all of the observable cascade --- but at the cost of requiring a total mass in big bodies of $M_{\rm b,tot} \simeq 600$--2200 $M_\oplus$, at the upper bound of what is cosmogonically plausible \citep{krivov21}. The collision frequency of big bodies can be remarkably high, occurring on average once every $\sim$0.6--50 years (this is $t_{\rm age}/N_{\rm col}$, not to be confused with $t_{\rm col}$ which is the timescale over which all big bodies collide). As these are strongly gravitationally focussed collisions, they result in mergers with $f_{\rm debris}$ of just a few percent, not catastrophic disruptions, and are consequently harder to detect.
%(see footnote \ref{foot:au}).

%%%%%%%%%%%%%%%%%%%%%%%%%%%%%%%%%%%%%%%%%%%%%%%%%%%%%%%%%%%%
\begingroup % to localize the effect of the length modification below
\setlength{\medmuskip}{0mu} % to reduce the spacing around binary operators (\times)
\begin{deluxetable*}{cccccccc}[ht]
  \tablecaption{Big body properties that can source collisional cascades in giant impact candidate disks}\label{tab:match}
  \tablecolumns{8}
  \tablehead{
    \colhead{(1)} &
    \colhead{(2)} &
    \colhead{(3)} &
    \colhead{(4)} &
    \colhead{(5)} &
    \colhead{(6)} &
    \colhead{(7)} &
    \colhead{(8)} \\
    \colhead{{Disk ID}} &
    \colhead{$s_{\rm b}$} & 
    \colhead{$M_{\rm b,tot}$} & 
    \colhead{$N_{\rm col}$} & 
    \colhead{$t_{\rm age}/N_{\rm col}$} & 
    \colhead{$f_{\rm col}$} &
    \colhead{$M_{\rm big\,body\,debris}$} &
    \colhead{$M_{\rm cascade}$} \\
    \colhead{} &
    \colhead{(km)} & 
    \colhead{($m_\oplus$)} &
    \colhead{} &
    \colhead{(yr)} &
    \colhead{} &
    \colhead{($m_\oplus$)} &
    \colhead{($m_\oplus$)} 
  }
  \startdata
  \hline
  HD 15115 & 500  & $600$  & $9\times 10^5$ & $50$ & 46\% & 8 & 
  5.6--16  \\   
    HD 32297  & 200  & $2200$  & $5\times 10^7$ & $0.6$ & 47\% & 31 & 
  17--60   \\ 
        HD 197481 (AU Mic) & $100$   & $6$  & $5\times 10^5$ & $40$ & 18\% & 0.035 & 
  0.03--0.05   \\ 
  \enddata
  \tablecomments{Columns: 
  (1) Disk name. These disks are the subset of radially narrow \citetalias{zawadzki26} disks (Table \ref{tab:disks}) proposed to exhibit the remains of giant impacts as traced by non-axisymmetric shapes in scattered light \citep{jones23,chiang17}. (2) Individual big body radius, 
  %(in units of Pluto's mass $m_{\rm Pluto} = 0.002\, m_\oplus$), 
  chosen specifically to yield $M_{\rm big\,body\,debris} \sim M_{\rm cascade}$ (columns 7 and 8).  
  (3) Collective big body mass integrated over the entire disk.   
  (4) Number of big body collisions occurring over the system age $t_{\rm age}$.   
  (5) Average time between big body collisions.
  (6) Fraction of big bodies that collide over the system age. (7) Estimated mass in debris from big body collisions over the system age ($M_{\rm big\,body\,debris} = M_{\rm b,tot} f_{\rm col} f_{\rm debris}$), assuming each collision releases a mass fraction $f_{\rm debris} = 3\%$ in ejecta. 
  (8) Mass in the ``standard'' collisional cascade producing small bodies observable from $\mu$m to mm wavelengths, calculated by Jankovic et al.~(2026, their Figure 7). The range reflects different assumptions about the slope of the size distribution.
  }
\end{deluxetable*}
\endgroup

%\vspace{0.2in}
\section{Summary and discussion}\label{sec:sum}

As particles in a disk gravitationally scatter off one another (these can be solid particles in a circumstellar debris disk or planetary ring, or stars in a galactic disk), their median and rms eccentricities $e$ and inclinations $i$ grow. Working in the dispersion-dominated regime where bodies are on crossing orbits, with relative velocities controlled by $e$ and $i$ and not by the background shear, we have shown that when $e \gg i$, i.e.~when relative velocities between particles are much larger parallel to the disk midplane than perpendicular, particles are gravitationally (``viscously'') stirred into vertical density distributions resembling Lorentzians.

Vertical Lorentzians result from particles random walking in inclination $i$ such that their individual steps $\Delta i \propto \pm \,i$. Such a scaling, obviously true when $i=0$, arises because what changes $i$ in a scattering encounter between particles is the vertical component of the gravitational acceleration, and this component scales as $f_N$, the scattering angle between the line joining the bodies and the horizontal (Figure \ref{fig:picture}).  When $i$ is sufficiently small that vertical displacements between scattering bodies are smaller than scattering impact parameters, $f_N$ is small and scales with $i$ (Equation \ref{eq:kickoom}). When $i$ random walks with steps $\Delta i$ that scale with $i$ (i.e.~steps of fixed $\Delta i/i = \Delta \ln i$), the inclination distribution $dN/di$ grows a log normal tail at high $i$, leading to broad Lorentzian-like wings in the disk's vertical density profile $dN/dz$.

Given enough dispersion-dominated  scatterings, inclinations grow until vertical displacements during encounters become comparable to horizontal displacements. Then scattering impact angles $f_N$ become broadly distributed over an order-unity range, $i$ stops exponentiating, and the vertical density profile $dN/dz$ relaxes toward --- but may not actually reach, depending on the time elapsed, and how much mass and what size bodies are contained in the disk --- a Gaussian. The endpoint of stirring is equipartition, with $i \simeq e/2$. There are four regimes of inclination evolution, with the quasi-2D regime studied by \citet{rafikov10} at the extreme low-$i$ end of the evolutionary sequence, and equipartition on the other high-$i$ extreme. While inclinations grow in tandem with eccentricities in equipartition, inclinations grow faster than eccentricities in the three out-of-equipartition regimes, when $i < e/2$ (Section \ref{subsec:time}).

We have reached these conclusions informally and empirically, by semi-analytic reasoning and $N$-body integrations. Conceivably our results could be obtained formally. One could, for example, compute analytically the sky-projection integral that transforms an inclination distribution $dN/di$ to a vertical density profile $dN/dz$ (e.g.~section 5.1 of \citealt{matra19}). Or one could try to relate our findings to the generalized central limit theorem which shows how Gaussians and Lorentzians arise from sampling distributions with finite and infinite variance, respectively (e.g.~\citealt{feller71}; \citealt{bouchaud90}; see also the Holtsmark distribution reviewed by \citealt{chandrasekhar43}). Vertical density profiles, in real life and in our numerical computations, only approximate Gaussians or Lorentzians over limited ranges in height, so care needs to be taken when connecting reality to formal theorems which may only be true asymptotically, in the limit of infinite sample sizes. These explorations are left for the future.

In related work, \citet{collins06} and \citet{collins07} solved a Boltzmann-like equation to derive broad, non-Rayleigh distributions for eccentricity in planar, viscously stirred, shear-dominated particle disks. While our focus has been on the vertical dynamics of dispersion-dominated disks where relative velocities between bodies are not controlled by the background shear, we also found that log normal inclination distributions and by extension vertical Lorentzians can arise in shear-dominated disks, for non-crossing scattering bodies separated in semimajor axis by $|a-a_{\rm b}| \gg ia$ (Appendix \ref{sec:appendix}). In shear-dominated disks, $e$ grows faster than $i$ and therefore $e \gg i$ naturally (e.g.~\citealt{goldreich04}; \citealt{collins06}). But typical debris disks are not shear-dominated; for observed disk parameters, there is no set of perturbing ``big body'' masses and surface densities consistent with shear-dominated stirring explaining present-day vertical thicknesses (Jankovic et al.~2026).

%\vspace{0.5in}
\subsection{Anisotropic viscous stirring in \citetalias{zawadzki26} debris disks: dynamical and collisional masses}\label{subsec:ideas}

Most of the debris disks studied in ARKS III \citep{zawadzki26} have vertical Lorentzian profiles, which could imply they are out-of-equipartition, with $i < e/2$. 
%The doubling rates for $i$ are greater out-of-equipartition than in equipartition, implying less disk mass and/or smaller bodies to reproduce a given observed $i$. While equipartition stirring applied to ARKS III disks allows for Earth-sized to super-Earth-sized stirrers (Jankovic et al.~2026, their Figure 7, colored lines), 
Out-of-equipartition stirring (specifically under conditions we call 3D-anisotropic; Section \ref{subsec:time}) implies either single sub-Earths (a single Pluto in the case of AU Mic), or multiple sub-Plutos (Table \ref{tab:disks} and Figure \ref{fig:locus}).

The price for this out-of-equipartition interpretation is that $i$ and $e$ de-couple: while we can explain the observed $i$'s by viscous stirring with a population of big bodies, the $e$'s, which by assumption need to be larger than $2i$, require a separate explanation. While we do not have a definite theory for the larger $e$'s, we can outline possibilities. Some general considerations first: recall that the $e$'s can refer either to the small bodies or to the big-body stirrers. What matters is that in-plane relative velocities greatly exceed out-of-plane relative velocities, and that requirement can be met whether the stirrers or the objects being stirred are on highly eccentric orbits. Recognize also that $e$ is generally easier to excite than $i$ because bodies are born in disks where in-plane forces may be larger than out-of-plane forces. Imaged debris disks are located in the outermost regions of planetary systems, and here large eccentricities are especially plausible because local orbital velocities are small and easier to change. 

If there were a single big body viscously stirring the disk, one can imagine its eccentricity having been excited externally by another planet, either secularly or in a prior episode of planet-planet scattering. In our Solar System, a ``rogue'' planet might have been temporarily scattered by neighboring giant planets into the Kuiper belt (e.g.~\citealt{huang22}). A single eccentric big body crossing orbits with small-body disk particles may also force a global eccentricity on the disk (e.g.~Fig.~\ref{fig:params_28} in Appendix \ref{sec:extra_app}). While a global disk eccentricity would not be relevant for setting relative velocities (the global $e$ is secularly forced, not free; see the end of Section \ref{subsec:procrat}), it would serve as an observational signature of an eccentric perturber. In this regard it is encouraging that a number of disks show global eccentricities \citep[][ARKS VI]{lovell26}.

Another possibility is that large eccentricities of small bodies were excited by sweeping mean-motion resonances with non-orbit-crossing planets. The strongest mean-motion resonances are eccentricity resonances, not inclination resonances. For example, the 2:1 eccentricity resonance with a stochastically migrating Neptune may have swept across the Cold Classical Kuiper belt (semimajor axes of $\sim$40--50 au), exciting the belt's eccentricities but not its inclinations (\citealt{nesvorny15a}; \citealt{nesvorny18}). The Cold Classical belt presents the opposite problem in that its inclinations $i \sim 0.05$ rad are, to our knowledge, unexplained. They are too large to have been inherited from the era of planetesimal formation (\citealt{li25,li26}), and they also cannot have been generated by viscous self-stirring, as many Cold Classical objects are binaries having component relative velocities of $\lesssim 10$ m/s. The vertical kick velocities needed to viscously stir the heliocentric inclinations of Cold Classicals are $\sim$100 m/s and would have fissioned the binaries.

Viscous stirring is intrinsically degenerate in that the smaller the individual stirrers, the more of them are needed to achieve the same degree of stirring (Section \ref{subsec:time}).  Real-life disks could have their $i$'s self-stirred by a large number of small stirrers, with individual masses comparable to dwarf planets like Pluto (diameter $\sim$2400 km) or Varuna (diameter $\sim$700 km), or even smaller, and collective masses ranging from $\sim$1--$10^3 \, m_\oplus$ (Figure \ref{fig:locus}). Having many stirrers opens the door to having a fraction of them collide over the system age. Indeed half of the disks in our sample have been argued to exhibit the scattered-light signatures of recent giant impacts: AU Mic \citep{chiang17} and HD 15115 and HD 32297 \citep{jones23}. Taking these candidate giant impact systems as examples, we found that in certain corners of parameter space (combining small individual stirring size and large total mass), $\sim$20--50\% of the stirring population can collide over system ages, releasing enough debris to contribute toward, and perhaps even source entirely, the collisional cascades generating the observable small bodies (Table \ref{tab:match}). In other words, the time-averaged rate at which stirrers collide and inject debris into the disk can in principle match the rate at which cascades grind mass down into observable $\mu$m to mm-sized dust. For disks like HD 32297 and HD 61005, we do not favor this hypothesis as it requires so much mass in stirrers --- of order $\sim$$10^3 m_\oplus$ (diamonds in Figure \ref{fig:locus_Ncol}) --- as to be cosmogonically improbable  (e.g.~\citealt{krivov21}). The hypothesis is more plausible for a small-$i$ disk like AU Mic (Table \ref{tab:disks}), for which the required stirring mass would only be $\sim$$6 \,m_\oplus$ (Table \ref{tab:match}).

%Future work can model how freshly injected debris from giant impacts relaxes into a quasi-steady cascade, distinguishing between the injection size spectrum (typically cited with a differential size index $q=4$ which distributes mass equally across mass bins; e.g.~\citealt{gault63,takasawa11,leinhardt12,kral15}) and the collisionally relaxed spectrum ($q=3.5$; e.g.~\citealt{dohnanyi69,pan05,wyatt11}).

%To recap, it is unclear whether the big body stirrers are large (planet sized), few, and collisionless; or small (dwarf planet sized), many, and collisional. These possibilities are also not mutually exclusive. In addition to exploring origin scenarios for the big body stirrers, one can also try to divine their fate. Our self-stirring parameters imply that big body collisions are strongly gravitationally focussed and therefore largely accretionary --- only a small fraction, $f_{\rm debris} \sim$ a few percent, of the colliding mass is ejected into interplanetary space (we assumed $f_{\rm debris} = 3\%$ in our cascade sourcing statements above). The viscously stirring bodies in debris disks, if there are enough of them, may not be done growing.

%EC: think about "not enough time since giant impact to equilibrate" to explain Lorentzian. Actually doesn't work in our picture because we take all of the system age to create all the small bodies; the small bodies created from just the last giant impact would be way too few in number

Gas drag on particles damps inclinations and suppresses the formation of thick vertical tails, more so for smaller particles. Two disks out of our \citetalias{zawadzki26} subsample have CO gas detections, HD 32297 and HD 131488. For HD 32297 which exhibits a vertically Lorentzian profile at mm wavelengths, we hypothesize that despite its CO gas detection, there is insufficient total gas to drag the mm-sized grains presumed responsible for the ALMA emission out of their Lorentzian tails. 
\citet{olofsson22} showed that for gas masses $\lesssim 0.1 \, m_\oplus$ --- such modest masses may characterize gas that is collisionally generated (second-generation) --- mm-sized grains can be relatively immune to gas drag (the same is not necessarily true for $\mu$m-sized grains, which could be dragged into a thin layer at the midplane). This interpretation is consistent with 
current estimates for the amount of gas in HD 32297, which overlap with $\sim$$0.1 \, m_\oplus$ (\citealt{cataldi23}, their Figures 8, 9, and 11).

\subsection{Alternative explanations of\\ non-Gaussian vertical profiles}\label{subsec:alt}

Out-of-equipartition viscous stirring is not the only explanation for vertical density profiles having thick non-Gaussian tails. Having a dynamically ``hot'' population implanted by planet migration on top of a ``cold'' population remains viable --- assuming both populations have inclination distributions that can be explained by (presumably equipartition) stirring \citep{matra19}. The Classical belt at heliocentric semimajor axes of 40--50 au is thought to represent a superposition of hot and cold objects, with hot objects implanted by a migrating Neptune (\citealt{gomes03}, \citealt{gomes05}, \citealt{nesvorny15b}). But as we noted above, the inclinations of the Cold Classicals, small as they are, lack  explanation.

Degeneracies between vertical and radial structures are a perennial worry. Many of the \citetalias{zawadzki26} disks are radially extended, and line-of-sight projection of radial structures can mimic vertically extended structures. We found that even a vertically Gaussian disk with a constant vertical-to-horizontal aspect ratio (constant aspect ratios are the default assumption in ARKS fitting routines) can exhibit vertical density distributions with broad non-Gaussian tails, if such disks span more than a factor of a few in radius and are viewed edge-on (see also \citealt{olofsson22}). The \citetalias{zawadzki26} fitting procedure explicitly accounts for such 3D projection effects, and still finds evidence for non-Gaussian vertical profiles. Nevertheless vertical Gaussians with radially varying aspect ratios (e.g.~\citealt{sefilian25}) could be lurking in the data.

Still another possibility is a particle size distribution where different sized particles have different scale heights to yield a non-Gaussian emissivity (\citealt{terrill23}, their Figure 14). More flexible methods available for mapping disk substructures are \texttt{frank} \citep{jennings20, terrill23}, and \texttt{rave} (\citealt{han22}, \citealt{han25}), which fit for radial structures non-parametrically. Current implementations of these algorithms assume vertical Gaussians, a restriction that could be relaxed for future work. Measuring how disk images vary with wavelength from the sub-mm through the radio, uncomplicated by the effects of radiation pressure (cf.~\citealt{han26new}, \citealt{rebollido24}), can also help to sort between hypotheses (\citealt{vizgan22}).

\vspace{0.2in}
%\begin{acknowledgements}
\noindent We thank Tuhin Ghosh, Rixin Li, Yoram Lithwick, Miguel Martinez, David Nesvorn\'y, and Roman Rafikov for discussions, and acknowledge the use of A.I.~tools (Claude.ai, Gemini, ChatGPT) for assistance in coding, pointers to the literature, and conversation partners for exploring ideas. The authors take full responsibility for the content of this paper; no text was written by an A.I. An anonymous referee provided a helpful report. EC is supported by the Simons Investigator program and the Miller Institute for Basic Research in Science, University of California, Berkeley. TDP is supported by a UKRI Stephen Hawking Fellowship and a Warwick Prize Fellowship, the latter made possible by a generous philanthropic donation. MRJ acknowledges funding provided by the Institute of Physics
Belgrade, through the grant by the Ministry of Science, Technological
Development, and Innovations of the Republic of Serbia.  YH is supported by a Barr Fellowship at Caltech. 
%MP 
Support for BZ was provided by The Brinson Foundation.
AMH gratefully acknowledges support from the National Science Foundation under Grant No. AST-2307920. 
JBL acknowledges the Smithsonian Institution for funding via a CfA J.C.~Ryan Fellowship. 
SM acknowledges funding by the Royal Society through a Royal Society University Research Fellowship (URF-R1-221669) and the European Union through the FEED ERC project (grant number 101162711). Views and opinions expressed are however those of the authors only and do not necessarily reflect those of the European Union or the European Research Council Executive Agency. Neither the European Union nor the granting authority can be held responsible for them. 
%AVK 
AAS is supported by the Heising-Simons Foundation through a 51 Pegasi b Fellowship. 
%DW
MW was supported by the Science and Technology Facilities Council grant UKRI1198. 
PA received funding from the Hungarian NKFIH project No. K-147380. 
AK is supported by the NKFIH NKKP grant ADVANCED 149943, which has been implemented with the support provided by the Ministry of Culture and Innovation of Hungary from the National Research, Development and Innovation Fund, financed under the NKKP ADVANCED funding scheme. 
%\end{acknowledgements}

\bibliographystyle{aasjournal}
\bibliography{vertical_profiles}

@article{ida90,
	adsurl = {https://ui.adsabs.harvard.edu/abs/1990Icar...88..129I},
	author = {{Ida}, Shigeru},
	doi = {10.1016/0019-1035(90)90182-9},
	journal = {\icarus},
	month = nov,
	number = {1},
	pages = {129-145},
	title = {{Stirring and dynamical friction rates of planetesimals in the solar gravitational field}},
	volume = {88},
	year = 1990}

@article{petrovich14,
	adsurl = {https://ui.adsabs.harvard.edu/abs/2014ApJ...786..101P},
	archiveprefix = {arXiv},
	author = {{Petrovich}, Cristobal and {Tremaine}, Scott and {Rafikov}, Roman},
	doi = {10.1088/0004-637X/786/2/101},
	eid = {101},
	eprint = {1401.4457},
	journal = {\apj},
	month = may,
	number = {2},
	pages = {101},
	primaryclass = {astro-ph.EP},
	title = {{Scattering Outcomes of Close-in Planets: Constraints on Planet Migration}},
	volume = {786},
	year = 2014}

@article{rafikov10,
	adsurl = {https://ui.adsabs.harvard.edu/abs/2010AJ....139..565R},
	archiveprefix = {arXiv},
	author = {{Rafikov}, Roman R. and {Slepian}, Zachary S.},
	doi = {10.1088/0004-6256/139/2/565},
	eprint = {0908.1398},
	journal = {\aj},
	month = feb,
	number = {2},
	pages = {565-579},
	primaryclass = {astro-ph.EP},
	title = {{Dynamical Evolution of Thin Dispersion-Dominated Planetesimal Disks}},
	volume = {139},
	year = 2010}

@article{gomes05,
	adsurl = {https://ui.adsabs.harvard.edu/abs/2005CeMDA..91..109G},
	author = {{Gomes}, Rodney S. and {Gallardo}, Tabar{\'e} and {Fern{\'a}ndez}, Julio A. and {Brunini}, Adri{\'a}n},
	doi = {10.1007/s10569-004-4623-y},
	journal = {Celestial Mechanics and Dynamical Astronomy},
	month = jan,
	number = {1-2},
	pages = {109-129},
	title = {{On The Origin of The High-Perihelion Scattered Disk: The Role of The Kozai Mechanism And Mean Motion Resonances}},
	volume = {91},
	year = 2005}

@article{gomes03,
	adsurl = {https://ui.adsabs.harvard.edu/abs/2003Icar..161..404G},
	author = {{Gomes}, Rodney S.},
	doi = {10.1016/S0019-1035(02)00056-8},
	journal = {\icarus},
	month = feb,
	number = {2},
	pages = {404-418},
	title = {{The origin of the Kuiper Belt high-inclination population}},
	volume = {161},
	year = 2003}

@article{nesvorny15b,
	adsurl = {https://ui.adsabs.harvard.edu/abs/2015AJ....150...73N},
	archiveprefix = {arXiv},
	author = {{Nesvorn{\'y}}, David},
	doi = {10.1088/0004-6256/150/3/73},
	eid = {73},
	eprint = {1504.06021},
	journal = {\aj},
	month = sep,
	number = {3},
	pages = {73},
	primaryclass = {astro-ph.EP},
	title = {{Evidence for Slow Migration of Neptune from the Inclination Distribution of Kuiper Belt Objects}},
	volume = {150},
	year = 2015}

@article{li25,
	adsurl = {https://ui.adsabs.harvard.edu/abs/2025ApJ...995..214L},
	archiveprefix = {arXiv},
	author = {{Li}, Rixin and {Chiang}, Eugene},
	doi = {10.3847/1538-4357/ae18c3},
	eid = {214},
	eprint = {2508.04776},
	journal = {\apj},
	month = dec,
	number = {2},
	pages = {214},
	primaryclass = {astro-ph.EP},
	title = {{In Situ Formation of the Cold Classical Kuiper Belt}},
	volume = {995},
	year = 2025}

@article{li26,
	adsurl = {https://ui.adsabs.harvard.edu/abs/2026arXiv260614704L},
	archiveprefix = {arXiv},
	author = {{Li}, Rixin and {Chiang}, Eugene},
	doi = {10.48550/arXiv.2606.14704},
	eid = {arXiv:2606.14704},
	eprint = {2606.14704},
	journal = {arXiv e-prints},
	month = jun,
	pages = {arXiv:2606.14704},
	primaryclass = {astro-ph.EP},
	title = {{The Edges of Planetary Systems: Falling Off the Kuiper Cliff in a Dissipating Gas Disk}},
	year = 2026}

@article{nesvorny15a,
	adsurl = {https://ui.adsabs.harvard.edu/abs/2015AJ....150...68N},
	archiveprefix = {arXiv},
	author = {{Nesvorn{\'y}}, David},
	doi = {10.1088/0004-6256/150/3/68},
	eid = {68},
	eprint = {1506.06019},
	journal = {\aj},
	month = sep,
	number = {3},
	pages = {68},
	primaryclass = {astro-ph.EP},
	title = {{Jumping Neptune Can Explain the Kuiper Belt Kernel}},
	volume = {150},
	year = 2015}

@article{emsenhuber24,
	adsurl = {https://ui.adsabs.harvard.edu/abs/2024PSJ.....5...59E},
	archiveprefix = {arXiv},
	author = {{Emsenhuber}, Alexandre and {Asphaug}, Erik and {Cambioni}, Saverio and {Gabriel}, Travis S.~J. and {Schwartz}, Stephen R. and {Melikyan}, Robert E. and {Denton}, C. Adeene},
	doi = {10.3847/PSJ/ad2178},
	eid = {59},
	eprint = {2401.17356},
	journal = {\psj},
	month = mar,
	number = {3},
	pages = {59},
	primaryclass = {astro-ph.EP},
	title = {{A New Database of Giant Impacts over a Wide Range of Masses and with Material Strength: A First Analysis of Outcomes}},
	volume = {5},
	year = 2024}

@article{emsenhuber20,
	adsurl = {https://ui.adsabs.harvard.edu/abs/2020ApJ...891....6E},
	archiveprefix = {arXiv},
	author = {{Emsenhuber}, Alexandre and {Cambioni}, Saverio and {Asphaug}, Erik and {Gabriel}, Travis S.~J. and {Schwartz}, Stephen R. and {Furfaro}, Roberto},
	doi = {10.3847/1538-4357/ab6de5},
	eid = {6},
	eprint = {2001.00951},
	journal = {\apj},
	month = mar,
	number = {1},
	pages = {6},
	primaryclass = {astro-ph.EP},
	title = {{Realistic On-the-fly Outcomes of Planetary Collisions. II. Bringing Machine Learning to N-body Simulations}},
	volume = {891},
	year = 2020}

@article{kalas05,
	adsurl = {https://ui.adsabs.harvard.edu/abs/2005Natur.435.1067K},
	archiveprefix = {arXiv},
	author = {{Kalas}, Paul and {Graham}, James R. and {Clampin}, Mark},
	doi = {10.1038/nature03601},
	eprint = {astro-ph/0506574},
	journal = {\nat},
	month = jun,
	number = {7045},
	pages = {1067-1070},
	primaryclass = {astro-ph},
	title = {{A planetary system as the origin of structure in Fomalhaut's dust belt}},
	volume = {435},
	year = 2005}

@article{macgregor17,
	adsurl = {https://ui.adsabs.harvard.edu/abs/2017ApJ...842....8M},
	archiveprefix = {arXiv},
	author = {{MacGregor}, Meredith A. and {Matr{\`a}}, Luca and {Kalas}, Paul and {Wilner}, David J. and {Pan}, Margaret and {Kennedy}, Grant M. and {Wyatt}, Mark C. and {Duchene}, Gaspard and {Hughes}, A. Meredith and {Rieke}, George H. and {Clampin}, Mark and {Fitzgerald}, Michael P. and {Graham}, James R. and {Holland}, Wayne S. and {Pani{\'c}}, Olja and {Shannon}, Andrew and {Su}, Kate},
	doi = {10.3847/1538-4357/aa71ae},
	eid = {8},
	eprint = {1705.05867},
	journal = {\apj},
	month = jun,
	number = {1},
	pages = {8},
	primaryclass = {astro-ph.EP},
	title = {{A Complete ALMA Map of the Fomalhaut Debris Disk}},
	volume = {842},
	year = 2017}

@article{lovell26,
	adsurl = {https://ui.adsabs.harvard.edu/abs/2026A&A...705A.200L},
	archiveprefix = {arXiv},
	author = {{Lovell}, J.~B. and {Hales}, A.~S. and {Kennedy}, G.~M. and {Marino}, S. and {Olofsson}, J. and {Hughes}, A.~M. and {Mansell}, E. and {Matthews}, B.~C. and {Pearce}, T.~D. and {Sefilian}, A.~A. and {Wilner}, D.~J. and {Zawadzki}, B. and {Booth}, M. and {Bonduelle}, M. and {Brennan}, A. and {del Burgo}, C. and {Carpenter}, J.~M. and {Cataldi}, G. and {Chiang}, E. and {Fehr}, A. and {Han}, Y. and {Henning}, Th. and {Krivov}, A.~V. and {Luppe}, P. and {Marshall}, J.~P. and {Mac Manamon}, S. and {Milli}, J. and {Mo{\'o}r}, A. and {Wyatt}, M.~C. and {Ertel}, S. and {Jankovic}, M.~R. and {K{\'o}sp{\'a}l}, {\'A}. and {MacGregor}, M.~A. and {Matr{\`a}}, L. and {P{\'e}rez}, S. and {Weber}, P.},
	doi = {10.1051/0004-6361/202556568},
	eid = {A200},
	eprint = {2601.11766},
	journal = {\aap},
	month = jan,
	pages = {A200 (ARKS VI)},
	primaryclass = {astro-ph.EP},
	title = {{The ALMA survey to Resolve exoKuiper belt Substructures (ARKS): VI. Asymmetries and offsets}},
	volume = {705},
	year = 2026}

@article{lovell25,
	adsurl = {https://ui.adsabs.harvard.edu/abs/2025ApJ...990..145L},
	archiveprefix = {arXiv},
	author = {{Lovell}, Joshua B. and {Lynch}, Elliot M. and {Chittidi}, Jay and {Sefilian}, Antranik A. and {Andrews}, Sean M. and {Kennedy}, Grant M. and {MacGregor}, Meredith and {Wilner}, David J. and {Wyatt}, Mark C.},
	doi = {10.3847/1538-4357/adfadc},
	eid = {145},
	eprint = {2509.02884},
	journal = {\apj},
	month = sep,
	number = {2},
	pages = {145},
	primaryclass = {astro-ph.EP},
	title = {{ALMA Reveals an Eccentricity Gradient in the Fomalhaut Debris Disk}},
	volume = {990},
	year = 2025}

@article{gladman93,
	adsurl = {https://ui.adsabs.harvard.edu/abs/1993Icar..106..247G},
	author = {{Gladman}, Brett},
	doi = {10.1006/icar.1993.1169},
	journal = {\icarus},
	month = nov,
	number = {1},
	pages = {247-263},
	title = {{Dynamics of Systems of Two Close Planets}},
	volume = {106},
	year = 1993}

@article{chambers96,
	adsurl = {https://ui.adsabs.harvard.edu/abs/1996Icar..119..261C},
	author = {{Chambers}, J.~E. and {Wetherill}, G.~W. and {Boss}, A.~P.},
	doi = {10.1006/icar.1996.0019},
	journal = {\icarus},
	month = feb,
	number = {2},
	pages = {261-268},
	title = {{The Stability of Multi-Planet Systems}},
	volume = {119},
	year = 1996}

@article{rebollido24,
	adsurl = {https://ui.adsabs.harvard.edu/abs/2024AJ....167...69R},
	archiveprefix = {arXiv},
	author = {{Rebollido}, Isabel and {Stark}, Christopher C. and {Kammerer}, Jens and {Perrin}, Marshall D. and {Lawson}, Kellen and {Pueyo}, Laurent and {Chen}, Christine and {Hines}, Dean and {Girard}, Julien H. and {Worthen}, Kadin and {Ingerbretsen}, Carl and {Betti}, Sarah and {Clampin}, Mark and {Golimowski}, David and {Hoch}, Kielan and {Lewis}, Nikole K. and {Lu}, Cicero X. and {van der Marel}, Roeland P. and {Rickman}, Emily and {Seager}, Sara and {Soummer}, R{\'e}mi and {Valenti}, Jeff A. and {Ward-Duong}, Kimberly and {Mountain}, C. Matt},
	doi = {10.3847/1538-3881/ad1759},
	eid = {69},
	eprint = {2401.05271},
	journal = {\aj},
	month = feb,
	number = {2},
	pages = {69},
	primaryclass = {astro-ph.EP},
	title = {{JWST-TST High Contrast: Asymmetries, Dust Populations, and Hints of a Collision in the {\ensuremath{\beta}} Pictoris Disk with NIRCam and MIRI}},
	volume = {167},
	year = 2024}

@article{cataldi23,
	adsurl = {https://ui.adsabs.harvard.edu/abs/2023ApJ...951..111C},
	archiveprefix = {arXiv},
	author = {{Cataldi}, Gianni and {Aikawa}, Yuri and {Iwasaki}, Kazunari and {Marino}, Sebastian and {Brandeker}, Alexis and {Hales}, Antonio and {Henning}, Thomas and {Higuchi}, Aya E. and {Hughes}, A. Meredith and {Janson}, Markus and {Kral}, Quentin and {Matr{\`a}}, Luca and {Mo{\'o}r}, Attila and {Olofsson}, G{\"o}ran and {Redfield}, Seth and {Roberge}, Aki},
	doi = {10.3847/1538-4357/acd6f3},
	eid = {111},
	eprint = {2305.12093},
	journal = {\apj},
	month = jul,
	number = {2},
	pages = {111},
	primaryclass = {astro-ph.EP},
	title = {{Primordial or Secondary? Testing Models of Debris Disk Gas with ALMA}},
	volume = {951},
	year = 2023}

@article{wyatt11,
	adsurl = {https://ui.adsabs.harvard.edu/abs/2011CeMDA.111....1W},
	archiveprefix = {arXiv},
	author = {{Wyatt}, M.~C. and {Clarke}, C.~J. and {Booth}, M.},
	doi = {10.1007/s10569-011-9345-3},
	eprint = {1103.5499},
	journal = {Celestial Mechanics and Dynamical Astronomy},
	month = oct,
	number = {1-2},
	pages = {1-28},
	primaryclass = {astro-ph.EP},
	title = {{Debris disk size distributions: steady state collisional evolution with Poynting-Robertson drag and other loss processes}},
	volume = {111},
	year = 2011}

@article{pan05,
	adsurl = {https://ui.adsabs.harvard.edu/abs/2005Icar..173..342P},
	archiveprefix = {arXiv},
	author = {{Pan}, Margaret and {Sari}, Re'em},
	doi = {10.1016/j.icarus.2004.09.004},
	eprint = {astro-ph/0402138},
	journal = {\icarus},
	month = feb,
	number = {2},
	pages = {342-348},
	primaryclass = {astro-ph},
	title = {{Shaping the Kuiper belt size distribution by shattering large but strengthless bodies}},
	volume = {173},
	year = 2005}

@article{strubbe06,
	adsurl = {https://ui.adsabs.harvard.edu/abs/2006ApJ...648..652S},
	archiveprefix = {arXiv},
	author = {{Strubbe}, Linda E. and {Chiang}, Eugene I.},
	doi = {10.1086/505736},
	eprint = {astro-ph/0510527},
	journal = {\apj},
	month = sep,
	number = {1},
	pages = {652-665},
	primaryclass = {astro-ph},
	title = {{Dust Dynamics, Surface Brightness Profiles, and Thermal Spectra of Debris Disks: The Case of AU Microscopii}},
	volume = {648},
	year = 2006}

@article{olofsson22,
	adsurl = {https://ui.adsabs.harvard.edu/abs/2022MNRAS.513..713O},
	archiveprefix = {arXiv},
	author = {{Olofsson}, Johan and {Th{\'e}bault}, Philippe and {Kral}, Quentin and {Bayo}, Amelia and {Boccaletti}, Anthony and {Godoy}, Nicol{\'a}s and {Henning}, Thomas and {van Holstein}, Rob G. and {Mauc{\'o}}, Karina and {Milli}, Julien and {Montesinos}, Mat{\'\i}as and {Rein}, Hanno and {Sefilian}, Antranik A.},
	doi = {10.1093/mnras/stac455},
	eprint = {2202.08313},
	journal = {\mnras},
	month = jun,
	number = {1},
	pages = {713-734},
	primaryclass = {astro-ph.EP},
	title = {{The vertical structure of debris discs and the impact of gas}},
	volume = {513},
	year = 2022}

@article{pearce25,
	adsurl = {https://ui.adsabs.harvard.edu/abs/2025MNRAS.544.1447P},
	archiveprefix = {arXiv},
	author = {{Pearce}, Tim D. and {L{\"o}hne}, Torsten and {Krivov}, Alexander V.},
	doi = {10.1093/mnras/staf1735},
	eprint = {2510.07187},
	journal = {\mnras},
	month = dec,
	number = {2},
	pages = {1447-1462},
	primaryclass = {astro-ph.EP},
	title = {{Fomalhaut's debris disc is not dominated by primordial Plutos}},
	volume = {544},
	year = 2025}

@article{han25,
	adsurl = {https://ui.adsabs.harvard.edu/abs/2025MNRAS.537.3839H},
	archiveprefix = {arXiv},
	author = {{Han}, Yinuo and {Wyatt}, Mark C. and {Marino}, Sebastian},
	doi = {10.1093/mnras/staf282},
	eprint = {2502.08584},
	journal = {\mnras},
	month = mar,
	number = {4},
	pages = {3839-3860},
	primaryclass = {astro-ph.EP},
	title = {{Recovering the structure of debris discs non-parametrically from images}},
	volume = {537},
	year = 2025}

@article{han22,
	adsurl = {https://ui.adsabs.harvard.edu/abs/2022MNRAS.511.4921H},
	archiveprefix = {arXiv},
	author = {{Han}, Yinuo and {Wyatt}, Mark C. and {Matr{\`a}}, Luca},
	doi = {10.1093/mnras/stac373},
	eprint = {2202.04475},
	journal = {\mnras},
	month = apr,
	number = {4},
	pages = {4921-4936},
	primaryclass = {astro-ph.EP},
	title = {{RAVE: a non-parametric method for recovering the surface brightness and height profiles of edge-on debris discs}},
	volume = {511},
	year = 2022}

@article{vizgan22,
	adsurl = {https://ui.adsabs.harvard.edu/abs/2022ApJ...935..131V},
	archiveprefix = {arXiv},
	author = {{Vizgan}, David and {Hughes}, A. Meredith and {Carter}, Evan S. and {Flaherty}, Kevin M. and {Pan}, Margaret and {Chiang}, Eugene and {Schlichting}, Hilke and {Wilner}, David J. and {Andrews}, Sean M. and {Carpenter}, John M. and {Mo{\'o}r}, Attila and {MacGregor}, Meredith A.},
	doi = {10.3847/1538-4357/ac80b8},
	eid = {131},
	eprint = {2207.05277},
	journal = {\apj},
	month = aug,
	number = {2},
	pages = {131},
	primaryclass = {astro-ph.EP},
	title = {{Multiwavelength Vertical Structure in the AU Mic Debris Disk: Characterizing the Collisional Cascade}},
	volume = {935},
	year = 2022}

@article{jennings20,
	adsurl = {https://ui.adsabs.harvard.edu/abs/2020MNRAS.495.3209J},
	archiveprefix = {arXiv},
	author = {{Jennings}, Jeff and {Booth}, Richard A. and {Tazzari}, Marco and {Rosotti}, Giovanni P. and {Clarke}, Cathie J.},
	doi = {10.1093/mnras/staa1365},
	eprint = {2005.07709},
	journal = {\mnras},
	month = jul,
	number = {3},
	pages = {3209-3232},
	primaryclass = {astro-ph.EP},
	title = {{frankenstein: protoplanetary disc brightness profile reconstruction at sub-beam resolution with a rapid Gaussian process}},
	volume = {495},
	year = 2020}

@article{terrill23,
	adsurl = {https://ui.adsabs.harvard.edu/abs/2023MNRAS.524.1229T},
	archiveprefix = {arXiv},
	author = {{Terrill}, James and {Marino}, Sebastian and {Booth}, Richard A. and {Han}, Yinuo and {Jennings}, Jeff and {Wyatt}, Mark C.},
	doi = {10.1093/mnras/stad1847},
	eprint = {2306.09715},
	journal = {\mnras},
	month = sep,
	number = {1},
	pages = {1229-1245},
	primaryclass = {astro-ph.EP},
	title = {{Deprojecting and constraining the vertical thickness of exoKuiper belts}},
	volume = {524},
	year = 2023}

@article{han26new,
	adsurl = {https://ui.adsabs.harvard.edu/abs/2026arXiv260303540H},
	archiveprefix = {arXiv},
	author = {{Han}, Yinuo and {Wyatt}, Mark C. and {Jankovic}, Marija R. and {Zhang}, Andrew and {Dent}, William R.~F. and {Hughes}, A Meredith and {Matr{\`a}}, Luca},
	doi = {10.48550/arXiv.2603.03540},
	eid = {arXiv:2603.03540},
	eprint = {2603.03540},
	journal = {arXiv e-prints},
	month = mar,
	pages = {arXiv:2603.03540},
	primaryclass = {astro-ph.EP},
	title = {{The multi-wavelength vertical structure of the archetypal $β$ Pictoris debris disk}},
	year = 2026}

@article{ahmic09,
	adsurl = {https://ui.adsabs.harvard.edu/abs/2009ApJ...705..529A},
	archiveprefix = {arXiv},
	author = {{Ahmic}, Mirza and {Croll}, Bryce and {Artymowicz}, Pawel},
	doi = {10.1088/0004-637X/705/1/529},
	eprint = {0909.0730},
	journal = {\apj},
	month = nov,
	number = {1},
	pages = {529-542},
	primaryclass = {astro-ph.SR},
	title = {{Dust Distribution in the {\ensuremath{\beta}} Pictoris Circumstellar Disks}},
	volume = {705},
	year = 2009}

@book{frank02,
	adsurl = {https://ui.adsabs.harvard.edu/abs/2002apa..book.....F},
	author = {{Frank}, Juhan and {King}, Andrew and {Raine}, Derek J.},
	publisher = {Cambridge University Press},
	title = {{Accretion Power in Astrophysics: Third Edition}},
	year = 2002}

@article{sefilian25,
	adsurl = {https://ui.adsabs.harvard.edu/abs/2025MNRAS.543.3123S},
	archiveprefix = {arXiv},
	author = {{Sefilian}, Antranik A. and {Kratter}, Kaitlin M. and {Wyatt}, Mark C. and {Petrovich}, Cristobal and {Th{\'e}bault}, Philippe and {Malhotra}, Renu and {Faramaz-Gorka}, Virginie},
	doi = {10.1093/mnras/staf1555},
	eprint = {2505.09578},
	journal = {\mnras},
	month = nov,
	number = {4},
	pages = {3123-3151},
	primaryclass = {astro-ph.EP},
	title = {{The vertical structure of debris discs and the role of disc gravity: a primer using a simplified model}},
	volume = {543},
	year = 2025}

@inproceedings{lissauer93,
	adsurl = {https://ui.adsabs.harvard.edu/abs/1993prpl.conf.1061L},
	author = {{Lissauer}, Jack J. and {Stewart}, Glen R.},
	booktitle = {Protostars and Planets III},
	editor = {{Levy}, Eugene H. and {Lunine}, Jonathan I.},
	month = jan,
	pages = {1061},
	title = {{Growth of Planets from Planetesimals}},
	year = 1993}

@article{krivov21,
	adsurl = {https://ui.adsabs.harvard.edu/abs/2021MNRAS.500..718K},
	archiveprefix = {arXiv},
	author = {{Krivov}, Alexander V. and {Wyatt}, Mark C.},
	doi = {10.1093/mnras/staa2385},
	eprint = {2008.07406},
	journal = {\mnras},
	month = jan,
	number = {1},
	pages = {718-735},
	primaryclass = {astro-ph.EP},
	title = {{Solution to the debris disc mass problem: planetesimals are born small?}},
	volume = {500},
	year = 2021}

@article{huang22,
	adsurl = {https://ui.adsabs.harvard.edu/abs/2022ApJ...938L..23H},
	archiveprefix = {arXiv},
	author = {{Huang}, Yukun and {Gladman}, Brett and {Beaudoin}, Matthew and {Zhang}, Kevin},
	doi = {10.3847/2041-8213/ac9480},
	eid = {L23},
	eprint = {2209.09399},
	journal = {\apjl},
	month = oct,
	number = {2},
	pages = {L23},
	primaryclass = {astro-ph.EP},
	title = {{A Rogue Planet Helps to Populate the Distant Kuiper Belt}},
	volume = {938},
	year = 2022}

@article{nesvorny18,
	adsurl = {https://ui.adsabs.harvard.edu/abs/2018ARA&A..56..137N},
	archiveprefix = {arXiv},
	author = {{Nesvorn{\'y}}, David},
	doi = {10.1146/annurev-astro-081817-052028},
	eprint = {1807.06647},
	journal = {\araa},
	month = sep,
	pages = {137-174},
	primaryclass = {astro-ph.EP},
	title = {{Dynamical Evolution of the Early Solar System}},
	volume = {56},
	year = 2018}

@article{bouchaud90,
	adsurl = {https://ui.adsabs.harvard.edu/abs/1990PhR...195..127B},
	author = {{Bouchaud}, Jean-Philippe and {Georges}, Antoine},
	doi = {10.1016/0370-1573(90)90099-N},
	journal = {\physrep},
	month = nov,
	number = {4-5},
	pages = {127-293},
	title = {{Anomalous diffusion in disordered media: Statistical mechanisms, models and physical applications}},
	volume = {195},
	year = 1990}

@article{chandrasekhar43,
	adsurl = {https://ui.adsabs.harvard.edu/abs/1943RvMP...15....1C},
	author = {{Chandrasekhar}, S.},
	doi = {10.1103/RevModPhys.15.1},
	journal = {Reviews of Modern Physics},
	month = jan,
	number = {1},
	pages = {1-89},
	title = {{Stochastic Problems in Physics and Astronomy}},
	volume = {15},
	year = 1943}

@book{feller71,
	adsurl = {https://ui.adsabs.harvard.edu/abs/1971aitp.book.....F},
	author = {{Feller}, William},
	publisher = {Wiley},
	title = {{An introduction to probability theory and its applications}},
	year = 1971}

@article{collins07,
	adsurl = {https://ui.adsabs.harvard.edu/abs/2007AJ....133.2389C},
	archiveprefix = {arXiv},
	author = {{Collins}, Benjamin F. and {Schlichting}, Hilke E. and {Sari}, Re'em},
	doi = {10.1086/513718},
	eprint = {astro-ph/0609801},
	journal = {\aj},
	month = may,
	number = {5},
	pages = {2389-2392},
	primaryclass = {astro-ph},
	title = {{The Self-Similarity of Shear-dominated Viscous Stirring}},
	volume = {133},
	year = 2007}

@article{collins06,
	adsurl = {https://ui.adsabs.harvard.edu/abs/2006AJ....132.1316C},
	archiveprefix = {arXiv},
	author = {{Collins}, Benjamin F. and {Sari}, Re'em},
	doi = {10.1086/506388},
	eprint = {astro-ph/0604078},
	journal = {\aj},
	month = sep,
	number = {3},
	pages = {1316-1321},
	primaryclass = {astro-ph},
	title = {{Protoplanet Dynamics in a Shear-dominated Disk}},
	volume = {132},
	year = 2006}

@article{jones23,
	adsurl = {https://ui.adsabs.harvard.edu/abs/2023ApJ...948..102J},
	archiveprefix = {arXiv},
	author = {{Jones}, Joshua W. and {Chiang}, Eugene and {Duch{\^e}ne}, Gaspard and {Kalas}, Paul and {Esposito}, Thomas M.},
	doi = {10.3847/1538-4357/acc466},
	eid = {102},
	eprint = {2303.10189},
	journal = {\apj},
	month = may,
	number = {2},
	pages = {102},
	primaryclass = {astro-ph.EP},
	title = {{Giant Impacts and Debris Disk Morphology}},
	volume = {948},
	year = 2023}

@article{chiang17,
	adsurl = {https://ui.adsabs.harvard.edu/abs/2017ApJ...848....4C},
	archiveprefix = {arXiv},
	author = {{Chiang}, Eugene and {Fung}, Jeffrey},
	doi = {10.3847/1538-4357/aa89e6},
	eid = {4},
	eprint = {1707.08970},
	journal = {\apj},
	month = oct,
	number = {1},
	pages = {4},
	primaryclass = {astro-ph.EP},
	title = {{Stellar Winds and Dust Avalanches in the AU Mic Debris Disk}},
	volume = {848},
	year = 2017}

@book{rybicki86,
	adsurl = {https://ui.adsabs.harvard.edu/abs/1986rpa..book.....R},
	author = {{Rybicki}, George B. and {Lightman}, Alan P.},
	publisher = {Wiley},
	title = {{Radiative Processes in Astrophysics}},
	year = 1986}

@book{murray99,
	adsurl = {https://ui.adsabs.harvard.edu/abs/1999ssd..book.....M},
	author = {{Murray}, Carl D. and {Dermott}, Stanley F.},
	doi = {10.1017/CBO9781139174817},
	publisher = {Cambridge University Press},
	title = {{Solar System Dynamics}},
	year = 1999}

@article{rein12,
	adsurl = {https://ui.adsabs.harvard.edu/abs/2012A&A...537A.128R},
	archiveprefix = {arXiv},
	author = {{Rein}, H. and {Liu}, S.-F.},
	doi = {10.1051/0004-6361/201118085},
	eid = {A128},
	eprint = {1110.4876},
	journal = {\aap},
	month = jan,
	pages = {A128},
	primaryclass = {astro-ph.EP},
	title = {{REBOUND: an open-source multi-purpose N-body code for collisional dynamics}},
	volume = {537},
	year = 2012}

@article{tamayo20,
	adsurl = {https://ui.adsabs.harvard.edu/abs/2020MNRAS.491.2885T},
	archiveprefix = {arXiv},
	author = {{Tamayo}, Daniel and {Rein}, Hanno and {Shi}, Pengshuai and {Hernandez}, David M.},
	doi = {10.1093/mnras/stz2870},
	eprint = {1908.05634},
	journal = {\mnras},
	month = jan,
	number = {2},
	pages = {2885-2901},
	primaryclass = {astro-ph.EP},
	title = {{REBOUNDx: a library for adding conservative and dissipative forces to otherwise symplectic N-body integrations}},
	volume = {491},
	year = 2020}

@article{rein19,
	adsurl = {https://ui.adsabs.harvard.edu/abs/2019MNRAS.485.5490R},
	archiveprefix = {arXiv},
	author = {{Rein}, Hanno and {Hernandez}, David M. and {Tamayo}, Daniel and {Brown}, Garett and {Eckels}, Emily and {Holmes}, Emma and {Lau}, Michelle and {Leblanc}, R{\'e}jean and {Silburt}, Ari},
	doi = {10.1093/mnras/stz769},
	eprint = {1903.04972},
	journal = {\mnras},
	month = jun,
	number = {4},
	pages = {5490-5497},
	primaryclass = {astro-ph.EP},
	title = {{Hybrid symplectic integrators for planetary dynamics}},
	volume = {485},
	year = 2019}

@article{jankovic24,
	adsurl = {https://ui.adsabs.harvard.edu/abs/2024A&A...691A.302J},
	archiveprefix = {arXiv},
	author = {{Jankovic}, Marija R. and {Wyatt}, Mark C. and {L{\"o}hne}, Torsten},
	doi = {10.1051/0004-6361/202451080},
	eid = {A302},
	eprint = {2411.13991},
	journal = {\aap},
	month = nov,
	pages = {A302},
	primaryclass = {astro-ph.EP},
	title = {{Collisional damping in debris discs: Only significant if collision velocities are low}},
	volume = {691},
	year = 2024}

@article{pearce24,
	adsurl = {https://ui.adsabs.harvard.edu/abs/2024MNRAS.527.3876P},
	archiveprefix = {arXiv},
	author = {{Pearce}, Tim D. and {Krivov}, Alexander V. and {Sefilian}, Antranik A. and {Jankovic}, Marija R. and {L{\"o}hne}, Torsten and {Morgner}, Tobias and {Wyatt}, Mark C. and {Booth}, Mark and {Marino}, Sebastian},
	doi = {10.1093/mnras/stad3462},
	eprint = {2311.04265},
	journal = {\mnras},
	month = jan,
	number = {2},
	pages = {3876-3899},
	primaryclass = {astro-ph.EP},
	title = {{The effect of sculpting planets on the steepness of debris-disc inner edges}},
	volume = {527},
	year = 2024}

@article{wisdom80,
	adsurl = {https://ui.adsabs.harvard.edu/abs/1980AJ.....85.1122W},
	author = {{Wisdom}, J.},
	doi = {10.1086/112778},
	journal = {\aj},
	month = aug,
	pages = {1122-1133},
	title = {{The resonance overlap criterion and the onset of stochastic behavior in the restricted three-body problem}},
	volume = {85},
	year = 1980}

@article{goldreich04,
	adsurl = {https://ui.adsabs.harvard.edu/abs/2004ARA&A..42..549G},
	archiveprefix = {arXiv},
	author = {{Goldreich}, Peter and {Lithwick}, Yoram and {Sari}, Re'em},
	doi = {10.1146/annurev.astro.42.053102.134004},
	eprint = {astro-ph/0405215},
	journal = {\araa},
	month = sep,
	number = {1},
	pages = {549-601},
	primaryclass = {astro-ph},
	title = {{Planet Formation by Coagulation: A Focus on Uranus and Neptune}},
	volume = {42},
	year = 2004}

@article{matra19,
	adsurl = {https://ui.adsabs.harvard.edu/abs/2019AJ....157..135M},
	archiveprefix = {arXiv},
	author = {{Matr{\`a}}, L. and {Wyatt}, M.~C. and {Wilner}, D.~J. and {Dent}, W.~R.~F. and {Marino}, S. and {Kennedy}, G.~M. and {Milli}, J.},
	doi = {10.3847/1538-3881/ab06c0},
	eid = {135},
	eprint = {1902.04081},
	journal = {\aj},
	month = apr,
	number = {4},
	pages = {135},
	primaryclass = {astro-ph.EP},
	title = {{Kuiper Belt-like Hot and Cold Populations of Planetesimal Inclinations in the {\ensuremath{\beta}} Pictoris Belt Revealed by ALMA}},
	volume = {157},
	year = 2019}

@article{marino26,
	adsurl = {https://ui.adsabs.harvard.edu/abs/2026A&A...705A.195M},
	archiveprefix = {arXiv},
	author = {{Marino}, S. and {Matr{\`a}}, L. and {Hughes}, A.~M. and {Ehrhardt}, J. and {Kennedy}, G.~M. and {del Burgo}, C. and {Brennan}, A. and {Han}, Y. and {Jankovic}, M.~R. and {Lovell}, J.~B. and {Mac Manamon}, S. and {Milli}, J. and {Weber}, P. and {Zawadzki}, B. and {Bendahan-West}, R. and {Fehr}, A. and {Mansell}, E. and {Olofsson}, J. and {Pearce}, T.~D. and {Bayo}, A. and {Matthews}, B.~C. and {L{\"o}hne}, T. and {Wyatt}, M.~C. and {{\'A}brah{\'a}m}, P. and {Bonduelle}, M. and {Booth}, M. and {Cataldi}, G. and {Carpenter}, J.~M. and {Chiang}, E. and {Ertel}, S. and {Hales}, A.~S. and {Henning}, Th. and {K{\'o}sp{\'a}l}, {\'A}. and {Krivov}, A.~V. and {Luppe}, P. and {MacGregor}, M.~A. and {Marshall}, J.~P. and {Mo{\'o}r}, A. and {P{\'e}rez}, S. and {Sefilian}, A.~A. and {Sepulveda}, A.~G. and {Wilner}, D.~J.},
	doi = {10.1051/0004-6361/202556489},
	eid = {A195},
	eprint = {2601.11708},
	journal = {\aap},
	month = jan,
	pages = {A195 (ARKS I)},
	primaryclass = {astro-ph.EP},
	title = {{The ALMA survey to Resolve exoKuiper belt Substructures (ARKS): I. Motivation, sample, data reduction, and results overview}},
	volume = {705},
	year = 2026}

@article{han26,
	adsurl = {https://ui.adsabs.harvard.edu/abs/2026A&A...705A.196H},
	archiveprefix = {arXiv},
	author = {{Han}, Yinuo and {Mansell}, Elias and {Jennings}, Jeff and {Marino}, Sebastian and {Hughes}, A. Meredith and {Zawadzki}, Brianna and {Fehr}, Anna and {Kittling}, Jamar and {Hou}, Catherine and {Nurmohamed}, Aliya and {Lee}, Junu and {Cheruiyot}, Allan and {Mpofu}, Yamani and {Booth}, Mark and {Booth}, Richard and {Bonduelle}, Myriam and {Brennan}, Aoife and {del Burgo}, Carlos and {Carpenter}, John M. and {Cataldi}, Gianni and {Chiang}, Eugene and {Ertel}, Steve and {Henning}, Thomas and {Jankovic}, Marija R. and {K{\'o}sp{\'a}l}, {\'A}gnes and {Krivov}, Alexander V. and {Lovell}, Joshua B. and {Luppe}, Patricia and {MacGregor}, Meredith A. and {Mac Manamon}, Sorcha and {Marshall}, Jonathan P. and {Matr{\`a}}, Luca and {Milli}, Julien and {Mo{\'o}r}, Attila and {Olofsson}, Johan and {Pearce}, Tim and {P{\'e}rez}, Sebasti{\'a}n and {Sefilian}, Antranik A. and {Weber}, Philipp and {Wilner}, David J. and {Wyatt}, Mark C.},
	doi = {10.1051/0004-6361/202556450},
	eid = {A196},
	eprint = {2601.13670},
	journal = {\aap},
	month = jan,
	pages = {A196 (ARKS II)},
	primaryclass = {astro-ph.EP},
	title = {{The ALMA survey to Resolve exoKuiper belt Substructures (ARKS): II. The radial structure of debris discs}},
	volume = {705},
	year = 2026}

@article{zawadzki26,
	adsurl = {https://ui.adsabs.harvard.edu/abs/2026A&A...705A.197Z},
	archiveprefix = {arXiv},
	author = {{Zawadzki}, B. and {Fehr}, A. and {Hughes}, A.~M. and {Mansell}, E. and {Kittling}, J. and {Han}, Y. and {Hou}, C. and {Pan}, M. and {Milli}, J. and {Olofsson}, J. and {Pearce}, T. and {Sefilian}, A.~A. and {Nurmohamed}, A. and {Lee}, J. and {Mpofu}, Y. and {Bonduelle}, M. and {Booth}, M. and {Brennan}, A. and {del Burgo}, C. and {Carpenter}, J.~M. and {Cataldi}, G. and {Chiang}, E. and {Ertel}, S. and {Henning}, Th. and {Jankovic}, M.~R. and {Kennedy}, G.~M. and {K{\'o}sp{\'a}l}, {\'A}. and {Krivov}, A.~V. and {Lovell}, J.~B. and {Luppe}, P. and {MacGregor}, M.~A. and {Mac Manamon}, S. and {Marino}, S. and {Marshall}, J.~P. and {Matr{\`a}}, L. and {Mo{\'o}r}, A. and {P{\'e}rez}, S. and {Weber}, P. and {Wilner}, D.~J. and {Wyatt}, M.~C.},
	doi = {10.1051/0004-6361/202556505},
	eid = {A197},
	eprint = {2601.12128},
	journal = {\aap},
	month = jan,
	pages = {A197 (ARKS III)},
	primaryclass = {astro-ph.EP},
	title = {{The ALMA survey to Resolve exoKuiper belt Substructures (ARKS): III. The vertical structure of debris disks}},
	volume = {705},
	year = 2026}

@article{ida92,
	adsurl = {https://ui.adsabs.harvard.edu/abs/1992Icar...96..107I},
	author = {{Ida}, Shigeru and {Makino}, Junichiro},
	doi = {10.1016/0019-1035(92)90008-U},
	journal = {\icarus},
	month = mar,
	number = {1},
	pages = {107-120},
	title = {{N-Body simulation of gravitational interaction between planetesimals and a protoplanet . I. velocity distribution of planetesimals}},
	volume = {96},
	year = 1992}

@article{ida93,
	adsurl = {https://ui.adsabs.harvard.edu/abs/1993Icar..106..210I},
	author = {{Ida}, Shigeru and {Makino}, Junichiro},
	doi = {10.1006/icar.1993.1167},
	journal = {\icarus},
	month = nov,
	number = {1},
	pages = {210-227},
	title = {{Scattering of Planetesimals by a Protoplanet: Slowing Down of Runaway Growth}},
	volume = {106},
	year = 1993}

\appendix
In Appendix \ref{sec:appendix} we show how viscous stirring in shear-dominated disks should lead to vertical Lorentzians. Appendix \ref{sec:extra_app} contains some results from the single big body, dispersion-dominated numerical integrations described in Section \ref{sec:num}. Although these $N$-body runs support our findings of vertical Lorentzians, they also exhibit a few artifacts resulting from our assumption of a single big body. 
%In Appendix \ref{sec:eb} we derive how small body eccentricities $e$ and inclinations $i$ grow with time when the perturbing big bodies have eccentricities $e_{\rm b} > e$, and how our results in Figs.~\ref{fig:locus} and \ref{fig:locus_Ncol} may change as a consequence.

%%%%%%%%%%%%%%%%%%%%%%%%%%%%%%%%%%%%%%%%%%%%%%%%%%%%%%%%%%%%%%%%%%%%%%%%%%%%%%%%
\vspace{0.2in}
\section{Vertical scattering in shear-dominated particle disks}\label{sec:appendix}
%Following Pearce et al.~(P24) we measure chaotic zone dimensions in units of the big body Hill radius. 
%We work in dimensionless units where $G = M_\star = a_{\rm b} = \Omega_{\rm b} = 1$ (subscript b for big body perturber). 

\begin{figure}[H]
\centering
\includegraphics[width=1.0\linewidth]{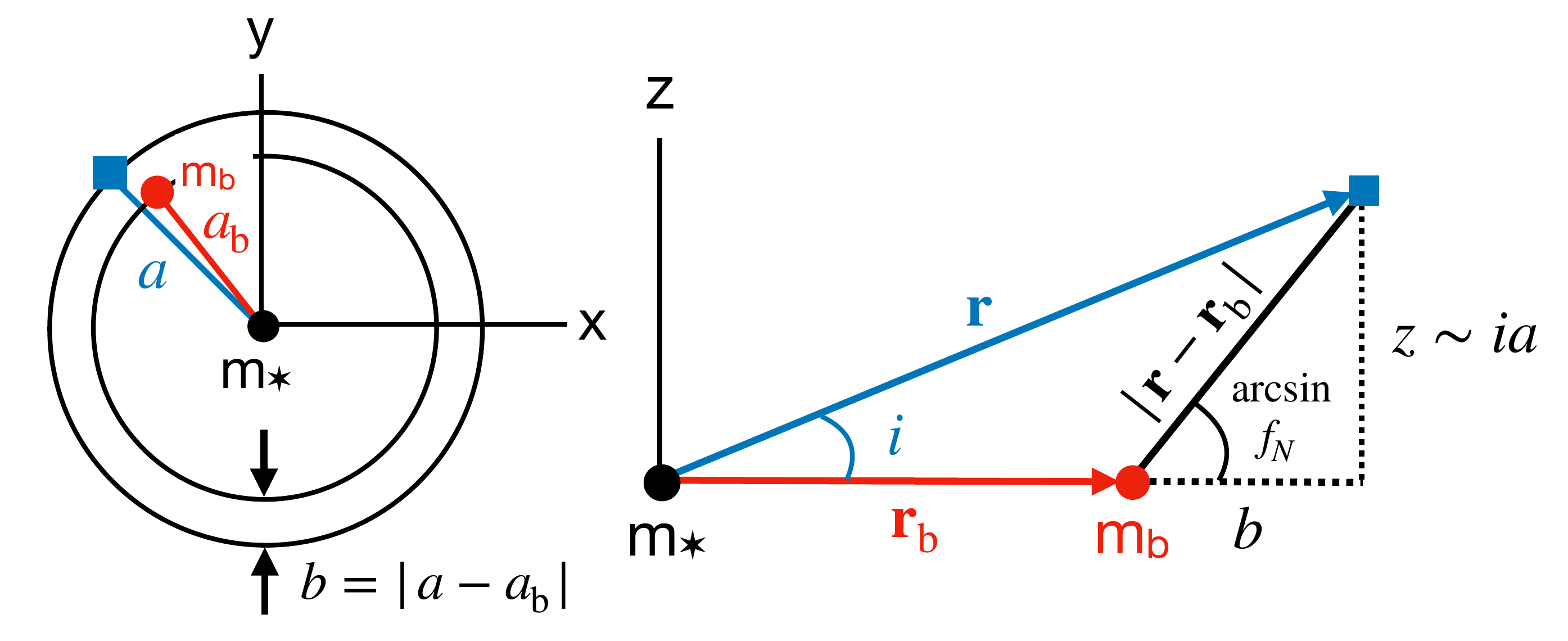}
  \caption{Same as Figure \ref{fig:picture}, but now for a shear-dominated configuration between a big body of mass $m_{\rm b}$ (red circle) and a small body (test particle, blue square), both on circular orbits separated in semimajor axis by $b = |a-a_{\rm b}|$. The bodies have similar semimajor axes $a \gg b$. At the moment of encounter (conjunction), the small body feels a vertical gravitational acceleration which is less than the full gravitational acceleration by a factor of $|f_N|$, by definition. To order of magnitude, $|f_N| \sim ia/b$. \label{fig:appendix}}
\end{figure}

We show that the same $|\Delta i| \propto i$ scaling for scatterings in a dispersion-dominated, out-of-equipartition disk holds in a shear-dominated disk, where bodies are not on crossing orbits and their relative velocities are controlled by the background Keplerian shear instead of by eccentricities and inclinations. Figure \ref{fig:appendix} shows the setup: a small body (test particle) and a big body perturber of mass $m_{\rm b}$, separated in semimajor axis by $b \equiv |a_{\rm b} - a| \ll a$. We measure $b \equiv FR_{\rm H}$ in units of the big body's Hill radius $R_{\rm H} = [m_{\rm b}/(3m_\star)]^{1/3} a_{\rm b} = (\mu_{\rm b}/3)^{1/3} a_{\rm b}$ with $F > 1$, and  define a Hill eccentricity and inclination,  $e_{\rm H} = i_{\rm H} \equiv (\mu_{\rm b}/3)^{1/3}$.

%The small body (test particle) is assumed to be at the edge of the big body's chaotic zone, displaced in semimajor axis from the big body of mass $\mu$ by $|a-a_{\rm b}| \equiv x = F R_{\rm H} \ll 1$, where the Hill radius $R_{\rm H} = (\mu/3)^{1/3}$ and $F > 1$ is an order-unity constant ($F \sim 3$ from P24). The time between successive conjunctions is
%\begin{equation}
%t_{\rm conj} = \frac{2\pi}{(3/2)x} = \frac{4\pi}{3x} \gg 1\,.
%\end{equation}

The relative shearing velocity between the bodies is $v_{\rm rel} \simeq (3/2) n b$, where $n$ is the local mean motion (orbital angular frequency). Under the same impulse approximation that we used to write down eq.~(\ref{eq:kickoom}), every close encounter (conjunction) changes the small body eccentricity by:
\begin{align}
\Delta e \sim \pm \frac{1}{v_{\rm K}} \cdot \frac{Gm_{\rm b}}{b^2} \cdot \frac{2b}{v_{\rm rel}} \sim \pm \frac{4}{F^2} e_{\rm H}\,.
\end{align}
The inclination change is derived the same way except that it is downweighted by the factor $f_N$, whose magnitude is of order $ia$ (vertical epicyclic excursion) divided by $b$ (horizontal separation; see Fig.~\ref{fig:appendix}):
\begin{align}
\Delta i & \sim \pm \,\Delta e \cdot f_{N} \sim \pm \frac{4}{F^3} i
\end{align}
valid for $ia < b$ or equivalently $i < F i_{\rm H}$ (sub-Hill inclinations). Thus $\Delta i \propto \pm i$, and repeated encounters (assuming they are uncorrelated; for a single big body this requires $F \lesssim 3$ for the small body to stay within the big body's chaotic zone) should lead to diffusion in $\log i$ and by extension a vertical Lorentzian profile. Eccentricities should grow faster than inclinations insofar as $\Delta e > \Delta i$, as appears borne out at early times for our $N$-body \texttt{Runs E} and \texttt{EE} (Table \ref{tab:paras}).
We could go further to explore shear-dominated stirring rates (e.g.~\citealt{ida92,ida93}; \citealt{goldreich04}), but are unmotivated to do so as Jankovic et al.~(2026) have shown that debris disks could not have been stirred to their present states under shear-dominated conditions.

{\section{Numerical integrations with a single big body}\label{sec:extra_app}

Figures \ref{fig:dNdz_twopanel_28}--\ref{fig:params_28} document \texttt{Run A}, whose results parallel those of \texttt{Run BB} (Figs.~\ref{fig:dNdz_twopanel_17}--\ref{fig:params_17}; see Table \ref{tab:paras} for parameters, and captions for details). Note how in Fig.~\ref{fig:params_28} the test particle disk is becoming globally eccentric, conforming to the eccentric orbit of the big body. This is a consequence of our using only a single eccentric big body in \texttt{Run A} to perturb the small bodies  (if there were instead many big bodies on crossing orbits with randomly oriented apsides, the small bodies would be perturbed into an axisymmetric disk).

In the other single big body runs \texttt{C}, \texttt{D}, and \texttt{E}, test particle inclinations are anti-correlated with eccentricities (data not shown). This follows from the Jacobi constant being conserved when there is only a single big body with $e_{\rm b}=0$ \citep{murray99}. Such an anti-correlation is not present in our other runs where small bodies are stirred by eccentric or multiple big bodies. The anti-correlation does not affect the main finding of all our numerical integrations, that vertical Lorentzians manifest when $e, e_{\rm b} \gg 2i$, and vertical Gaussians when $e, e_{\rm b} \simeq 2i$.

\begin{figure}[H]
\centering
\includegraphics[width=0.5\linewidth]{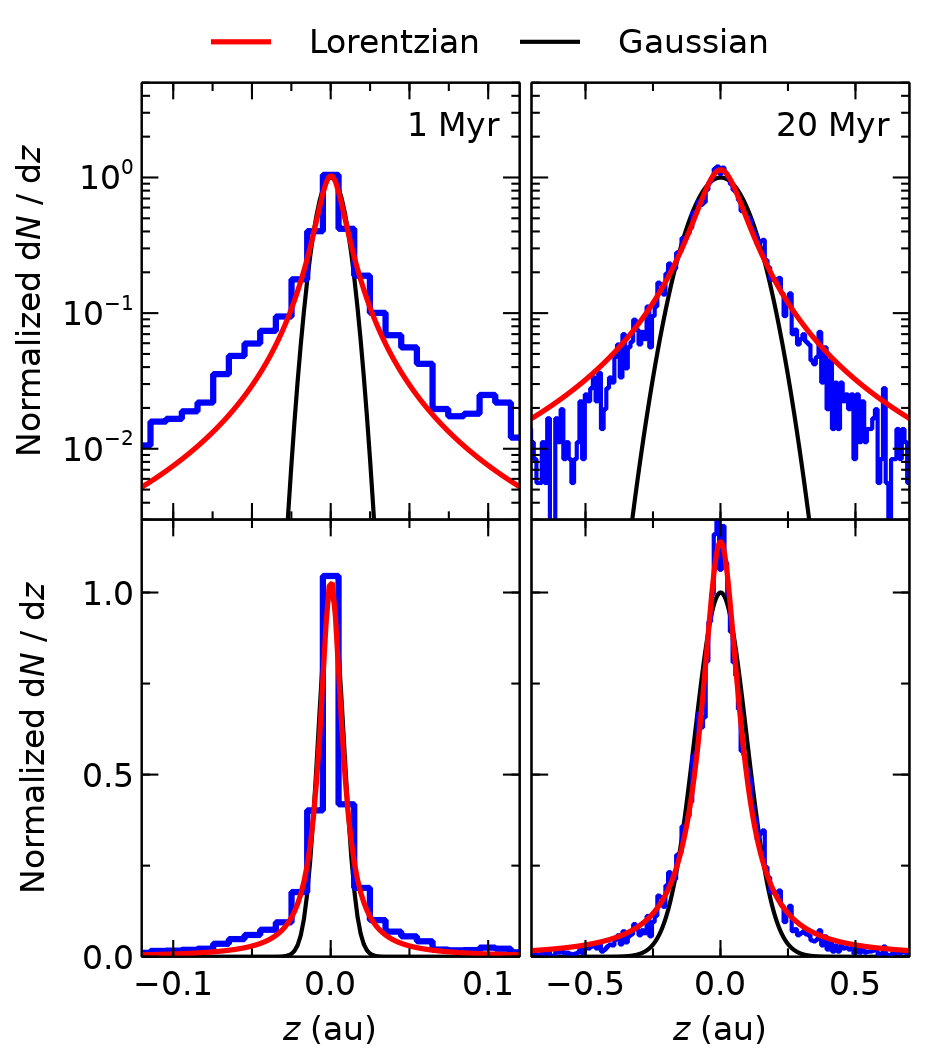}
\caption{Vertical density profiles of test particles (a.k.a.~small bodies) simulated with \texttt{REBOUND} in \texttt{Run A} at $t = 1$ Myr (left panel) and $20$ Myr (right). \texttt{Run A} features a single big body with a large eccentricity ($e_{\rm b} = 0.1$) and small initial mutual inclination with test particles ($10^{-5}$ rad); for most of the evolution, in-plane relative velocities between the big body and small bodies well exceed out-of-plane relative velocities. We expect under such circumstances that small bodies take inclination steps $\Delta i \propto \pm \,i$ (Section \ref{sec:theory}) and that their vertical distributions (blue histograms) resemble Lorentzians (red curves) more than Gaussians (black curves). 
Test particles having $i < 10^{-4}$ (constituting 55\% of all test particles at $t = 1$ Myr, and 0.15\% at 20 Myr) have been omitted as these reflect too strongly our assumed initial inclinations of $10^{-5}$ rad). The horizontal axis changes scale by a factor of 5 between left and right columns as the disk thickens vertically with time.
  \label{fig:dNdz_twopanel_28}}
\end{figure}

\begin{figure}[H]
\centering
\includegraphics[width=0.5\linewidth]{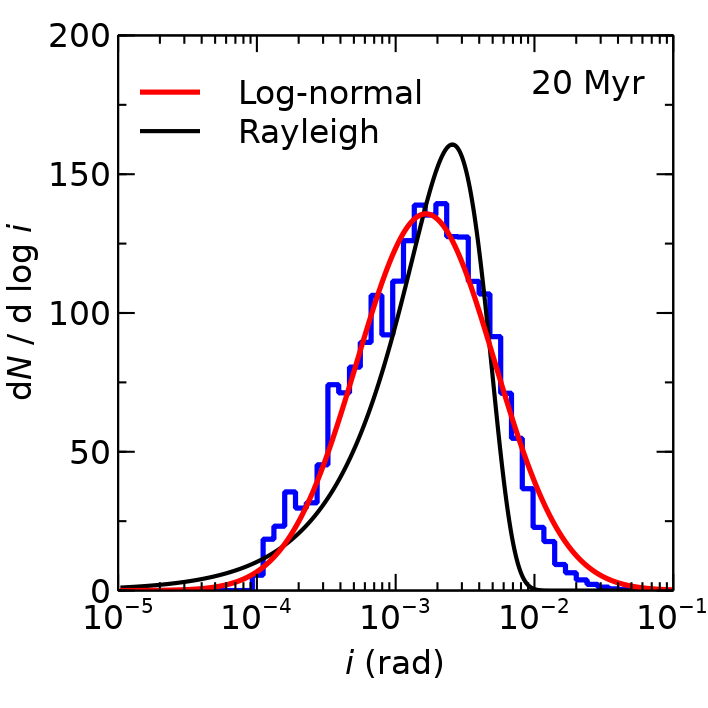}
\caption{Inclination distribution of test particles for \texttt{Run A} (single big body with $e_{\rm b} = 0.1$, test particles on initially circular orbits) at $t = 20$ Myr (blue histogram, unnormalized). Overlaid are a best-fit log normal distribution (red curve) and a Rayleigh distribution (black curve). The log normal fits better, as expected when in-plane relative velocities between the big body and test particles greatly exceed out-of-plane relative velocities, and small bodies random walk with fixed steps in $\Delta \ln i$ (Section \ref{sec:theory}). 
%Particle inclinations at $t = 0$ are $i_{\rm init} = 10^{-5}$, and the low-$i$ tail at $t = 20$ Myr reflects this initial condition which has not been completely forgotten.
  \label{fig:dNdi_28}}
\end{figure}

\begin{figure}[H]
\centering
\includegraphics[width=0.5\linewidth]{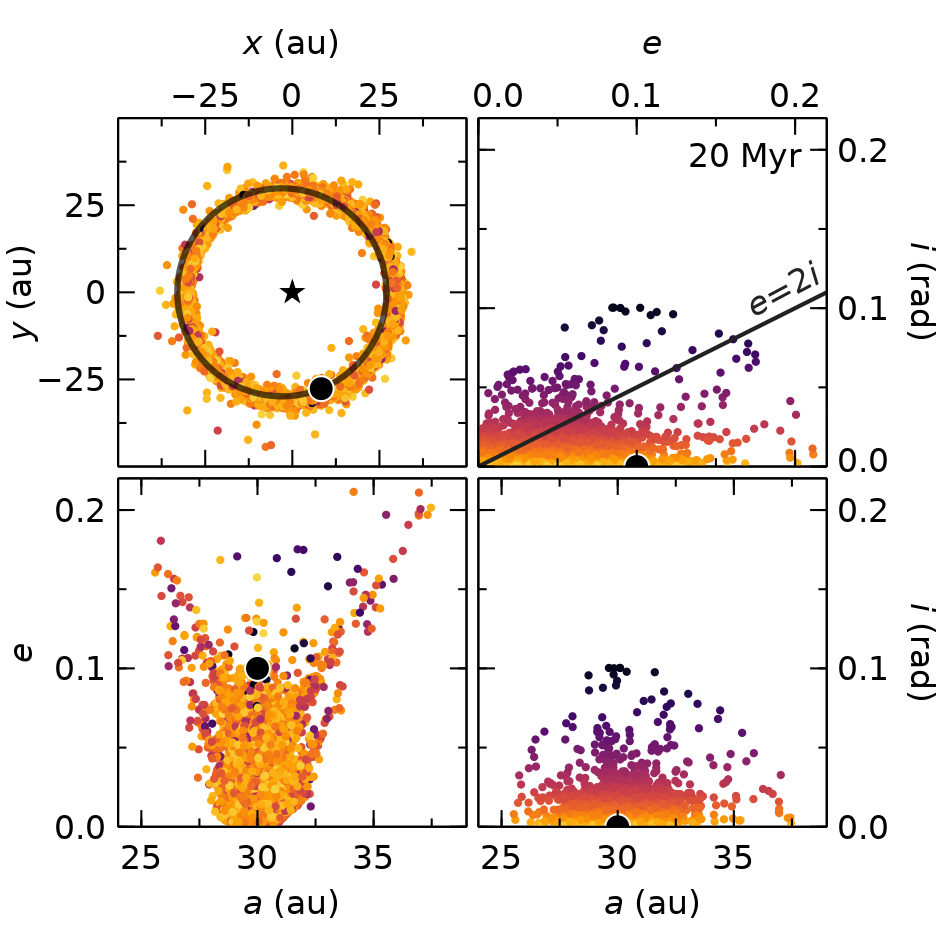}
\caption{Orbital elements for \texttt{Run A} (single big body with $e_{\rm b} = 0.1$, test particles on initially circular orbits) at $t = 20$ Myr. In all panels, test particles are represented by colored points, and the big body is represented by a black disc. {\it Top left}: Snapshot of test particles, the big body and its orbit, and the host star in the $x$-$y$ plane. Note how the test particles have been sculpted into a globally eccentric disk, apsidally aligned with the single eccentric big body. {\it Top right}: Inclinations $i$ vs.~eccentricities $e$ for the big body and test particles. The test particle points are colored according to their inclinations; the same color scheme is used for all panels. The $e=2i$ line represents equipartition between in-plane and out-of-plane motions; most bodies have $i < e/2$, a requirement for $|\Delta i| \propto i$ (Section \ref{sec:theory}). {\it Bottom left}: Eccentricities $e$ vs.~semimajor axes $a$. {\it Bottom right}: Inclinations $i$ vs.~semimajor axes $a$.
  \label{fig:params_28}}
\end{figure}

\end{document}